\documentclass[aps,showpacs,nofootinbib,superscriptaddress]{revtex4}
\usepackage{epsf}
\usepackage{graphicx}
\usepackage{amsmath}
\def\slashchar#1{\setbox0=\hbox{$#1$}
   \dimen0=\wd0 \setbox1=\hbox{/} \dimen1=\wd1
   \ifdim\dimen0>\dimen1 \rlap{\hbox to \dimen0{\hfil/\hfil}} #1
   \else  \rlap{\hbox to \dimen1{\hfil$#1$\hfil}} / \fi}

\begin{document}

\title{ Weak Pion Production off the Nucleon } 

\author{ E. Hern\'andez} \affiliation{Grupo de F\'\i sica Nuclear,
Departamento de F\'\i sica Fundamental e IUFFyM,\\ 
Facultad de Ciencias, E-37008 Salamanca, Spain.}
\author{J.~Nieves} 
\affiliation{Departamento de
 F{\'\i}sica At\'omica, Molecular y Nuclear,\\ Universidad de Granada,
 E-18071 Granada, Spain}
\author{M.~Valverde} 
\affiliation{Departamento de
 F{\'\i}sica At\'omica, Molecular y Nuclear,\\ Universidad de Granada,
 E-18071 Granada, Spain}


\begin{abstract}
  We develop a model for the weak pion production off the nucleon,
  which besides the Delta pole mechanism (weak excitation of the
  $\Delta(1232)$ resonance and its subsequent decay into $N\pi$),
  includes also some background terms required by chiral symmetry.  We
  re-fit the $C_5^A(q^2)$ form factor to the flux averaged $\nu_\mu p
  \to \mu^-p\pi^+$ ANL $q^2-$differential cross section data, finding
  a substantially smaller contribution of the Delta pole mechanism
  than traditionally assumed in the literature.  Within this scheme,
  we calculate several differential and integrated cross sections,
  including pion angular distributions, induced by neutrinos and
  antineutrinos and driven both by charged and neutral currents.  In
  all cases we find that the background terms produce quite
  significant effects and that they lead to an overall improved
  description of the data, as compared to the case where only the Delta pole
  mechanism is considered. We also show that the interference
  between the Delta pole and the background terms produces
  parity-violating contributions to the pion angular differential
  cross section, which are intimately linked to $T-$odd correlations
  in the contraction between the leptonic and hadronic
  tensors. However, these latter correlations do not imply a genuine
  violation of time reversal invariance because of the existence of
  strong final state interaction effects.

\end{abstract}

\maketitle

\section{Introduction}

The pion production processes from nucleons and nuclei at intermediate
energies are important tools to study the hadronic structure and have
become very important in the analysis of the neutrino oscillation
experiments with atmospheric neutrinos. The energy spectrum of
atmospheric neutrino at Kamioka~\cite{kamio} is such that the weak
pion production contributes about 20\% of the quasielastic lepton
production and it is a major source of uncertainty in the
identification of electron and muon events. In particular, the Neutral
Current (NC) $\pi^0$ production contributes to the background of
$e^{\pm}$ production while $\pi^{\pm}$ contributes to the background
of $\mu^{\pm}$ production. This is because both particles, i.e.
$\pi^0$ and $e^{\pm}$ or $\pi^{\pm}$ and $\mu^{\pm}$ produce similar
single--ring events in Cherenkov detectors, commonly used in neutrino
oscillation experiments. Moreover, the neutral current $\pi^0$
production might play an important role in distinguishing between the
two oscillation mechanisms $\nu_\mu \to \nu_\tau$ and $\nu_\mu \to
\nu_{\rm sterile}$~\cite{ViS98}. 
These comments apply also for $\nu_e$ appearance experiments such as
K2K~\cite{K2K} and MiniBooNE~\cite{Boone}.

The neutrino oscillation experiments are generally performed with
detectors which use material with nuclei like $^{12}$C, $^{16}$O, etc.
as targets. It is therefore desirable that nuclear medium effects be
studied in the production of leptons and pions induced by the
atmospheric as well as accelerator neutrinos used in these oscillation
experiments. To this end, the starting point should be a correct
understanding of the reaction mechanisms in the free space. In this
context we study in this work, the weak pion production off the
nucleon driven both by Charged Currents (CC) and NC at intermediate
energies. The model derived here will allow us to extend the results of
Refs.~\cite{ccjuan1,ccjuan2} for CC and Ref.~\cite{ncjuan} for NC
driven neutrino-nucleus reactions in the quasielastic region, to
higher excitation energies above the pion production threshold 
up to the $\Delta(1232)$ peak.

In the past, there have been several studies of the weak pion
production off the nucleon at intermediate
energies~\cite{adler}--\cite{LPP06}. Most of them describe the pion
production process at intermediate energies\footnote{Higher resonance
effects, which are certainly important for energies larger than those
considered in this work, are carefully discussed in
Ref.~\cite{LPP06}.}  by means of the Delta pole ($\Delta P$) mechanism
(weak excitation of the $\Delta(1232)$ resonance and its subsequent
decay into $N\pi$) and do not incorporate any background terms.  Here,
we have also included some background terms, required by chiral
symmetry.  Starting from a SU(2) non-linear $\sigma$ model involving
pions and nucleons, which implements the pattern of spontaneous chiral
symmetry breaking of QCD, we derive the corresponding vector and axial
currents (up to order ${\cal O}(1/f_\pi^3)$) which determine the
structure of the chiral non-resonant terms. Some background terms were
also considered in Refs.~\cite{FN79,FN80} and~\cite{SUL03}. In the
latter reference, the chiral counting was broken to account explicitly
for $\rho$ and $\omega$ exchanges in the $t-$channel, while  the
first two works are not consistent with the chiral counting either,
since contact terms were not included. Moreover in  ~\cite{FN79,FN80} 
a rather small axial mass ($\approx 0.65$ GeV) was used.

We will show that the background terms produce quite significant
effects, which will require to re-adjust the $C_5^A(q^2)$ form--factor
that controls the largest term of the $\Delta-$axial contribution. 
We will find corrections of the order of 30\% to the off diagonal
Goldberger-Treiman relation when the Argonne  bubble chamber
cross section~\cite{anl} are fitted. Such corrections would be smaller if the
Brookhaven bubble chamber data~\cite{bnlviejo} were
considered.  We will also show that interference between the $\Delta
P$ and the background terms produces parity-violating contributions to
the pion angular differential cross section, which are intimately
linked to $T-$odd correlations in the contraction between the leptonic
and hadronic tensors. However, these $T-$odd correlations do not imply
a genuine violation of time reversal invariance because of the
existence of strong final state interaction effects.

The paper is organized as follows. After this introduction, in
Sect.~\ref{sec:ccna} the model for CC neutrino and antineutrino
induced reactions is presented. There, some general definitions
involving kinematics and differential cross sections are given
(Subsect.~\ref{sec:kin}). The consequences of isospin symmetry are
exploited in Subsect.~\ref{sec:isospin}, while in the next subsection
the model for the $W N \to N' \pi$ reaction is presented. In
Sect.~\ref{sec:nc-an}, the findings of the latter section are extended
to the case of NC driven processes. Results are presented and
discussed in Sect.~\ref{sec:resul} and the main conclusions of this
work can be found in Sect.~\ref{sec:concl}. Besides in
Appendix~\ref{sec:phi}, the cross section dependence on the pion
azimuthal angle is discussed in terms of Lorentz, parity and
time--reversal invariances, and finally in Appendix~\ref{app:phases}, 
we discuss in some detail the
effects on the neutrino and antineutrino induced cross sections of
different relative signs between the axial and vector $W^{\pm}\Delta N$
form-factors, and between the resonant and chiral non-resonant contributions.

\section{CC Neutrino and Antineutrino induced reactions}

\label{sec:ccna}
\subsection{Kinematics and differential cross section}

\label{sec:kin}

We will focus on the neutrino--pion production reaction off the
nucleon driven by charged currents,
\begin{equation}
  \nu_l (k) +\, N(p)  \to l^- (k^\prime) + N(p^\prime) +\, \pi(k_\pi) 
\label{eq:reac}
\end{equation}
though the generalization of the obtained expressions to antineutrino
induced reactions is straightforward.

\begin{figure}[tbh]
\centerline{\includegraphics[height=4.05cm]{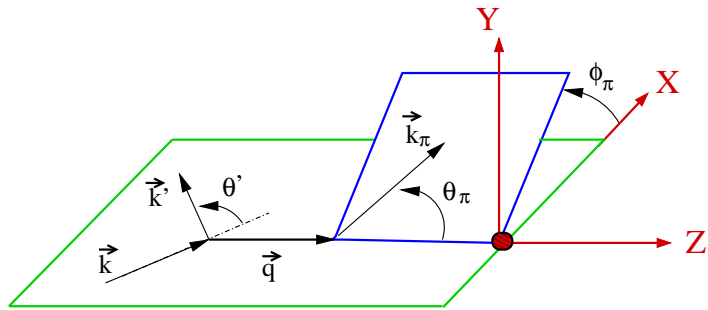}}
\caption{\footnotesize Definition of the different kinematical variables
used through this work.}\label{fig:coor}
\end{figure}

The unpolarized differential cross section, with respect to the
outgoing lepton and pion kinematical variables, is given in the
Laboratory (LAB) frame (kinematics is sketched in Fig~\ref{fig:coor})
by\footnote{To obtain Eq.~(\ref{eq:sec}) we have neglected the
four-momentum carried out by the intermediate $W-$boson with respect
to its mass $(M_W)$, and have used the existing relation between the
gauge weak coupling constant, $g = e/\sin \theta_W$, and the Fermi
constant: $G/\sqrt 2 = g^2/8M^2_W$, with $e$ the electron charge,
$\theta_W$ the Weinberg angle and $M_W$ the
$W-$boson mass.}
\begin{equation}
\frac{d^{\,5}\sigma_{\nu_l
    l}}{d\Omega(\hat{k^\prime})dE^\prime d\Omega(\hat{k}_\pi) } =
    \frac{|\vec{k}^\prime|}{|\vec{k}~|}\frac{G^2}{4\pi^2}
    \int_0^{+\infty}\frac{d|\vec{k}_\pi| |\vec{k}_\pi|^2}{E_\pi}
    L_{\mu\sigma}^{(\nu)}\left(W^{\mu\sigma}_{{\rm CC}
    \pi}\right)^{(\nu)} \label{eq:sec}
\end{equation}
with $\vec{k}$ and $\vec{k}^\prime~$ the LAB lepton momenta,
$E^{\prime} = (\vec{k}^{\prime\, 2} + m_l^2 )^{1/2}$ and $m_l$ the
energy and the mass of the outgoing lepton ($m_\mu = 105.65$ MeV, $m_e
= 0.511$ MeV ), $G=1.1664\times 10^{-11}$ MeV$^{-2}$, the Fermi
constant, $\vec{k}_\pi$ and $E_\pi= (\vec{k}^{2}_\pi + m_\pi^2
)^{1/2}$ the LAB momentum and energy of the outgoing pion\footnote{For
$m_\pi$, we use the isospin averaged pion mass.}, and $L$ and $W$ the
leptonic and hadronic tensors, respectively. The leptonic tensor is
given by (in our convention, we take $\epsilon_{0123}= +1$ and the
metric $g^{\mu\nu}=(+,-,-,-)$):
\begin{eqnarray}
L_{\mu\sigma}^{(\nu)}&=& (L^{(\nu)}_s)_{\mu\sigma}+ {\rm i}
 (L^{(\nu)}_a)_{\mu\sigma} =
 k^\prime_\mu k_\sigma +k^\prime_\sigma k_\mu
- g_{\mu\sigma} k\cdot k^\prime + {\rm i}
\epsilon_{\mu\sigma\alpha\beta}k^{\prime\alpha}k^\beta \label{eq:lep}
\end{eqnarray}
and it is not orthogonal to $q^\mu$ even for massless neutrinos, i.e,
$L_{\mu\sigma}^{(\nu)} q^\mu = -m^2_l k_\sigma$.

The hadronic tensor includes all sort of non-leptonic
vertices and it reads
\begin{eqnarray}
(W^{\mu\sigma}_{{\rm CC} \pi})^{(\nu)} &=&
 \frac{1}{4M}\overline{\sum_{\rm spins}}
 \int\frac{d^3p^\prime}{(2\pi)^3} \frac{1}{2E^\prime_N}
  \delta^4(p^\prime+k_\pi-q-p) \langle N^\prime \pi |
 j^\mu_{\rm cc+}(0) | N \rangle \langle N^\prime \pi | j^\sigma_{\rm
 cc+}(0) | N \rangle^*
\label{eq:wmunu}
\end{eqnarray}
with $M$ the nucleon mass\footnote{We take the average of the neutron
and proton masses.}, $q=k-k^\prime$ and $E^\prime_N$ the energy of the
outgoing nucleon.  The bar over the sum of initial and final spins,
denotes the average on the initial ones. As for the one particle
states, they are normalized so that $\langle \vec{p}\, |
\vec{p}^{\,\prime} \rangle = (2\pi)^3 2p_0
\delta^3(\vec{p}-\vec{p}^{\,\prime})$, and finally for the charged
current which couples to the $W^+$ we take
\begin{equation}
j^\mu_{\rm cc+} = \bar{\Psi}_u\gamma^\mu(1-\gamma_5)(\cos\theta_C \Psi_d +
\sin\theta_C \Psi_s) 
\end{equation}
with $\Psi_u$, $\Psi_d$ and $\Psi_s$ quark fields, and $\theta_C$ the
Cabibbo angle ($\cos\theta_C= 0.974$). Note that with all these
definitions, the matrix element $\langle N^\prime \pi | j^\mu_{\rm
cc}(0) | N \rangle$ is dimensionless.  After performing the
$d^3p^\prime$ integration, there will still be left 
an energy conserving Dirac's delta function
($\delta(p^{\prime\,0}+k_\pi^0-q^0-p^0) $) in the
hadronic tensor, which can be used to perform the $d|\vec{k}_\pi|$ integration
in Eq.~(\ref{eq:sec}).  Since the quantity $\int d\Omega_\pi
L_{\mu\sigma}^{(\nu)}\left(W^{\mu\sigma}_{{\rm CC} \pi}\right)$ is a
scalar to evaluate it we take for convenience $\vec{q}$ in the $Z$
direction. Referring now the pion variables to the outgoing $\pi N$
pair Center of Mass (CM) frame (as it is usual in pion electroproduction)
would be readily done by means of a boost in the $Z$ direction. Note
that the azimuthal angle $\phi_\pi$ is left unchanged by such a boost.

By construction, the hadronic tensor accomplishes
\begin{eqnarray}
\left (W^{\mu\sigma}_{{\rm CC} \pi}\right)^{(\nu)}= 
\left (W^{\mu\sigma}_{{\rm CC} \pi}\right)^{(\nu)}_s + {\rm i}
\left(W^{\mu\sigma}_{{\rm CC} \pi}\right)^{(\nu)}_a 
\end{eqnarray}
with $\left(W^{\mu\sigma}_{{\rm CC} \pi}\right)^{(\nu)}_s$
and $\left (W^{\mu\sigma}_{{\rm CC} \pi}\right)^{(\nu)}_a$ real
symmetric  and antisymmetric parts, respectively.

As it is explicitly shown in  Appendix~\ref{sec:phi}, Lorentz
invariance restricts the $\phi_\pi$ dependence,
\begin{eqnarray}
\frac{d^{\,5}\sigma_{\nu_l
    l}}{d\Omega(\hat{k^\prime})dE^\prime d\Omega(\hat{k}_\pi) } &=&
\frac{|\vec{k}^\prime|}{|\vec{k}~|}\frac{G^2}{4\pi^2}
     \left \{ A +
    B \cos\phi_\pi + C \cos 2\phi_\pi+
    D \sin\phi_\pi + E \sin 2\phi_\pi \right\} \label{eq:phipi}
\end{eqnarray}
with $A,B,C,D$ and $E$  real, structure functions,
which depend on $q^2,\, p\cdot q$, $k_\pi\cdot q$ and $k_\pi\cdot p $.

For antineutrino induced reactions we have 
\begin{equation}
L_{\mu\sigma}^{(\bar\nu)} = L_{\sigma\mu}^{(\nu)} \label{eq:anti}
\end{equation}
and we will discuss below the existing relation between the matrix
elements of $j^\mu_{\rm cc+}$ and $j^\mu_{\rm cc-}=
j^{\mu\dagger}_{\rm cc+}$, charged currents which couple to the $W^+$
and $W^-$ bosons, respectively.

\subsection{Isospin relations}
\label{sec:isospin}
The non-strange parts
of $j^\mu_{cc\pm}$ behave as the spherical $\pm 1$ component of an
isovector, since 
\begin{equation}
\bar \Psi_u \Psi_d = -\bar \Psi_q\, \frac{\tau_{+1}^1}{\sqrt 2}\, 
\Psi_q, \qquad
\bar \Psi_ d \Psi_u
= \bar \Psi_q\, \frac{\tau_{-1}^1}{\sqrt 2}\, \Psi_q,   \qquad \Psi_q=
\left (\begin{array}{c}\Psi_u\cr \Psi_d\end{array}\right ), \qquad
  \tau_0^1=\tau_z, \tau^1_{\pm 1} = \mp \frac{\tau_x  \pm {\rm i}
    \tau_y}{\sqrt{2}}
\end{equation}
with $\vec\tau$ the Pauli matrices.  Thanks to the Wigner-Eckart's theorem, we find that all $\langle
N^\prime \pi | j^\mu_{\rm cc\pm}(0) | N \rangle$ matrix elements are
just determined by two of them, f.i. $\langle
p \pi^+ | j^\mu_{\rm cc+}(0) | p \rangle$ and $\langle
n \pi^+ | j^\mu_{\rm cc+}(0) | n \rangle$,
\begin{eqnarray}
\langle p \pi^0 | j^\mu_{\rm cc+}(0) | n \rangle &=& - \frac{1}{\sqrt 2}
\left [ \langle
p \pi^+ | j^\mu_{\rm cc+}(0) | p \rangle -\langle
n \pi^+ | j^\mu_{\rm cc+}(0) | n \rangle\right ] \\
\langle p \pi^- | j^\mu_{\rm cc-}(0) | p \rangle &=& 
\langle n \pi^+ | j^\mu_{\rm cc+}(0) | n \rangle \label{eq:anti1}\\
\langle n \pi^- | j^\mu_{\rm cc-}(0) | n \rangle &=& 
\langle p \pi^+ | j^\mu_{\rm cc+}(0) | p \rangle \label{eq:anti2}\\
\langle n \pi^0 | j^\mu_{\rm cc-}(0) | p \rangle &=& 
-\langle p \pi^0 | j^\mu_{\rm cc+}(0) | n \rangle = \frac{1}{\sqrt 2}
\left [ \langle
p \pi^+ | j^\mu_{\rm cc+}(0) | p \rangle -\langle
n \pi^+ | j^\mu_{\rm cc+}(0) | n \rangle\right ] \label{eq:anti3}
\end{eqnarray}
Thus, Eqs.~(\ref{eq:anti}) and (\ref{eq:anti1})--(\ref{eq:anti3})
allow us to determine  all CC antineutrino cross sections 
 from the neutrino induced amplitudes.

Besides, the vector contribution of the matrix elements of the weak CC 
between $N$ and $N'\pi$ states is related to that of the
electromagnetic current $s^\mu_{\rm em}(0)$,
\begin{eqnarray}
s^\mu_{\rm em} &=& \frac23 \bar{\Psi}_u\gamma^\mu \Psi_u 
- \frac13 \bar{\Psi}_d\gamma^\mu \Psi_d 
- \frac13 \bar{\Psi}_s\gamma^\mu \Psi_s\\ 
&=& \frac16 
\bar \Psi_q\gamma^\mu\Psi_q - \frac13 \bar\Psi_s \gamma^\mu\Psi_s 
+ \frac{1}{\sqrt  2} \bar\Psi_q\,\gamma^\mu \frac{\tau_0^1}{\sqrt 2} \,\Psi_q
\end{eqnarray}
The matrix elements of the isovector part ($\tau_0^1$) are related to
 those of $j_{cc\pm}$, while the first two terms are isoscalar
 operators. One easily finds,
\begin{eqnarray}
\frac{1}{\cos\theta_C} \langle p \pi^+ | V^\mu_{\rm cc+}(0) | p \rangle &=&  \sqrt 2  \langle
n \pi^0 | s^\mu_{\rm em}(0) | n \rangle + \langle
p \pi^- | s^\mu_{\rm em}(0) | n \rangle  \\
\frac{1}{\cos\theta_C} \langle n \pi^+ | V^\mu_{\rm cc+}(0) | n \rangle &=&  \sqrt 2  \langle
p \pi^0 | s^\mu_{\rm em}(0) | p \rangle -\langle
p \pi^- | s^\mu_{\rm em}(0) | n \rangle 
\end{eqnarray}

\subsection{Model for the $W N \to N' \pi$ reaction}

\subsubsection{SU(2) non-linear $\sigma$ model}

\label{sec:sigma}

Let us start with the effective lagrangian of the SU(2) non-linear $\sigma$
model. It implements the pattern of spontaneous chiral symmetry
breaking  of QCD and it is given by
\begin{equation}
{\cal L}_{N\pi} = \bar\Psi {\rm i} \gamma^\mu \left [ \partial_\mu +
  {\cal V}_\mu \right] \Psi -M \bar\Psi \Psi + g_A
  \bar\Psi\gamma^\mu\gamma_5 {\cal A}_\mu \Psi + \frac12 {\rm Tr}
  \left [ \partial_\mu U^\dagger \partial^\mu U\right ]  \label{eq:sigma}
\end{equation}
where $\Psi= \left (\begin{array}{c}p\cr n\end{array}\right )$ is the
nucleon field. The fields ${\cal V}_\mu$ and
${\cal A}_\mu$ are given in terms of the matrix field $\xi$ derived
from the pion fields\footnote{We use a convention such that
$\phi=(\phi_x-{\rm i}\, \phi_y)/\sqrt{2}$ creates a $\pi^-$ from the
vacuum or annihilates a $\pi^+$ and the $\phi_z$ field creates or
annihilates a $\pi^0$.} $\vec{\phi}$,
\begin{equation}
{\cal V}_\mu = \frac12 \left ( \xi \partial_\mu \xi^\dagger + \xi^\dagger
\partial_\mu \xi \right )\qquad  {\cal A}_\mu = \frac{\rm i}{2} \left ( \xi
\partial_\mu \xi^\dagger - \xi^\dagger \partial_\mu \xi \right )
\end{equation}

The pions $\vec{\phi}$ are the Goldstone bosons associated to the
spontaneous breaking of the SU(2)$_V\times$SU(2)$_A$ chiral
symmetry. We describe their dynamics in terms of $2\times 2$ matrix
field $U$ given by
\begin{equation}
U = \frac{f_\pi}{\sqrt 2} e^{{\rm
    i}\,\vec{\tau}\cdot\vec{\phi}/f_\pi} = \frac{f_\pi}{\sqrt 2} \xi^2
\end{equation}
with $f_\pi \simeq 93$ MeV the pion weak decay constant. The matrix field
\begin{equation}
\xi = e^{{\rm
    i}\,\vec{\tau}\cdot\vec{\phi}/(2f_\pi)}
\end{equation}
transforms under SU(2)$_V\times$SU(2)$_A$ as
\begin{equation}
\begin{array}{c}\cr\xi~\cr \end{array}  \begin{array}{c}{\rm SU(2)}_V\cr\Longrightarrow\cr\end{array} ~
 \begin{array}{c}\cr T_V\, \xi \,T_V^\dagger\cr \end{array} \qquad  
\begin{array}{c}\cr \xi~ \cr \end{array}  \begin{array}{c}{\rm SU(2)}_A\cr\Longrightarrow\cr\end{array} ~
 \begin{array}{c}\cr T_A^\dagger\, \xi \,\Lambda^\dagger = \Lambda\, \xi \,T_A^\dagger\cr\end{array}  \label{eq:trans}
\end{equation}
where $T_V= \exp(-{\rm i}\, \vec{\tau}\cdot\vec{\theta_V})/2$ and $T_A=
\exp(-{\rm i}\, \vec{\tau}\cdot\vec{\theta_A})/2$ are global
transformations belonging  to  SU(2)$_V$ and SU(2)$_A$
respectively. As for $\Lambda= \exp(-{\rm i}\,
\vec{\tau}\cdot\vec{\theta_\Lambda})/2 $, it is a unitary matrix field that
depends on the axial transformation $T_A$ and the $\vec{\phi}$
Goldstone boson fields. 

On the other hand the nucleon field, $\Psi$ transforms as
\begin{equation}
\begin{array}{c}\cr\Psi~\cr \end{array}  \begin{array}{c}{\rm SU(2)}_V\cr\Longrightarrow\cr\end{array} ~
 \begin{array}{c}\cr T_V\, \Psi \end{array} \qquad  
\begin{array}{c}\cr \Psi~ \cr \end{array}  \begin{array}{c}{\rm SU(2)}_A\cr\Longrightarrow\cr\end{array} ~
 \begin{array}{c}\cr \Lambda\, \Psi \cr\end{array}  
\end{equation}
Each term of the effective Lagrangian of Eq.~(\ref{eq:sigma}) is
separately invariant under the chiral group
SU(2)$_V\times$SU(2)$_A$. This is why one can introduce an axial nucleon
coupling $g_A\ne 1$ in the model without violating chiral symmetry. We
will use $g_A=1.26$ through this work.

Explicit SU(2)$_A$ breaking terms are included in the model as 
\begin{equation}
  m_\pi^2\frac{f_\pi}{\sqrt 2}\frac12{\rm
  Tr}(U+U^\dagger-\sqrt{2}f_\pi) \label{eq:mpi}
\end{equation}
to give mass to the pions. Neglecting ${\cal O}(1/f_\pi^4)$, the
effective Lagrangian of Eqs.~(\ref{eq:sigma}) and~(\ref{eq:mpi}) reads
\begin{eqnarray}
{\cal L}& =& \bar\Psi [{\rm i} \slashchar{\partial}-M] \Psi+\frac12
\partial_\mu\vec{\phi}\partial^\mu\vec{\phi} -\frac12 m_\pi^2
\vec{\phi}^{\,2} + {\cal L}_{\rm int}^\sigma \label{eq:lsigma}\\ 
{\cal L}_ {\rm
int}^\sigma&=& \frac{g_A}{f_\pi} \bar\Psi \gamma^\mu\gamma_5
\frac{\vec{\tau}}{2}(\partial_\mu \vec{\phi})\Psi
-\frac{1}{4f_\pi^2}\bar\Psi \gamma_\mu \vec{\tau}\left
(\vec{\phi}\times\partial^\mu\vec{\phi}\right)\Psi -
\frac{1}{6f_\pi^2} \left ( \vec{\phi}^{\,2}
\partial_\mu\vec{\phi}\partial^\mu \vec{\phi}-(\vec{\phi}\partial_\mu
\vec{\phi} )(\vec{\phi}\partial^\mu
\vec{\phi}) \right) + \frac{m_\pi^2}{24f_\pi^2}(\vec{\phi}^{\,2})^2
\nonumber \\
\nonumber &&\\
&&-\frac{g_A}{6f_\pi^3} \bar\Psi \gamma^\mu\gamma_5
\left[\vec{\phi}^{\,2} \frac{\vec{\tau}}{2}\partial_\mu \vec{\phi} -
(\vec{\phi}\partial_\mu\vec{\phi})\frac{\vec{\tau}}{2} \vec{\phi}
\right]\Psi+ {\cal O}(\frac{1}{f_\pi^4})\label{eq:lint}
\end{eqnarray}
In contrast to the linear $\sigma$ model, the coupling between the $p$
and $n$ and the pions is of the pseudo-vector type. Writing the
coupling constant in the usual way as $g_{\pi NN}/2M = f/m_\pi$, we
recover the Goldberger--Treiman relation
\begin{equation}
f = \frac{m_\pi}{2f_\pi}g_A \label{eq:gtN}
\end{equation}
which phenomenologically is satisfied at the level of 5\%.

The vector and axial currents that we derive from the lagrangian in
Eq.~(\ref{eq:sigma}) and the transformation properties of the
fields\footnote{For infinitesimal vector and axial transformations, we
get from Eq.~(\ref{eq:trans})
\begin{equation}
\vec{\tau}\cdot \vec{\theta}_\Lambda  =
\frac{\vec{\phi}\times\vec{\tau}}{2f_\pi} \vec{\theta}_A + {\cal O}(\frac{1}{f_\pi^3})
\end{equation}
\begin{equation}
  \begin{array}{c}\cr\delta \vec{\phi}~\cr \end{array}
  \begin{array}{c}{\rm SU(2)}_V\cr~~=~~~\cr
\end{array}~\begin{array}{c}\cr (\vec{\theta}_V\times \vec{\phi}) +  {\cal
  O}(1/f_\pi^3) \cr \end{array} \qquad
  \begin{array}{c}\cr\delta \vec{\phi}~\cr \end{array}
  \begin{array}{c}{\rm SU(2)}_A\cr~~=~~~\cr
\end{array}~\begin{array}{c}\cr f_\pi \vec{\theta}_A +
    \left(\vec{\phi}(\vec{\phi}\cdot\vec{\theta}_A)-
  \vec{\theta}_A \vec{\phi}^{\,2} \right)/(3f_\pi) +  {\cal
  O}(1/f_\pi^3) \cr \end{array}  
\end{equation}
\begin{equation}
  \begin{array}{c}\cr\delta \Psi~\cr \end{array}
  \begin{array}{c}{\rm SU(2)}_V\cr~~=~~~\cr
\end{array}~\begin{array}{c}\cr -{\rm i}\,
  \frac{\vec{\tau}\cdot\vec{\theta}_V}{2} \Psi\cr \end{array} \qquad
  \begin{array}{c}\cr\delta \Psi~ \end{array}
  \begin{array}{c}{\rm SU(2)}_A\cr~~=~~~
\end{array}~\begin{array}{c}\cr -{\rm i}\,
  \frac{\left(\vec{\phi}\times\vec{\tau}\right)\cdot\vec{\theta}_A}{ 4f_\pi}\Psi+  {\cal
  O}(1/f_\pi^3)  \end{array}
\end{equation}
where $\delta\vec{\phi}$ and $\delta \Psi$ denote the infinitesimal
variations of the fields.} 
are given by
\begin{eqnarray}
{\vec V}^\mu &=& \underbrace{\vec{\phi} \times \partial^\mu
  \vec{\phi}}_{{\vec V}^\mu_a} + \underbrace{\bar\Psi
\gamma^\mu \frac{\vec\tau}{2} \Psi}_{{\vec V}^\mu_b} 
+ \underbrace{\frac{g_A}{2f_\pi}\bar\Psi
\gamma^\mu \gamma_5 (\vec{\phi}\times\vec{\tau})\Psi}_{{\vec V}^\mu_c}
  \overbrace{-
\frac{1}{4f_\pi^2} \bar\Psi \gamma^\mu\left [
  \vec{\tau}\vec{\phi}^{\,2}-
  \vec{\phi}(\vec{\tau}\cdot\vec{\phi})\right]\Psi -
\frac{\vec{\phi}^{\,2}}{3f_\pi^2}(\vec{\phi}\times \partial^\mu
\vec{\phi})}^{{\vec V}^\mu_d} + {\cal O}(\frac{1}{f_\pi^3})
  \label{eq:vcurrent}\\
{\vec A}^\mu &=& \underbrace{f_\pi \partial^\mu
  \vec{\phi}}_{{\vec A}^\mu_a} + \underbrace{g_A\bar\Psi
\gamma^\mu \gamma_5\frac{\vec\tau}{2} \Psi}_{{\vec A}^\mu_b} 
+ \underbrace{\frac{1}{2f_\pi}\bar\Psi
\gamma^\mu (\vec{\phi}\times\vec{\tau})\Psi}_{{\vec A}^\mu_c}
 + \overbrace{
\frac{2}{3f_\pi}\left[\vec{\phi} (\vec{\phi}\cdot\partial^\mu\vec{\phi})-
\vec{\phi}^{\,2}\partial^\mu\vec{\phi}\,\right]
-\frac{g_A}{4f_\pi^2} \bar\Psi \gamma^\mu\gamma_5\left [
  \vec{\tau}\vec{\phi}^{\,2}-
  \vec{\phi}(\vec{\tau}\cdot\vec{\phi})\right]\Psi }^{{\vec
  A}^\mu_d}\nonumber \\
 &+& {\cal O}(\frac{1}{f_\pi^3}) \label{eq:acurrent}
\end{eqnarray}
and determine the weak transition vertex where the $W-$boson is
absorbed. This is because these currents, up to a factor of
$\cos\theta_C$, are the hadronic realization of the electroweak quark
current $j^\mu_{cc}$ for a system of interacting pions and
nucleons. Thus, ${\vec A}^\mu_a$ and ${\vec V}^\mu_a$ account for the
$W-$decay into one and two pions, respectively, while ${\vec
A}^\mu_b$ and ${\vec V}^\mu_b$ provide the $WNN$ vector and axial
vector couplings. Besides, ${\vec A}^\mu_c$ and ${\vec V}^\mu_c$ lead
to contact $WNN\pi$ vertices and finally ${\vec A}^\mu_d$ and ${\vec
V}^\mu_d$ either contribute to processes with more than one pion in
the final state or provide loop corrections to the
leading order amplitude for one pion production.

The overall normalization is fixed by the $W^+np$ vertex
\begin{equation}
\langle p; \vec{p}^{~\prime}=\vec{p}+\vec{q}~ | j^\alpha_{cc+}(0) | n;
\vec{p}~\rangle =
\cos\theta_C\
\bar{u}(\vec{p}{~^\prime})(V^\alpha_N(q)-A^\alpha_N(q))u(\vec{p}\,) \label{eq:np}
\end{equation}
where the $u$'s are Dirac spinors for the neutron and proton,
normalized such that $\bar u u=2M$, and vector and axial nucleon
currents are given by
\begin{equation}
V_N^\alpha(q) = 2\times  \left ( F_1^V(q^2)\gamma^\alpha + {\rm
i}\mu_V \frac{F_2^V(q^2)}{2M}\sigma^{\alpha\nu}q_\nu\right), \qquad
A_N^\alpha (q)=  G_A(q^2) \times \left (  \gamma^\alpha\gamma_5 + 
\frac{\slashchar{q}}{m_\pi^2-q^2}q^\alpha\gamma_5 \right) \label{eq:axial1}
\end{equation}
being all form factors real thanks to invariance under time
reversal. Invariance under G-parity has been assumed to discard a term
of the form $(p^\mu+p^{\prime \mu})\gamma_5$  and
we have only considered the pion pole-contribution in the pseudoscalar
form factor. Isospin symmetry relates the vector form factors to the
electromagnetic ones\footnote{We use the parameterization of Galster
and collaborators~\protect\cite{Ga71}
\begin{equation}
F_1^N = \frac{G_E^N+\tau G_M^N}{1+\tau}, \qquad \mu_N F_2^N =
\frac{G_M^N- G_E^N}{1+\tau}, \quad G_E^p = \frac{G_M^p}{\mu_p}=
\frac{G_M^n}{\mu_n} = -(1+\lambda_n\tau) \frac{G_E^n}{\mu_n \tau} =
\left(\frac{1}{1-q^2/M^2_D}\right)^2 \label{eq:f1n}
\end{equation}
with $\tau=-q^2/4M^2$, $M_D=0.843$ GeV, $\mu_p=2.792847$, $\mu_n=-1.913043$ and
$\lambda_n=5.6$.} 
\begin{equation}
 F_1^V(q^2) =  \frac12 \left (F_1^p(q^2)-F_1^n(q^2)\right),\qquad
 \mu_V F_2^V(q^2) = \frac12 \left ( \mu_p F_2^p(q^2) - \mu_n F_2^n(q^2)\right) 
\end{equation}
and  the axial form-factor is given by~\cite{eweise}
\begin{equation} 
G_A(q^2) = \frac{g_A}{(1-q^2/M_A^2)^2},\quad g_A=1.26, \quad M_A =
1.05~ {\rm GeV} \label{eq:axial}
\end{equation}
Thus, one realizes that, up to an overall $-\sqrt{2}\cos\theta_C$ factor, the
$+1$ spherical component\footnote{Note,
\begin{equation}
-\sqrt 2\tau_{+1}|n\rangle =  \left(\tau_x+{\rm
 i}\tau_y\right)|n\rangle = 2 |p\rangle
\end{equation}
.}
 ($[A^\mu_b]_{+1} = - \left([A^\mu_b]_x+{\rm
i}[A^\mu_b]_y\right)/\sqrt 2$ ) of $\vec{A}^\mu_b$ gives the axial
vector contribution, at $q^2=0$, of the $W^+n\to p$ weak
transition. Besides, $-\sqrt{2}[A^\mu_a]_{+1}$ and the $\pi NN$
coupling in ${\cal L}^\sigma_{\rm int}$ lead to the $q^\mu \gamma_5$
term in Eq.~(\ref{eq:axial1}). Analogously,
$-\sqrt{2}\cos\theta_C[V^\mu_b]_{+1}$, provides the Dirac part of the
vector contribution, at $q^2=0$, of the $W^+n\to p$ weak
transition. The magnetic part in Eq.~(\ref{eq:axial1}) is not provided
by the non-linear sigma model constructed here which assumes
structureless nucleons.

From the above discussion, we conclude that
$-\sqrt{2}\cos\theta_C\left ( [V^\mu]_{+1} - [A^\mu]_{+1}\right)$
provides the $W^+-$ absorption vertex, with the appropriate
normalization, in the $\langle N^\prime \pi | j^\mu_{\rm cc+}(0)
| N \rangle$ matrix element. We will improve on that by including the
$q^2$ dependence induced by the form factors in Eq.~(\ref{eq:axial1})
and adding the magnetic contribution ($F_2^V$ term) to the vector part
of the $W^+ N \to N$ amplitude.

\subsubsection{The $WN\Delta$ and $N\Delta\pi$ vertices}

\label{sec:delta}

At intermediate energies, the weak excitation of the $\Delta(1232)$
resonance and its subsequent decay into $N\pi$ dominates 
the $W N \to N' \pi$ reaction. A convenient parameterization for the
$W^+n\to\Delta^+$ hadron matrix element is the
following~\cite{LS72}
\begin{equation} \label{eq:delta}
\langle \Delta^+; p_\Delta= p+q\, | j^\mu_{cc+}(0) | n;
p \rangle =\bar u_\alpha(\vec{p}_\Delta ) \Gamma^{\alpha\mu}\left(p,q \right)
u(\vec{p}\,)\cos\theta_C, \qquad {\rm where} 
\end{equation}
\begin{eqnarray}
\Gamma^{\alpha\mu} (p,q) &=&
\left [ \frac{C_3^V}{M}\left(g^{\alpha\mu} \slashchar{q}-
q^\alpha\gamma^\mu\right) + \frac{C_4^V}{M^2} \left(g^{\alpha\mu}
q\cdot p_\Delta- q^\alpha p_\Delta^\mu\right)
+ \frac{C_5^V}{M^2} \left(g^{\alpha\mu}
q\cdot p- q^\alpha p^\mu\right) + C_6^V g^{\mu\alpha}
\right ]\gamma_5 \nonumber\\
&+& \left [ \frac{C_3^A}{M}\left(g^{\alpha\mu} \slashchar{q}-
q^\alpha\gamma^\mu\right) + \frac{C^A_4}{M^2} \left(g^{\alpha\mu}
q\cdot p_\Delta- q^\alpha p_\Delta^\mu\right)
+ C_5^A g^{\alpha\mu} + \frac{C_6^A}{M^2} q^\mu q^\alpha
\right ], \quad p_\Delta= p+q \label{eq:del_ffs}
 \end{eqnarray}
with $C_{3,4,5,6}^V$ and $C_{3,4,5,6}^A$ scalar and real vector and
axial form factors, which depend on $q^2$. Besides, $u_\alpha$ is a
Rarita Schwinger spinor for the $\Delta^+$. The $N\Delta\pi$ coupling
is given by
\begin{equation}
{\cal L}_{\pi N \Delta} = \frac{f^*}{m_\pi} \bar\Psi_\mu
\vec{T}^\dagger (\partial^\mu \vec{\phi}) \Psi + {\rm h.c.}
\label{eq:lagranpind}
\end{equation}
where $\Psi_\mu$ is a Rarita Schwinger $J^\pi = 3/2^+$ field,
$\vec{T}^\dagger$ is the isospin transition operator\footnote{It is a
vector under isospin rotations and its Wigner-Eckart irreducible
matrix element is taken to be one.} from isospin 1/2 to 3/2 and
$f^*=2.13\times f=2.14$. The Goldberger--Treiman relation implies
here\footnote{Note that the $C_5^A$ sign is quoted incorrectly in 
Ref.~\cite{LS72} (see comment in Ref.~\protect\cite{sh73}).}
\begin{equation}
C_5^A(0) = \sqrt\frac23\frac{f_\pi}{m_\pi}f^* \sim 1.15 \label{eq:gt}
\end{equation}
For the $\Delta-$propagator $G^{\mu\nu}(p_\Delta)$, we use in momentum
space 
\begin{equation}
G^{\mu\nu}(p_\Delta)= \frac{P^{\mu\nu}(p_\Delta)}{p_\Delta^2-M_\Delta^2+ i M_\Delta \Gamma_\Delta}
\end{equation}
with $M_\Delta$ the resonance mass ($\sim 1232$ MeV), $P^{\mu\nu}$ the
spin 3/2 projection operator
\begin{equation}
P^{\mu\nu}(p_\Delta)= - (\slashchar{p}_\Delta + M_\Delta) \left [ g^{\mu\nu}-
  \frac13 \gamma^\mu\gamma^\nu-\frac23\frac{p_\Delta^\mu
  p_\Delta^\nu}{M_\Delta^2}+ \frac13\frac{p_\Delta^\mu
  \gamma^\nu-p_\Delta^\nu \gamma^\mu }{M_\Delta}\right]
\end{equation}
and $\Gamma_\Delta$ the resonance width in its rest frame, given by
\begin{equation}
\Gamma_\Delta(s) = \frac{1}{6\pi} \left ( \frac{f^*}{m_\pi}\right )^2
 \frac{M}{\sqrt s} \left [ \frac{\lambda^\frac12
  (s,m_\pi^2,M^2)}{2\sqrt s} \right ]^3 \Theta(\sqrt s
-M-m_\pi),\qquad s= p_\Delta^2 
\end{equation}
with $\lambda(x,y,z) = x^2+y^2+z^2-2xy-2xz-2yz$ and $\Theta$ the step
function, as deduced from the lagrangian of Eq.~(\ref{eq:lagranpind}).

The determination of the form factors follows from general principles
and experimental results. The imposition of the conserved vector
current hypothesis implies $C_6^V=0$. The other three vector form
factors are then given in terms of the isovector electromagnetic form
factors in the $p-\Delta^+$  transition. The analysis
of photo and electroproduction data of $\Delta$ is done in terms of
multipole amplitudes~\cite{elDel}. Most of the previous papers on
neutrino production~\cite{ASV98,Pa05,LAM06} assume $M_{1^+}$
dominance\footnote{Recent data determine the contribution from the
electric multipole $E_{1^+}/M_{1^+}$ to be $\sim -2.5\% $ and from the
scalar multipole $S_{1^+}/M_{1^+} \sim -2.5\% $ \cite{CLAS02} .} of
the electroproduction amplitude, which implies $C_5^V=0$ and a
relation between $C_4^V$ and $C_3^V$. Here, we
take advantage of the recent work of Lalakulich, Paschos and
Piranishvili and improve on that by including the effect
of the subdominant multipoles~\cite{LPP06}
\begin{equation}
C_3^V = \frac{2.13}{(1-q^2/M_V^2)^2}\times
\frac{1}{1-\frac{q^2}{4M_V^2}}, \qquad C_4^V =
\frac{-1.51}{(1-q^2/M_V^2)^2}\times \frac{1}{1-\frac{q^2}{4M_V^2}},
\qquad C_5^V= \frac{0.48}{(1-q^2/M_V^2)^2}\times
\frac{1}{1-\frac{q^2}{0.776M_V^2}} \label{eq:c3v}
\end{equation}
with $M_V= 0.84$ GeV. Among the axial form factors the most important
contribution comes from $C_5^A$ whose numerical value is related to
the pseudoscalar form factor $C_6^A$ by PCAC. Since there are no other
theoretical constraints for $C_3^A(q^2)$, $C_4^A(q^2)$ and
$C_5^A(q^2)/C_5^A(0)$, they have to be fitted to neutrino scattering
data. The available information comes mainly from two bubble chamber
experiments, ANL~\cite{anlviejo,anl,anl-cam73} and BNL~\cite{bnlviejo,bnl}. We
adopt the Adler model \cite{adler}, as the ANL and BNL analyses did, where
\begin{equation}
C_4^A(q^2) = -\frac{C_5^A(q^2)}{4}, \qquad
C_3^A(q^2) = 0 
\end{equation}
and for $C_{5,6}^A$ we shall use~\cite{Pa04} 
\begin{equation}
C_5^A(q^2) = \frac{1.2}{(1-q^2/M^2_{A\Delta})^2}\times
\frac{1}{1-\frac{q^2}{3M_{A\Delta}^2}}, \qquad C_6^A(q^2) = C_5^A(q^2)
\frac{M^2}{m_\pi^2-q^2}, \quad {\rm with}\, M_{A\Delta}=1.05
\,{\rm GeV}. \label{eq:ca5old}
\end{equation}
 The value for $C_4^A$ was found to give a small contribution to the
cross section and setting the $C_3^A$ form factor to zero is
consistent with early dispersion calculations~\cite{adler,disp} and
recent lattice QCD results~\cite{negele07}. Note
that the contribution to the differential cross section of the $C_6^A$
term will be always proportional to the outgoing lepton mass.

Isospin symmetry implies that the transition matrix element
$W^+p\to\Delta^{++}$ is a factor $\sqrt{3}$ bigger than the
$W^+n\to\Delta^{+}$ one discussed above.
 
 For the weak transition $\Delta \to N$, we have
\begin{eqnarray}
\langle  N; p'| j^\mu_{cc+}(0) | \Delta; p_\Delta= p'-q\, \rangle &=&
\langle \Delta; p_\Delta= p'-q | j^\mu_{cc-}(0) | N; p'\rangle^*\\
&=&-\frac{(\frac12, 1, \frac 32 | t_N, -1, t_\Delta)}{(\frac12, 1, \frac 32
  | -\frac12, 1, \frac12)} \left\{ \bar
u_\alpha(\vec{p}_\Delta=\vec{p}^{\,\prime}-\vec{q}\, )
\Gamma^{\alpha\mu} \left(p',-q \right)
u(\vec{p}^{\,\prime}\,)\cos\theta_C \right\}^*
\end{eqnarray}
with $(t_1,t_2,t| m_1, m_2, m)$ Clebsch--Gordan coefficients and $t_N$
and $t_\Delta$ the nucleon and delta isobar isospin third components,
respectively.

\subsubsection{ Explicit expressions for the $\langle p \pi^+ |
  j^\mu_{\rm cc+}(0) | p \rangle$ and $\langle n \pi^+ | j^\mu_{\rm
  cc+}(0) | n \rangle$  amplitudes}
 
\begin{figure}[tbh]
\centerline{\includegraphics[height=10cm]{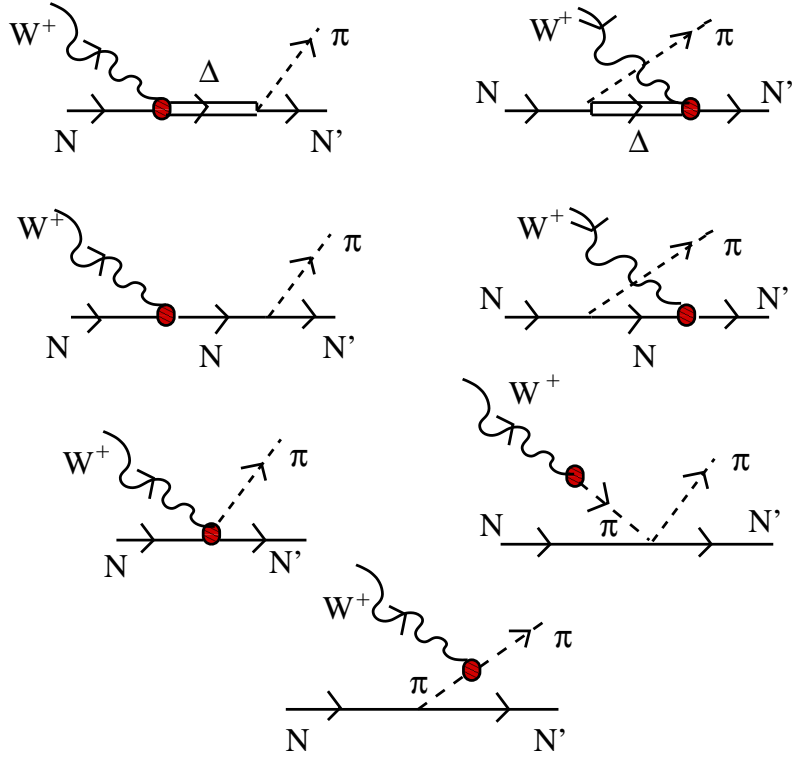}}
\caption{\footnotesize Model for the $W^+N\to N^\prime\pi$
  reaction. It consists of seven diagrams: Direct and crossed
  $\Delta(1232)-$ (first row) and nucleon (second row) pole terms,
  contact and pion pole contribution (third row) and finally the
  pion-in-flight term. Throughout this work, we will denote these
  contributions by: $\Delta P$, $C\Delta P$, $NP$, $CNP$, $CT$, $PP$ and
  $PF$, respectively. The circle in the diagrams stands for the weak
  transition vertex. }\label{fig:diagramas}
\end{figure}

In this subsection, we give explicit expressions for the $\langle p
  \pi^+ | j^\mu_{\rm cc+}(0) | p \rangle$ and $\langle n \pi^+ |
  j^\mu_{\rm cc+}(0) | n \rangle$ amplitudes, which we will denote by
  $(j^{\mu}_{cc+})_{p\pi^+}$ and $(j^{\mu}_{cc+})_{n\pi^+}$,
  respectively. All $\langle N^\prime \pi | j^\mu_{\rm cc\pm}(0) | N
  \rangle$ matrix elements can be expressed in terms of these two
  amplitudes (Subsect.~\ref{sec:isospin}). The model consists of seven
  Feynman diagrams, depicted in Fig.~\ref{fig:diagramas}, constructed
  out of the $W^+N\to N$, $W^+N\to \Delta$, $W^+N\to N\pi$ and the
  contact $W^+\pi\to \pi$ weak transition vertices
  (Eqs.~(\ref{eq:vcurrent}--\ref{eq:np}) and~(\ref{eq:delta}) ) and
  the $\pi NN$, $\pi\pi NN$ (Eq.~(\ref{eq:lint})) and $\pi N \Delta$
  (Eq.~(\ref{eq:lagranpind})) couplings, discussed in the
  Subsects.~\ref{sec:sigma} and \ref{sec:delta}. Since we have
  included a $q^2$ dependence ($F_1^V(q^2)$) on the Dirac part of the
  vector $WNN$ vertex and to preserve vector current conservation, we
  also include form--factors ($F_{PF}(q^2)$ and $F_{CT}^V(q^2)$) 
  in the $V^\mu_a$ and $V^\mu_c$  weak operators. This partially
  accounts for the nucleon structure.

We do not include loop corrections. This model is an extension of that
developed in Ref.~\cite{GNO97} for the $ e N \to e' N \pi$
reaction, though there exist some minor differences related to the
used form factors and a non--relativistic reduction was performed
in Ref.~\cite{GNO97}. 

 The amplitudes read,
\begin{eqnarray} 
j^\mu_{cc+}\Big|_{\Delta P} &=& {\rm
  i}\,C^\Delta\frac{f^*}{m_\pi}\sqrt{3}\cos\theta_C
  \frac{k_\pi^\alpha}{p_\Delta^2-M_\Delta^2+ i M_\Delta \Gamma_\Delta}
  \bar u(\vec{p}\,') P_{\alpha\beta}(p_\Delta) \Gamma^{\beta\mu}(p,q)
  u(\vec{p}\,),\quad p_\Delta=p+q, \nonumber \\\nonumber \\ 
&{\rm and} & C^\Delta = \left( \begin{array}{ccc} 1
  & {\rm for} & p\pi^+ \cr 1/3
  & {\rm for} & n\pi^+ \end{array} \right ) \nonumber \\  \nonumber \\
j^\mu_{cc+}\Big|_{C\Delta P} &=& {\rm
  i}\,C^{C\Delta}\frac{f^*}{m_\pi}\frac{1}{\sqrt 3}\cos\theta_C
  \frac{k_\pi^\beta}{p_\Delta^2-M_\Delta^2+ i M_\Delta \Gamma_\Delta}
  \bar u(\vec{p}\,')  {\hat \Gamma}^{\mu\alpha}(p',q) P_{\alpha\beta}(p_\Delta) 
  u(\vec{p}\,),\quad p_\Delta=p'-q, \nonumber \\\nonumber \\
&{\rm and} &  C^{C\Delta} = \left( \begin{array}{ccc} 1
  & {\rm for} & p\pi^+ \cr 3
  & {\rm for} & n\pi^+ \end{array} \right ), \quad {\hat \Gamma}^{\mu\alpha}(p',q) = \gamma^0\left
  [\Gamma^{\alpha\mu}\right(p', -q)]^\dagger \gamma^0  \nonumber
  \\\nonumber \\ 
j^\mu_{cc+}\Big|_{NP} &=& 
-{\rm i}\,C^{NP}\frac{g_A}{\sqrt 2 f_\pi}\cos\theta_C\
  \bar u(\vec{p}\,') 
 \slashchar{k}_\pi\gamma_5\frac{\slashchar{p}+\slashchar{q}+M}{(p+q)^2-M^2+ i\epsilon}\left [V^\mu_N(q)-A^\mu_N(q) \right]  
u(\vec{p}\,),\quad C^{NP} = \left( \begin{array}{ccc} 0
  & {\rm for} & p\pi^+ \cr 1
  & {\rm for} & n\pi^+ \end{array} \right )\nonumber \\\nonumber \\
j^\mu_{cc+}\Big|_{CNP} &=& 
-{\rm i}\,C^{CNP}\frac{g_A}{\sqrt 2 f_\pi}\cos\theta_C\
  \bar u(\vec{p}\,') \left [V^\mu_N(q)-A^\mu_N(q) \right]
\frac{\slashchar{p}'-\slashchar{q}+M}{(p'-q)^2-M^2+ i\epsilon} \slashchar{k}_\pi\gamma_5  u(\vec{p}\,),\quad C^{CNP} = \left( \begin{array}{ccc} 1
  & {\rm for} & p\pi^+ \cr 0
  & {\rm for} & n\pi^+ \end{array} \right )\nonumber \\\nonumber \\
j^\mu_{cc+}\Big|_{CT} &=& 
-{\rm i}\,C^{CT}\frac{1}{\sqrt 2 f_\pi}\cos\theta_C\
  \bar u(\vec{p}\,') \gamma^\mu\left (
  g_AF_{CT}^V(q^2)\gamma_5-F_\rho\left((q-k_\pi)^2\right) \right ) u(\vec{p}\,),\quad C^{CT} = \left( \begin{array}{ccc} \phantom{-}1
  & {\rm for} & p\pi^+ \cr -1
  & {\rm for} & n\pi^+ \end{array} \right )\nonumber \\\nonumber \\
j^\mu_{cc+}\Big|_{PP} &=& 
-{\rm i}\,C^{PP}F_\rho\left((q-k_\pi)^2\right)\frac{1}{\sqrt 2 f_\pi}\cos\theta_C \
  \frac{q^\mu}{q^2-m_\pi^2}
  \bar u(\vec{p}\,')\ \slashchar{q} \ u(\vec{p}\,),\quad C^{PP} = \left( \begin{array}{ccc} \phantom{-}1
  & {\rm for} & p\pi^+ \cr -1
  & {\rm for} & n\pi^+ \end{array} \right )\nonumber \\\nonumber \\
j^\mu_{cc+}\Big|_{PF} &=& 
-{\rm i}\,C^{PF}F_{PF}(q^2)\frac{g_A}{\sqrt 2 f_\pi}\cos\theta_C \
  \frac{(2k_\pi-q)^\mu}{(k_\pi-q)^2-m_\pi^2}
  2M\bar u(\vec{p}\,')  \gamma_5 u(\vec{p}\,),\quad C^{PF} = \left( \begin{array}{ccc} \phantom{-}1
  & {\rm for} & p\pi^+ \cr -1
  & {\rm for} & n\pi^+ \end{array} \right )\label{eq:eqscc}
\end{eqnarray}

Note that in the PF (PP) term   the weak transition is purely driven by
the vector (axial) CC. The contribution proportional to $g_A$ in
the CT diagram is the one due to the vector weak transition.
We impose
\begin{eqnarray}
F_{PF}(q^2) = F_{CT}^V(q^2) = 2 F_1^V(q^2) = F_1^p-F_1^n
\end{eqnarray}
to preserve conservation of vector current, as we required for the
$\Delta N$ weak transition. Besides, we have included a form factor
\begin{equation}
F_\rho(t) = \frac{1}{1-t/m^2_\rho}, \quad m_\rho=0.7758\, {\rm GeV}
\end{equation}
in the PP term to account for the $\rho-$meson dominance of the
$\pi\pi NN$ coupling. To preserve PCAC, the same form-factor has been
included in the CT axial contribution.

In the pion in flight term, we have the coupling $\pi NN$ with a virtual
pion. It is usual in the literature to use a form factor to account
for the off-shellness of the pion. To preserve vector current
conservation, if one includes this form--factor in this term, one
should also multiply by the same factor the CT term and  the $NP$
and $CNP$ $F_1^V$ contributions. This was the adopted scheme in the
study of the $ e N \to e' N \pi$ reaction carried out in
Ref.~\cite{GNO97}, where the induced changes by its inclusion
turned out to be moderately small. In the weak pion production case,
there are more important sources of uncertainties\footnote{For
instance: partial knowledge of the $\Delta$ resonance form factors, 
possible pion off-shell effects  in the weak transition vertex of the PP
diagram, etc\ldots} and the existing measurements are poorer.  Being this work
one of the first studies of weak pion production where background
terms are added to the dominant $\Delta$ contribution, and for the
sake of simplicity, we do not include this form factor.

The average and sum over the initial and final spins in Eq.~(\ref{eq:wmunu})
is readily done thanks to
\begin{equation}
\overline{\sum_{\rm spins}} \bar u(\vec{p}\,') {\cal S^\mu} u(\vec{p}\,) 
\left[\bar u(\vec{p}\,') {\cal S^\sigma} u(\vec{p}\,)\right]^* = \frac12
     {\rm Tr}\left((\slashchar{p}^\prime +M){\cal S^\mu}(\slashchar{p}
     +M) \gamma^0{\cal
       S^{\dagger\,\sigma}}\gamma^0 \right)
\end{equation}
where the spin dependence of the Dirac's spinors is understood and
${\cal S^\mu}$ is a matrix in the Dirac's space for each value of the
Lorentz index $\mu$.

\section{NC Neutrino and Antineutrino induced reactions}
\label{sec:nc-an}
The unpolarized differential cross section in the LAB frame for the 
process 
\begin{equation}
  \nu_l (k) +\, N(p)  \to \nu_l (k^\prime) + N(p^\prime) +\, \pi(k_\pi) 
\label{eq:neureac}
\end{equation}
reads

\begin{equation}
\frac{d^{\,5}\sigma_{\nu
    \nu}}{d\Omega(\hat{k^\prime})dE^\prime d\Omega(\hat{k}_\pi) } =
\frac{|\vec{k}^\prime|}{|\vec{k}~|}\frac{G^2}{16\pi^2} 
 \int_0^{+\infty}\frac{dk_\pi k_\pi^2}{E_\pi}
 L_{\mu\sigma}^{(\nu)}\left(W^{\mu\sigma}_{{\rm NC} \pi}\right)^{(\nu)} 
\end{equation}
with $\vec{k}^\prime~$, $E^{\prime}= |\vec{k}^{\prime\,}|$ the LAB
outgoing neutrino momentum and energy. The leptonic tensor is given in
Eq.~(\ref{eq:lep}) and it is now orthogonal to
$q^\mu=(k-k^\prime)^\mu$ for massless neutrinos, i.e,
$L_{\mu\sigma}^{(\nu)} q^\mu = L_{\mu\sigma}^{(\nu)} q^\sigma = 0$.

The hadronic tensor reads
\begin{eqnarray}
(W^{\mu\sigma}_{{\rm NC} \pi})^{(\nu)} &=& \frac{1}{4M}\overline{\sum_{\rm
 spins}} \int\frac{d^3p^\prime}{(2\pi)^3} \frac{1}{2E^\prime_N}
  \delta^4(p^\prime+k_\pi-q-p) \langle N^\prime \pi |
 j^\mu_{\rm nc}(0) | N \rangle \langle N^\prime \pi | j^\sigma_{\rm nc}(0) | N
 \rangle^*
\end{eqnarray}
where the neutral current at the quark level is
\begin{eqnarray}
j^\mu_{\rm nc} &=& \bar{\Psi}_u\gamma^\mu(1-\frac83 \sin^2\theta_W-
\gamma_5)\Psi_u  - \bar{\Psi}_d\gamma^\mu(1-\frac43 \sin^2\theta_W-
\gamma_5)\Psi_d  - \bar{\Psi}_s\gamma^\mu(1-\frac43 \sin^2\theta_W-
\gamma_5)\Psi_s  \nonumber\\
&=&\bar\Psi_q \gamma^\mu (1-\gamma_5) \tau^1_0 \Psi_q
- 4 \sin^2 \theta_W s^\mu_{\rm em} -\bar\Psi_s\gamma^\mu(1-\gamma_5)\Psi_s
\end{eqnarray}
where $\theta_W$ is the Weinberg angle ($\sin^2\theta_W= 0.231$). 

Both, lepton and hadron tensors
are independent of the neutrino lepton family, and therefore the cross
section for the reaction of Eq.~(\ref{eq:neureac}) is the same for
electron, muon or tau incident neutrinos. For antineutrino induced
reactions we have, besides the relation of Eq.~(\ref{eq:anti})  for
the leptonic tensor, 
\begin{equation}
(W^{\mu\sigma}_{{\rm NC} \pi})^{(\bar\nu)} = 
(W^{\mu\sigma}_{{\rm NC} \pi})^{(\nu)} \label{eq:neu-anti}
\end{equation}
As discussed above for CC induced process, Lorentz invariance here
also restricts the $\phi_\pi$ dependence, and the NC differential
cross section can be written as in Eq.~(\ref{eq:phipi}).  This
$\phi_\pi$ dependence has been carefully studied in Ref.~\cite{tau-nc}
as a potential tool to distinguish $\tau-$neutrinos from
antineutrinos, below the $\tau-$production threshold, but above the
pion production one.

The NC can be expressed as
\begin{eqnarray}
j^\mu_{\rm nc} &=& \bar\Psi_q \gamma^\mu (1-2\sin^2\theta_W-\gamma_5)
 \tau^1_0   \Psi_q  -4\sin^2\theta_W s^\mu_{{\rm em},IS} -\bar\Psi_s\gamma^\mu(1-\gamma_5)\Psi_s
\end{eqnarray}
where the isoscalar part of the electromagnetic current is given by 
\begin{equation}
s^\mu_{{\rm em}, IS} = \frac{1}{6}\bar\Psi_q\gamma^\mu\Psi_q
-\frac13 \bar\Psi_s\gamma^\mu\Psi_s
\end{equation}
Isospin symmetry relates the matrix elements of the isovector part
($\tau^1_0$ term) of $j^\mu_{\rm nc}$ with those of the vector
($V^\mu_{\rm cc+}$) and axial ($A^\mu_{\rm cc+}$) part of the current
$j^\mu_{\rm cc+}\, (=V^\mu_{\rm cc+}-A^\mu_{\rm cc+})$,

\begin{eqnarray}
\langle p \pi^0 \big| \bar\Psi_q \gamma^\mu
 (1-2\sin^2\theta_W-\gamma_5) \tau^1_0 \Psi_q \big | p \rangle &=&
  \frac{1}{\sqrt{2}\cos\theta_C}\Big \{ (1-2\sin^2\theta_W)
 \left[\langle p \pi^+ | V^\mu_{\rm cc+}(0) | p \rangle + \langle n
 \pi^+ | V^\mu_{\rm cc+}(0) | n \rangle \right]\nonumber\\
&& - \left[\langle p
 \pi^+ | A^\mu_{\rm cc+}(0) | p \rangle + \langle n \pi^+ | A^\mu_{\rm
 cc+}(0) | n \rangle \right] \Big\} \\ \nonumber\\
\langle n \pi^+ \big| \bar\Psi_q \gamma^\mu
 (1-2\sin^2\theta_W-\gamma_5) \tau^1_0 \Psi_q \big | p \rangle &=&
 -\frac{1}{\cos\theta_C}\Big \{ (1-2\sin^2\theta_W)
 \left[\langle p \pi^+ | V^\mu_{\rm cc+}(0) | p \rangle - \langle n
 \pi^+ | V^\mu_{\rm cc+}(0) | n \rangle \right]\nonumber\\
&& - \left[\langle p
 \pi^+ | A^\mu_{\rm cc+}(0) | p \rangle - \langle n \pi^+ | A^\mu_{\rm
 cc+}(0) | n \rangle \right] \Big\} \\ \nonumber\\
\langle n \pi^0 \big| \bar\Psi_q \gamma^\mu 
 (1-2\sin^2\theta_W-\gamma_5) \tau^1_0 \Psi_q \big | n \rangle
 &=& \langle p \pi^0 \big| \bar\Psi_q \gamma^\mu 
 (1-2\sin^2\theta_W-\gamma_5) \tau^1_0 \Psi_q \big | p \rangle
 \nonumber \\
\langle p \pi^- \big| \bar\Psi_q \gamma^\mu 
 (1-2\sin^2\theta_W-\gamma_5) \tau^1_0 \Psi_q \big | n \rangle
 &=& -\langle n \pi^+ \big| \bar\Psi_q \gamma^\mu 
 (1-2\sin^2\theta_W-\gamma_5) \tau^1_0 \Psi_q \big | p \rangle \label{eq:64}
\end{eqnarray}
For the isoscalar part of the electromagnetic current we have
\begin{eqnarray}
\langle n \pi^+ \big|  s^\mu_{{\rm em}, IS} \big | p \rangle &= &\langle p \pi^- \big|
  s^\mu_{{\rm em}, IS} \big | n \rangle =
 \sqrt 2 \langle p \pi^0
 \big|  s^\mu_{{\rm em}, IS} \big | p
 \rangle= -\sqrt 2 \langle n \pi^0 \big|  s^\mu_{{\rm em}, IS}
\big | n \rangle \label{eq:65}
\end{eqnarray}
with
\begin{equation}
\langle p \pi^0 \big|  s^\mu_{{\rm em}, IS} \big | p \rangle = -  \frac{\langle
n \pi^0 | s^\mu_{\rm em}(0) | n \rangle-\langle
p \pi^0 | s^\mu_{\rm em}(0) | p \rangle }{2} 
\end{equation}
To compute $\langle N \pi^0 | s^\mu_{\rm em}(0) | N \rangle$, we
derive the electromagnetic current associated to the lagrangian of the
SU(2) non-linear $\sigma$ model of Eq.~(\ref{eq:lsigma}),
\begin{equation}
s^\mu_{\rm em} = \bar\Psi \gamma^\mu \left ( \frac{1+\tau_z}{2}\right) \Psi
+ \frac{{\rm i}g_A}{2f_\pi} \bar\Psi \gamma^\mu \gamma_5 \left (
\tau^1_{-1} \phi^\dagger + \tau^1_{+1} \phi \right ) \Psi + {\rm i}
\left (\phi^\dagger \partial^\mu \phi -  \phi \partial^\mu
\phi^\dagger \right) + \cdots \label{eq:real_sem}
\end{equation}
where we have only kept those terms contributing to one pion
production in absence of chiral loop corrections. Thus, within our
framework, the model for the $\gamma N \to \pi N$ reaction would consist of
direct and crossed nucleon pole, contact and pion-in-flight terms. As
we did for the CC driven processes, such model should be supplemented
by including i) the $q^2$ dependence induced by the Dirac $F_1^{p,n}$
form factors, ii) the magnetic contribution in the $\gamma NN$ vertex
and iii) the direct and crossed $\Delta (1232)$ pole
terms~\cite{GNO97}. However, the $\Delta-$resonance diagrams cannot
contribute to the matrix elements of the isoscalar part of the
electromagnetic current. Besides, from Eq.~(\ref{eq:real_sem}) we see
that neither the pion-in-flight nor the contact terms contribute for
$\pi^0$ photoproduction. Hence, to compute $\langle n \pi^0 |
s^\mu_{\rm em}(0) | n \rangle-\langle p \pi^0 | s^\mu_{\rm em}(0) | p
\rangle$, we are just left with the direct and crossed nucleon pole
terms
\begin{eqnarray}
\frac{\langle
n \pi^0 | s^\mu_{\rm em}(0) | n \rangle-\langle
p \pi^0 | s^\mu_{\rm em}(0) | p \rangle }{2} &=& {\rm i}\,\frac{g_A}{
  2 f_\pi}  \bar u(\vec{p}\,') \Bigg \{
 \slashchar{k}_\pi\gamma_5\frac{\slashchar{p}+\slashchar{q}+M}{(p+q)^2-M^2+ i\epsilon}\left [F^{IS}_1(q^2)\gamma^\mu+{\rm i}\mu_{IS}\frac{F_2^{IS}(q^2)}{2M} \sigma^{\mu\nu}q_\nu 
\right]  \nonumber \\
&&+\left [F^{IS}_1(q^2)\gamma^\mu+{\rm i}\mu_{IS}\frac{F_2^{IS}(q^2)}{2M} \sigma^{\mu\nu}q_\nu 
\right] \frac{\slashchar{p}'-\slashchar{q}+M}{(p'-q)^2-M^2+ i\epsilon}
 \slashchar{k}_\pi\gamma_5 \Bigg\} u(\vec{p}\,)
\end{eqnarray}
with 
\begin{equation}
 F_1^{IS}(q^2) =  \frac12 \left (F_1^p(q^2)+F_1^n(q^2)\right),\qquad
 \mu_{IS} F_2^{IS}(q^2) = \frac12 \left ( \mu_p F_2^p(q^2) + \mu_n F_2^n(q^2)\right) 
\end{equation}

Finally, we pay attention to the matrix elements of the isoscalar
operator $\bar\Psi_s\gamma^\mu(1-\gamma_5)\Psi_s$ which are sensitive
to the strange content of the hadrons. Due to its isoscalar character
we have
\begin{eqnarray}
\langle n \pi^+ \big|  \left(\bar\Psi_s\gamma^\mu(1-\gamma_5)\Psi_s\right) (0) \big | p \rangle &= &\langle p \pi^- \big|
  \left(\bar\Psi_s\gamma^\mu(1-\gamma_5)\Psi_s\right) (0)\big | n \rangle =
 \sqrt 2 \langle p \pi^0
 \big| \left(\bar\Psi_s\gamma^\mu(1-\gamma_5)\Psi_s\right) (0) \big | p
 \rangle \nonumber \\
&=& -\sqrt 2 \langle n \pi^0
 \big| \left(\bar\Psi_s\gamma^\mu(1-\gamma_5)\Psi_s\right) (0) \big | n
 \rangle \label{eq:70}
\end{eqnarray}
This part of the NC operator can neither lead to $N\Delta$ transitions
 nor couple to a single pion. Thus, and assuming a model for the
$Z^0 N \to N'\pi$ reactions similar to that used for the CC driven
process, we should consider the contributions of the direct and
crossed nucleon pole, the contact and the pion-in-flight terms to the
matrix element of the $\bar\Psi_s\gamma^\mu(1-\gamma_5)\Psi_s$ quark operator.
The contribution of the first two terms ($NP$ and $CNP$) reads
\begin{eqnarray}
\langle p \pi^0
 \big| \left(\bar\Psi_s\gamma^\mu(1-\gamma_5)\Psi_s\right) (0) \big | p
 \rangle&=&-{\rm i}\,\frac{g_A}{
  2 f_\pi} \bar u(\vec{p}\,')\Bigg \{\slashchar{k}_\pi\gamma_5\frac{\slashchar{p}+\slashchar{q}+M}{(p+q)^2-M^2+
 i\epsilon}  \Big[F^{s}_1(q^2)\gamma^\mu+{\rm
 i}\mu_{s}\frac{F_2^{s}(q^2)}{2M} \sigma^{\mu\nu}q_\nu
  \nonumber \\
&& -G^s_A(q^2)\gamma^\mu \gamma_5-G_P^s q^\mu \gamma_5 
\Big] + \Big[F^{s}_1(q^2)\gamma^\mu+{\rm i}\mu_{s}\frac{F_2^{s}(q^2)}{2M} \sigma^{\mu\nu}q_\nu   \nonumber \\
&& -G^s_A(q^2)\gamma^\mu \gamma_5-G_P^s q^\mu \gamma_5 \Big] \frac{\slashchar{p}'-\slashchar{q}+M}{(p'-q)^2-M^2+ i\epsilon}
 \slashchar{k}_\pi\gamma_5 \Bigg\} u(\vec{p}\,)\label{eq:z0ss}
\end{eqnarray}
where $F_1^s, \mu_s F_2^s, G_A^s$ and $G_P^s$ are the strange vector
and axial nucleon form factors~\cite{Al96}. The pseudoscalar part of
the axial current does not contribute to the differential cross
section for massless neutrinos and for the rest of strange form
factors we use the results of the fit II of Ref.~\cite{Ga93},
\begin{equation}
G_A^s(q^2) = \frac{g_S}{(1-q^2/(M_A^s)^2)^2}, \quad F_1^s(q^2) =
\mu_s F_2^s(q^2) = 0 \label{eq:gs}
\end{equation}
with $g_S=-0.15$ and $M_A^s = M_A$.

The vector part in Eq.~(\ref{eq:z0ss}) is conserved, i.e., it is
orthogonal to $q^\mu$ independently of $F_1^s$.  Because of parity and
angular momentum conservation, a pion-in-flight type term can only
contribute to the vector part of the matrix element of the
$\bar\Psi_s\gamma^\mu(1-\gamma_5)\Psi_s$ operator and its contribution
should be proportional to $(2k_\pi-q)^\mu \bar u(\vec{p}\,') \gamma_5
u(\vec{p}\,)$, as in Eq~(\ref{eq:eqscc}). Assuming a structure of the
type $\bar u(\vec{p}\,') \gamma_5\gamma^\mu u(\vec{p}\,)$, as in
Eq~(\ref{eq:eqscc}), for the contact term vector contribution to the
matrix element of the $\bar\Psi_s\gamma^\mu(1-\gamma_5)\Psi_s$
operator, we will conclude that both types of contributions should be
exactly zero to preserve vector current conservation. Within our
scheme, we cannot say anything about a possible contact term axial
contribution to $\langle p \pi^0 \big|
\left(\bar\Psi_s\gamma^\mu(1-\gamma_5)\Psi_s\right) (0) \big | p
\rangle$, that for simplicity we will neglect. Thus, we will assume 
that this latter matrix element is given by the NP and CNP
contributions in Eq.~(\ref{eq:z0ss}).

\section{Results}
\label{sec:resul}

In this section, we will show differential and partially integrated
neutrino and antineutrino cross sections for pion production processes
driven by both CC and NC. As it is usual in pion electroproduction, we
will work with angular pion variables ($d\Omega^*(\hat{k}_\pi)$)
defined in the outgoing $\pi N$ pair CM frame, while the
incoming and outgoing lepton variables will be in the LAB frame. We will
pay here an special attention to the CC pion production cross section
dependence on the azimuthal pion angle $\phi_\pi^*$ (note that this
angle is not affected by the LAB$\to$CM boost), and thus we will 
show the different contributions to $\Sigma^* \left(q^2, p\cdot
q\,,\theta_\pi^*, \phi_\pi^*\right)$, defined by its relation to the
differential cross section,

\begin{eqnarray}
\frac{d^{\,5}\sigma_{\nu_l
    l}}{d\Omega(\hat{k^\prime})dE^\prime d\Omega^*(\hat{k}_\pi) } &=&
    \frac{|\vec{k}^\prime|}{|\vec{k}~|}\frac{G^2}{4\pi^2} \Sigma^*
      \\\nonumber \\
  \Sigma^*\left (q^2, p\cdot q\,,\theta_\pi^*, \phi_\pi^*\right) &=&  
\left \{ A^* + B^*
    \cos\phi_\pi^* + C^* \cos 2\phi_\pi^*+ D^* \sin\phi_\pi^* + E^*
    \sin 2\phi^*_\pi \right\} \label{eq:phipi*}
\end{eqnarray}
For the NC case this dependence has been already  discussed in
Ref.~\cite{tau-nc} with the aim of distinguishing between $\nu_\tau$ and
$\bar\nu_\tau$ below the $\tau-$production threshold, but above the
pion production one.

After integrating the pion solid angle, we will take as independent
variables the incoming neutrino energy $E=|\vec{k}\,|$, the
invariant mass $W$ of the outgoing pion-nucleon pair ($W^2=(p+q)^2$)
and the squared of the lepton four-momentum transfer $q^2$, 
\begin{eqnarray}
\frac{d^{\,3}\sigma_{a}}{dq^2 dW  } &=& 
\frac{d^{\,3}\sigma_{a}}{d\Omega(\hat{k^\prime})dE^\prime  } \times 
\frac{\pi W}{M E |\vec{k}^\prime\,|}, \qquad a=\nu_l l, \nu\nu.
\end{eqnarray}
where $W$ varies in the range $(m_\pi+M) \le W \le (\sqrt{S}-m_l)$,
with $S=(k+p)^2=M(M+2E)$. Thus, the incoming neutrino energy in the
LAB system, $E$, should be greater than 
$\left (m_\pi+m_l+(m_\pi+m_l)^2/2M \right )$ for the pion production
process to take place. Besides for a given outgoing $\pi N$ invariant
mass $W$, $q^2$ is comprised in the interval

\begin{equation}
q^2_{\rm min} (W) \equiv \left [m^2_l
  -2E_{CM}\left(E'_{CM}+\sqrt{E_{CM}^{'2\,}-m^2_l}\right )\right] \le
q^2 \le \left [m^2_l
  -2E_{CM}\left(E'_{CM}-\sqrt{E_{CM}^{'2\,}-m^2_l}\right )\right] \equiv q^2_{\rm
  max} (W)
\end{equation}
with $E_{CM}= (S-M^2)/2\sqrt{S}$ and $E'_{CM}=
(S-W^2+m_l^2)/2\sqrt{S}$, the incoming neutrino and outgoing lepton
energies in the neutrino--nucleon CM frame. It is also useful to
perform the phase space integrals in the other order around, which
allows one to find the $d\sigma/dq^2$ differential cross section. The
total range of $q^2$ is given by  $q^2_{\rm min} (W=m_\pi+M) \le q^2
\le q^2_{\rm max} (W=m_\pi+M)$ and for a given $q^2$,  the outgoing $\pi
N$ invariant mass $W$ varies
\begin{equation}
W_{\rm min} \equiv M+m_\pi \le
W \le \left [M^2+q^2+2M\left(\frac{q^2}{q^2-m^2_l}
E-\frac{m^2_l-q^2}{4E}\right)\right]^\frac12 \equiv W_{\rm max} (q^2)
\label{eq:wmax}
\end{equation}
where in all equations above, $m_l$ should be set to zero for NC
processes. 

\begin{figure}[tbh]
\begin{center}
\makebox[0pt]{\includegraphics[scale=0.7]{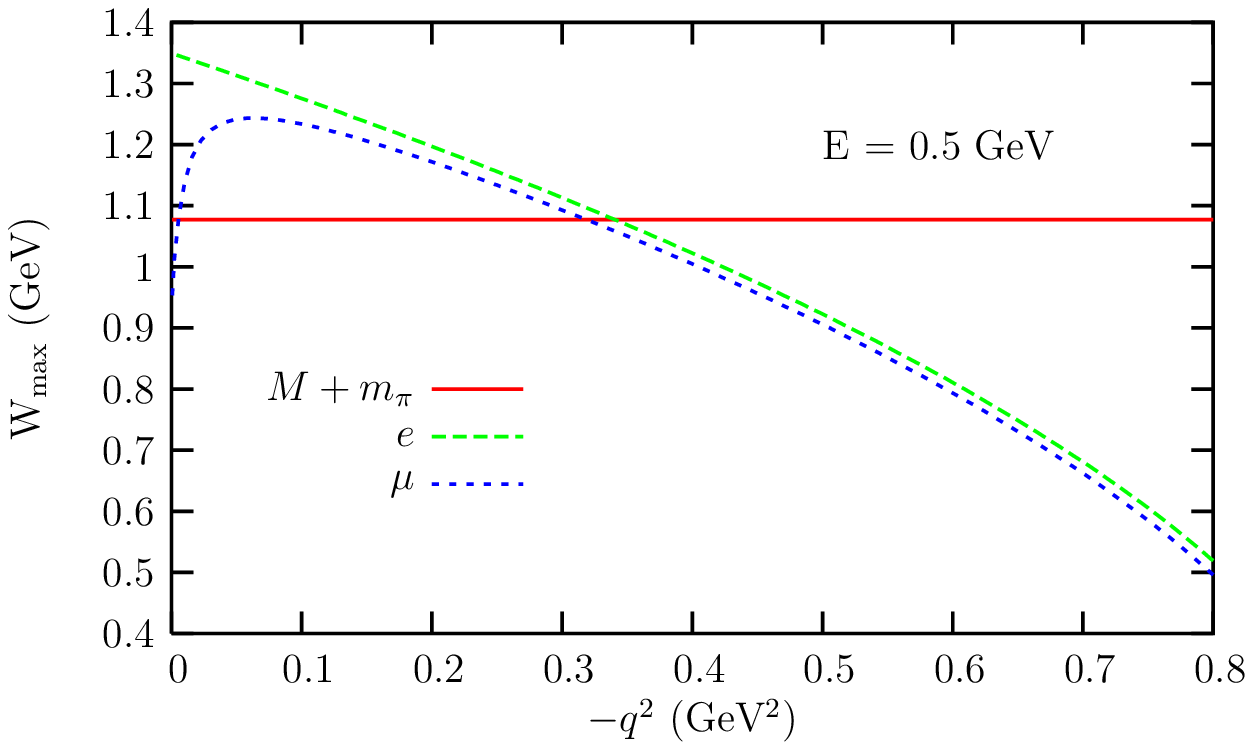}\hspace{1cm}\includegraphics[scale=0.7]{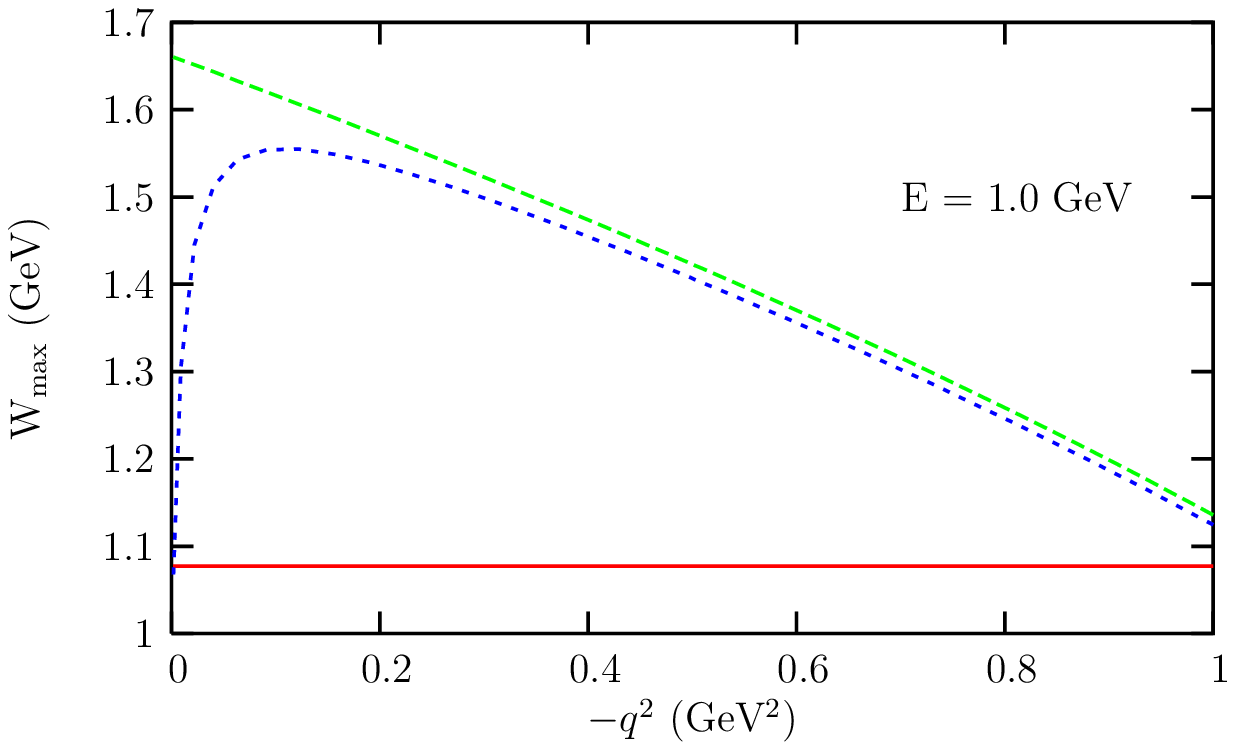}}\\\vspace{1cm}
\makebox[0pt]{\includegraphics[scale=0.7]{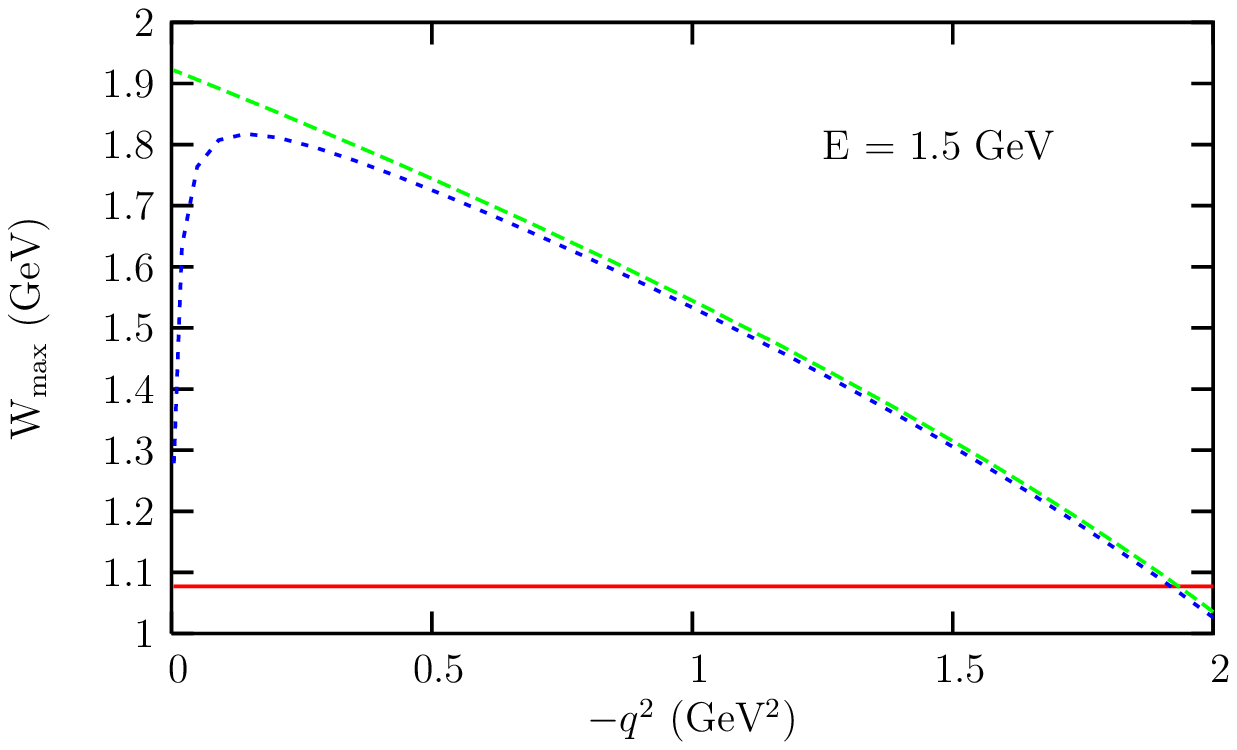}\hspace{1cm}\includegraphics[scale=0.7,bb= 124 568 480 779]{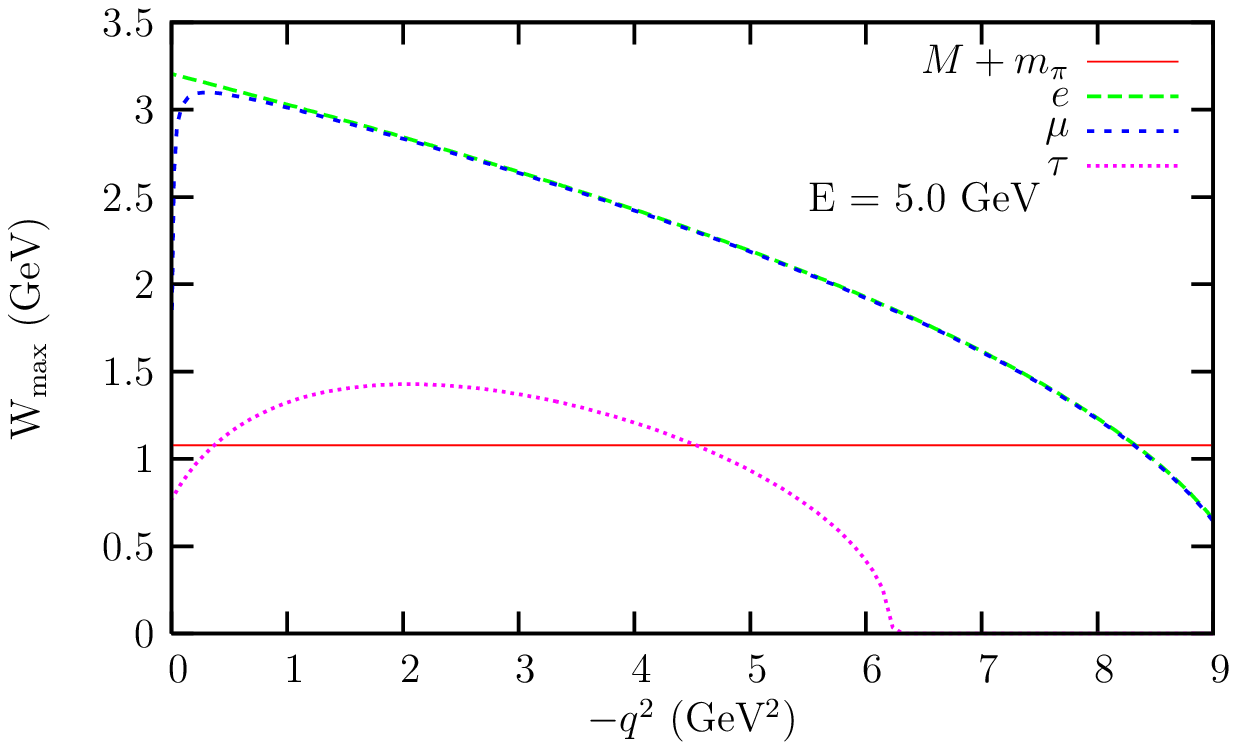}}
\end{center}
\caption{\footnotesize Upper integration limit $W_{\rm max}$ as a
  function of $q^2$ (Eq.~(\protect\ref{eq:wmax})) for incoming neutrino LAB
  energies $E=0.5, 1.0, 1.5$ and 5 GeV and different outgoing lepton
  masses. In all case the horizontal line stands for the phase space
  threshold $M+m_\pi$.  }\label{fig:wmax}
\end{figure}

In the outgoing $\pi N$ CM frame we will also use that
\begin{equation}
\int_0^{+\infty}\frac{d|\vec{k}_\pi| |\vec{k}_\pi|^2}{E_\pi}
\delta(p^{\prime\,0}+k_\pi^0-q^0-p^0)  =
\frac{|\vec{k}_\pi|E'_N}{W}\Big|_{CM} = 
\frac{W^2+M^2-m_\pi^2}{4W^3}\times \lambda^\frac12(W^2,M^2,m_\pi^2).
\end{equation}
Since the main dynamical ingredients of our model are the excitation
of the  $\Delta$ resonance and the non-resonant contributions deduced
from the leading SU(2) non-linear $\sigma$ lagrangian involving pions
and nucleons, we will concentrate in the $M+m_\pi \le W\le 1.3-1.4$
GeV region. For larger invariant masses, the chiral expansion will not
work, or at least the lowest order used here will not be
sufficient~\cite{IAM,IAMpipi}. Moreover, the effect of heavier
resonances will become much more important~\cite{LPP06}. Thus, we will
limit the available phase--space to guarantee that the invariant mass $W$
will lie in the above range. For a fixed incoming neutrino energy,
imposing an upper limit in $W$ will lead to different amounts of
phase-space reduction depending on $q^2$ (see
Fig.~\ref{fig:wmax}). For neutrino energies of about 1 GeV of
relevance in the CC  ANL~\cite{anl} and BNL~\cite{bnlviejo,bnl} bubble
chamber experiments, great part of the available phase space satisfies
the $W\le 1.3-1.4$ constrain. As the neutrino energy increases, the
$q^2$ interval which leads to CM $\pi N$ energies around the
$\Delta-$resonance pole gets reduced and the corresponding kinematic
cuts performed by the various experiments produces a significant
reduction of statistics.

We will see, as it also happens in the pion electroproduction
case~\cite{GNO97}, that the inclusion of non-resonant terms (background terms)
plays a crucial role close to the $M+m_\pi$ threshold.

\begin{figure}[tbh]
\begin{center}
\makebox[0pt]{\includegraphics[scale=0.74]{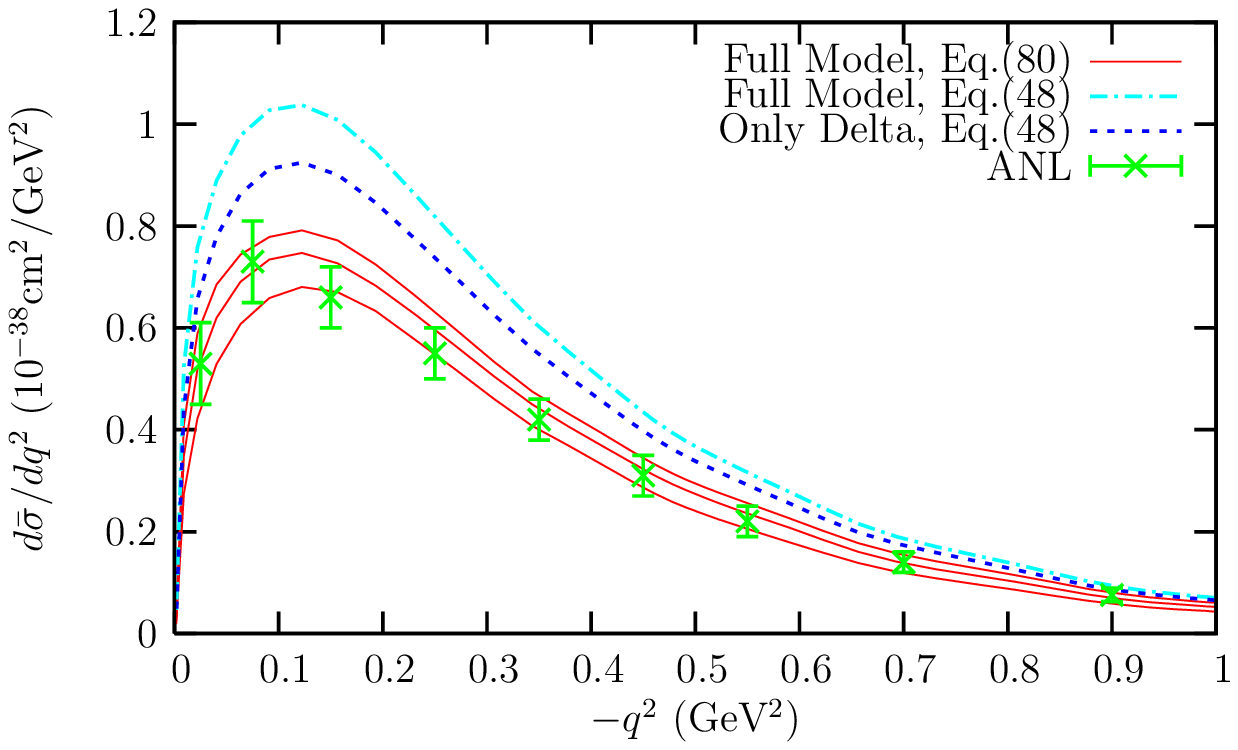}\hspace{-6.2cm}\includegraphics[scale=0.74]{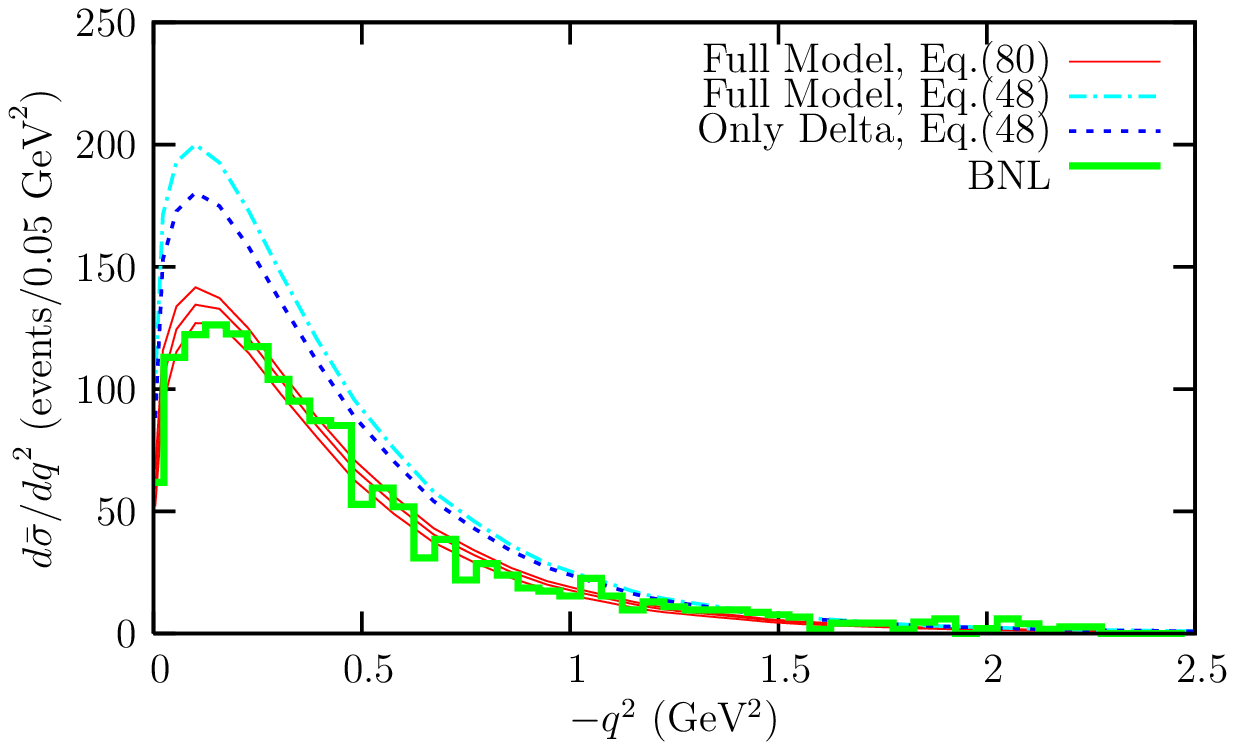}}
\end{center}
\vspace{-15cm}
\caption{\footnotesize Flux averaged $q^2-$differential $\nu_\mu p \to
  \mu^- p \pi^+$ cross section $\protect\int_{M+m_\pi}^{1.4\,{\rm
  GeV}}dW \frac{d\,\overline{\sigma}_{\nu_\mu \mu^-}}{dq^2dW} $
  compared with the ANL~\cite{anl} (left) and BNL~\cite{bnlviejo} (right)
  experiments. Dashed lines stand for the contribution of the
  excitation of the $\Delta^{++}$ resonance and its subsequent decay
  ($\Delta P$ mechanism) with $C_5^A(0)=1.2$ and $M_{A\Delta}= 1.05$
  GeV. Dashed--dotted and central solid lines are obtained when the
  full model of Fig.~\ref{fig:diagramas} is considered with
  $C_5^A(0)=1.2,\, M_{A\Delta}= 1.05$ GeV (dashed-dotted) and with our
  best fit parameters $C_5^A(0)=0.867,\, M_{A\Delta}= 0.985$ GeV
  (solid). In addition, we also show the 68\% CL bands (solid lines)
  deduced from the Gaussian correlated errors quoted in
  Eq.~(\protect\ref{eq:besfit}).}\label{fig:anl-bnlq2}
\end{figure}

\begin{figure}[t]
\begin{center}
\makebox[0pt]{\includegraphics[scale=0.74]{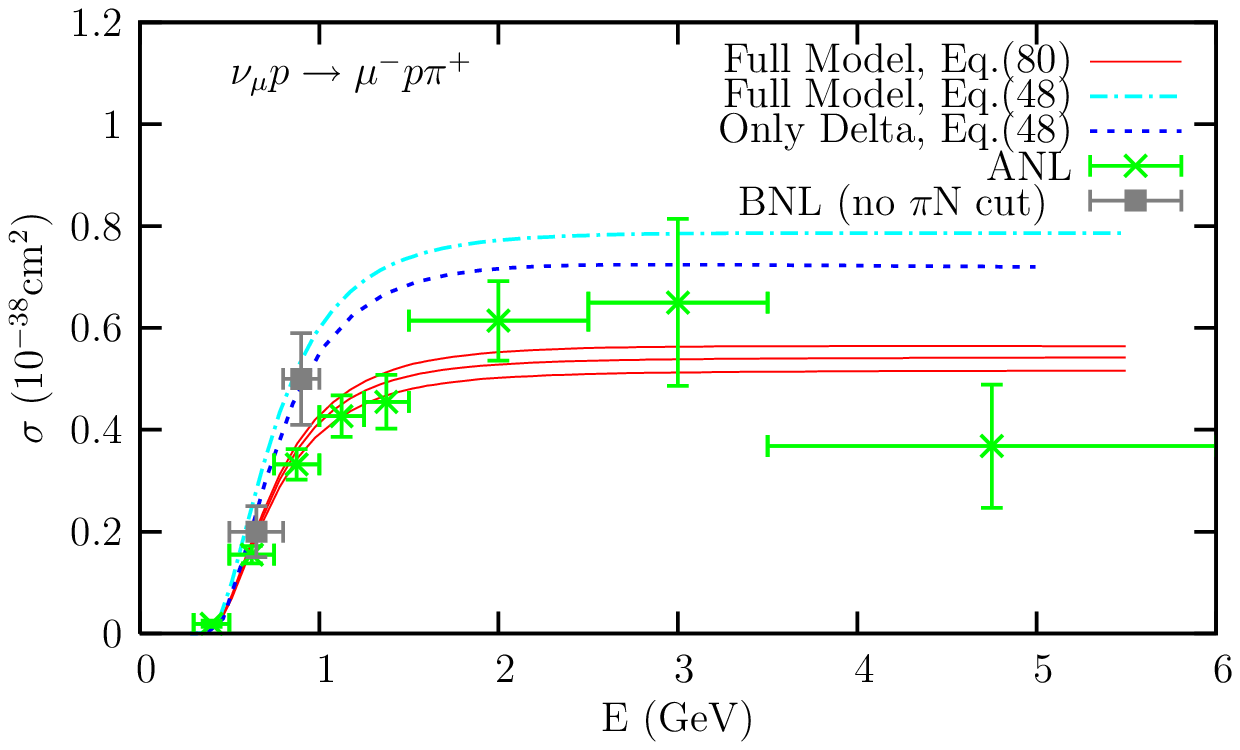}}\\\vspace{-16cm}
\makebox[0pt]{\includegraphics[scale=0.74]{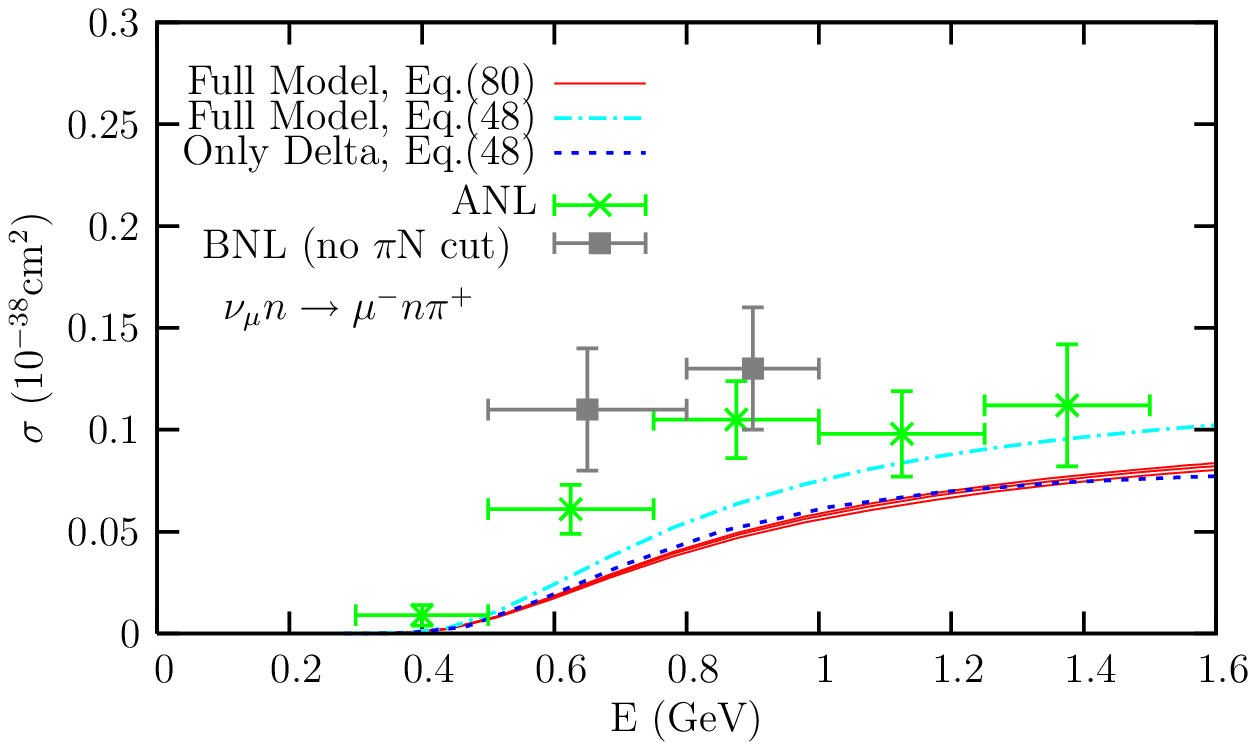}\hspace{-6.2cm}\includegraphics[scale=0.74]{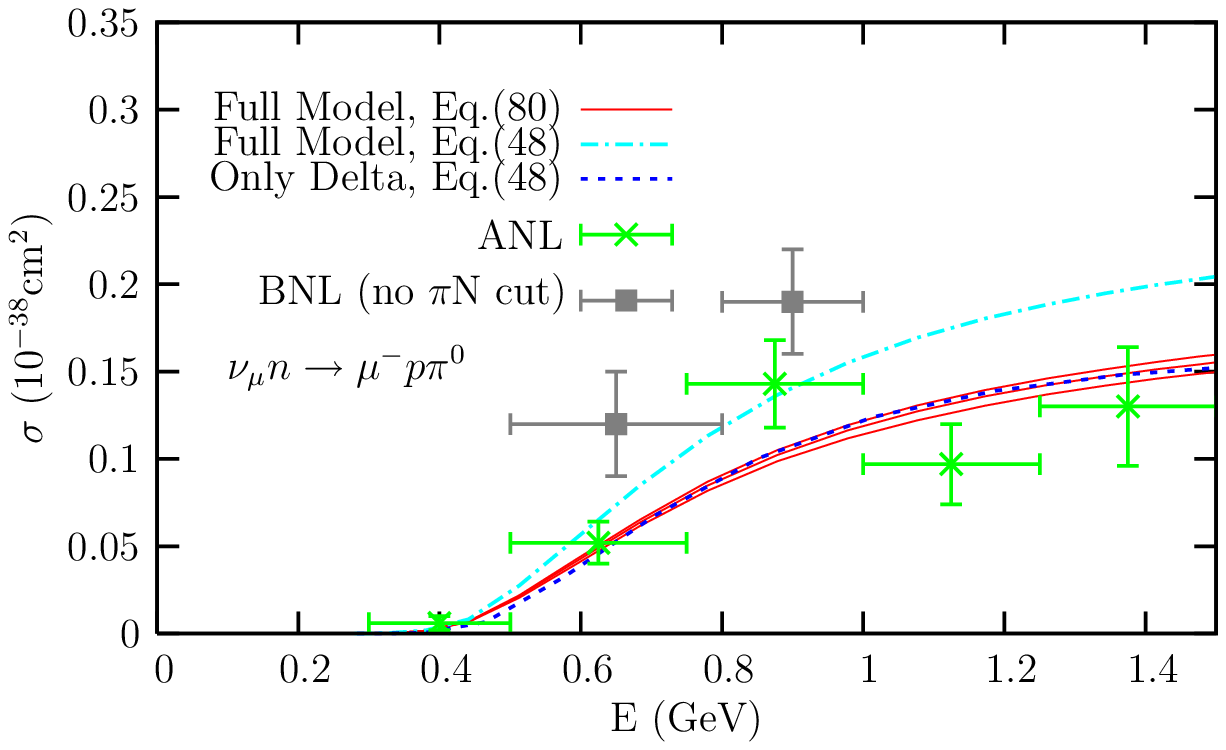}}
\end{center}
\vspace{-14.5cm}
\caption{\footnotesize Experimental and theoretical results for $\nu_\mu
p \to \mu^-p\pi^+$, $\nu_\mu n \to \mu^-p\pi^0$ and $\nu_\mu n \to
\mu^-n\pi^+$ cross sections, as a function of the neutrino
energy. The ANL results~\cite{anl} and theoretical cross sections incorporate
 the kinematical cut $W<1.4$ GeV. Dashed lines stand for the contribution
  of the excitation of the $\Delta$ resonance and its subsequent
  decay ($\Delta P$ mechanism) with $C_5^A(0)=1.2$ and $M_{A\Delta}=
  1.05$ GeV. Dashed--dotted and central solid lines are obtained when the
  full model of Fig.~\ref{fig:diagramas} is considered with
  $C_5^A(0)=1.2,\, M_{A\Delta}= 1.05$ GeV (dashed-dotted) and with our best
  fit parameters $C_5^A(0)=0.867,\, M_{A\Delta}= 0.985$ GeV
  (solid). In addition, we also show the 68\% CL bands (solid lines) deduced
  from the Gaussian correlated errors quoted in
  Eq.~(\protect\ref{eq:besfit}). We also display BNL cross
  section data from Ref.~\cite{bnlviejo} which do not include the
  $W<1.4$ GeV cut in the $\pi N$ invariant mass (see text). }
\label{fig:anl-integrated}
\end{figure}
\subsection{CC pion production cross sections}      
There exist several sets of data taken and analyzed in the late
seventies and early eighties. The most detailed studies, including
measurements, not only of the totally integrated neutrino cross
sections, but also of the neutrino flux--averaged $q^2$ and some
angular distributions were made in the ANL 12-foot bubble
chamber~\cite{anlviejo,anl} and in the BNL 7-foot deuterium-filled
bubble chamber~\cite{bnlviejo,bnl}. In both experiments the bubble
chambers were exposed to a wide-band of muon-type-neutrino beams with
average energies of approximately 1 GeV (ANL) and 1.6 GeV (BNL) and
events for the $\nu_\mu p \to \mu^-p\pi^+$, $\nu_\mu n \to
\mu^-p\pi^0$ and $\nu_\mu n \to \mu^-n\pi^+$ reactions, with and
without a $W\le 1.4$ GeV cut, were obtained. The ANL experiment used
hydrogen and deuterium targets, though most of data come from
deuterium exposure.  Incoming neutrino energy distributions can be
found in Figure 8 of Ref.~\cite{anl-flujo} and in Figure 7 of
Ref.~\cite{bnl-flujo} for the ANL and BNL experiments,
respectively\footnote{Flux--averaged $q^2$ differential cross sections
  read
\begin{equation}
\frac{d\,\overline{\sigma}}{dq^2}=
\frac{1}{{\cal N}}\int_{E^{\rm
min}}^{E^{\rm
max}}dE\frac{d\sigma(E)}{dq^2}\ \Phi(E),\quad
 {\cal N}= \int_{E^{\rm
min}}^{E^{\rm
max}}\Phi(E)\,dE,\label{eq:av}
\end{equation}
and similarly for other flux-averaged differential cross sections. }.

Muon-type-antineutrino (energy beam peaked around 1.5 GeV) induced
total cross sections off the proton (neutron) for final $\pi^- p$ and $\pi^0
n$ ($\pi^- n$) channels with and without the invariant mass cut
$W<1.4$ GeV were measured in the Gargamelle propane experiment at CERN
PS~\cite{cern1}.

There also exist experiments at higher neutrino energies carried out
at the FNAL 15 foot bubble chamber~\cite{FNAL} (neutrino energies from
5 to 100 GeV) and at CERN~\cite{cern2}. In this latter case, a
hydrogen target was illuminated with a wide band neutrino and
antineutrino beams (energies from 5 to 120 GeV), the mean event energy
being about 25 GeV. At such high energies the integrated cross section
remains constant with high accuracy, so the exact value of neutrino
energy is not important. The implementation of the invariant mass cut
$W<1.4$ GeV reduces significantly the statistics and we will not
consider these data sets in this work.

\begin{figure}[tbh]
\begin{center}
\makebox[0pt]{\includegraphics[scale=0.74]{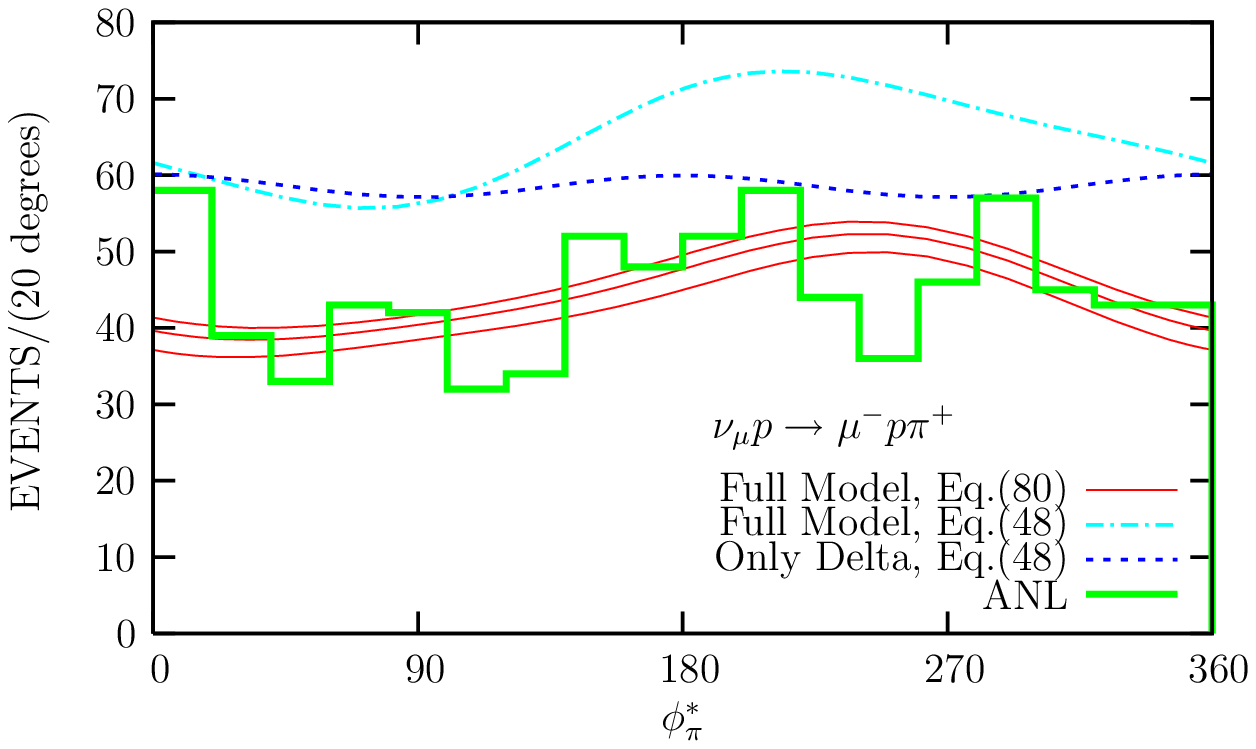}\hspace{-6.2cm}\includegraphics[scale=0.74]{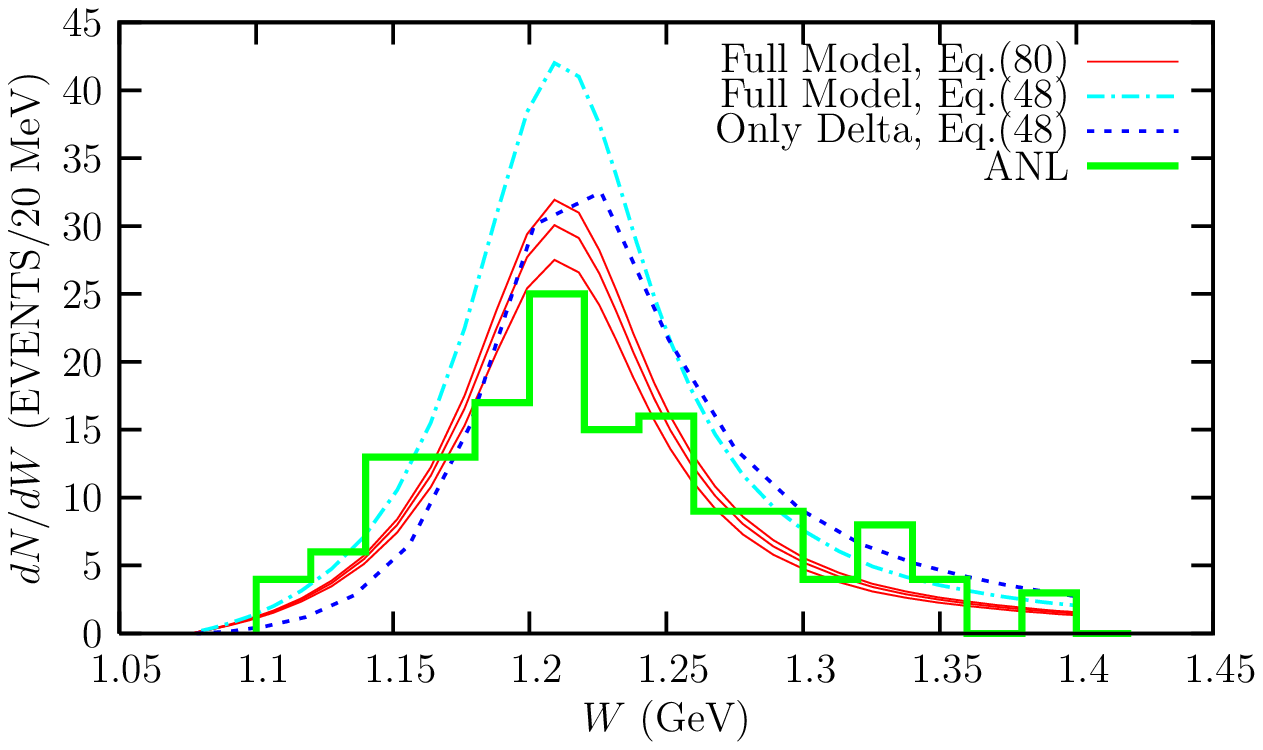}}
\end{center}
\vspace{-15cm}
\caption{\footnotesize Flux averaged ANL distribution of events in the
   pion azimuthal angle with $W<1.4$ GeV (left) and in the $\pi N$
   invariant mass (right) for $\mu^-p\pi^+$ final state. Data taken
   from Refs.~\cite{anl} and ~\cite{anl-cam73}, respectively. Dashed
   lines stand for the contribution of the excitation of the
   $\Delta^{++}$ resonance and its subsequent decay ($\Delta P$
   mechanism) with $C_5^A(0)=1.2$ and $M_{A\Delta}= 1.05$
   GeV. Dashed--dotted and central solid lines are obtained when the
   full model of Fig.~\ref{fig:diagramas} is considered with
   $C_5^A(0)=1.2,\, M_{A\Delta}= 1.05$ GeV (dashed-dotted) and with
   our best fit parameters $C_5^A(0)=0.867,\, M_{A\Delta}= 0.985$ GeV
   (solid). In addition, we also show the 68\% CL bands (solid lines)
   deduced from the Gaussian correlated errors quoted in
   Eq.~(\protect\ref{eq:besfit}).}\label{fig:phi}
\end{figure}

We start looking at the flux averaged $q^2$ differential cross
sections for the reaction $\nu_\mu p \to \mu^-p\pi^+$ measured by the
ANL and BNL experiments (Fig.~\ref{fig:anl-bnlq2}). In the latter
experiment, the cross section overall normalization is not provided.
The $\Delta$ vector form factors are fixed by electroproduction data,
while the axial weak $\Delta N$ transition form factors have been
adjusted in such a way that the $\Delta^{++}$ contribution alone would
lead to a reasonable description of the shape of the BNL data (see for
instance Ref.~\cite{Pa04}). Moreover, this set of axial form factors
also leads to a reasonable description~\cite{Pa05} of the ANL data
(dashed line in left panel of Fig.~\ref{fig:anl-bnlq2}).  The
agreement with the ANL data is certainly worsened when the background
terms, required by chiral symmetry, are considered (dashed-dotted
line).  Since the $C_5^A(q^2)$ controls the largest term of the axial
contribution, this strongly suggests the re-adjustment of this
form--factor. Assuming the same $q^2$ dependence as in
Eq.~({\ref{eq:ca5old}), a $\chi^2-$fit to the flux averaged ($W< 1.4$
GeV) ANL $\nu_\mu p \to \mu^-p\pi^+$ $q^2$ differential cross section
provides
\begin{equation}
C_5^A(0) = 0.867 \pm 0.075, \quad M_{A\Delta}=0.985\pm 0.082\,{\rm
  GeV}, \label{eq:besfit}
\end{equation}
with a Gaussian correlation coefficient $r=-0.85$ and a
$\chi^2/dof=0.4$. This fitted axial mass in the weak $N\Delta$ vertex
is in good agreement with the estimates of about 0.95 GeV and 0.84 GeV
given in the original ANL reference~\cite{anl} and in the work of
Ref.~\cite{Pa05}. On the other hand, we observe a correction of the
order of 30\% to the off diagonal Goldberger-Treiman relation
(Eq.~(\ref{eq:gt})). The lattice QCD results shown in Figure 4 of
Ref.~\cite{negele07} might support the ratio
$\sqrt\frac23\frac{f_\pi}{m_\pi}f^* /C_5^A(0)$ becoming significantly
larger than unity for realistic pion masses.

The ANL data come mostly from deuterium exposure,
and thus deuteron structure effects might affect/are included in this
determination of the $C_5^A$ form--factor.  Such effects were
investigated in Ref.~\cite{ASV99}, where it was estimated that they
were always less than 7\%. The solid line in the left panel of
Fig.~\ref{fig:anl-bnlq2} shows the quality of the fit. We also plot
the 68\% CL band deduced from the statistical errors quoted in
Eq.~(\ref{eq:besfit}). We do not fit to the BNL $q^2$ differential
cross section because this data set is given in arbitrary units. To
fix the overall BNL data scale, we normalize the area under the
theoretical curve, obtained when the full model of
Fig.~\ref{fig:diagramas} is considered with our best fit parameters
$C_5^A(0)=0.867,\, M_{A\Delta}= 0.985$ GeV, to that under the
experimental data. Here again it can be appreciated that, and despite
 the fact that the isospin factor of the $\Delta-$pole mechanism
(with the excitation of the $\Delta^{++}$ resonance and its subsequent
decay) in this channel is bigger than in the others, the effect of the
background terms is quite significant.
\begin{figure}[t]
\begin{center}
\makebox[0pt]{\includegraphics[scale=0.74]{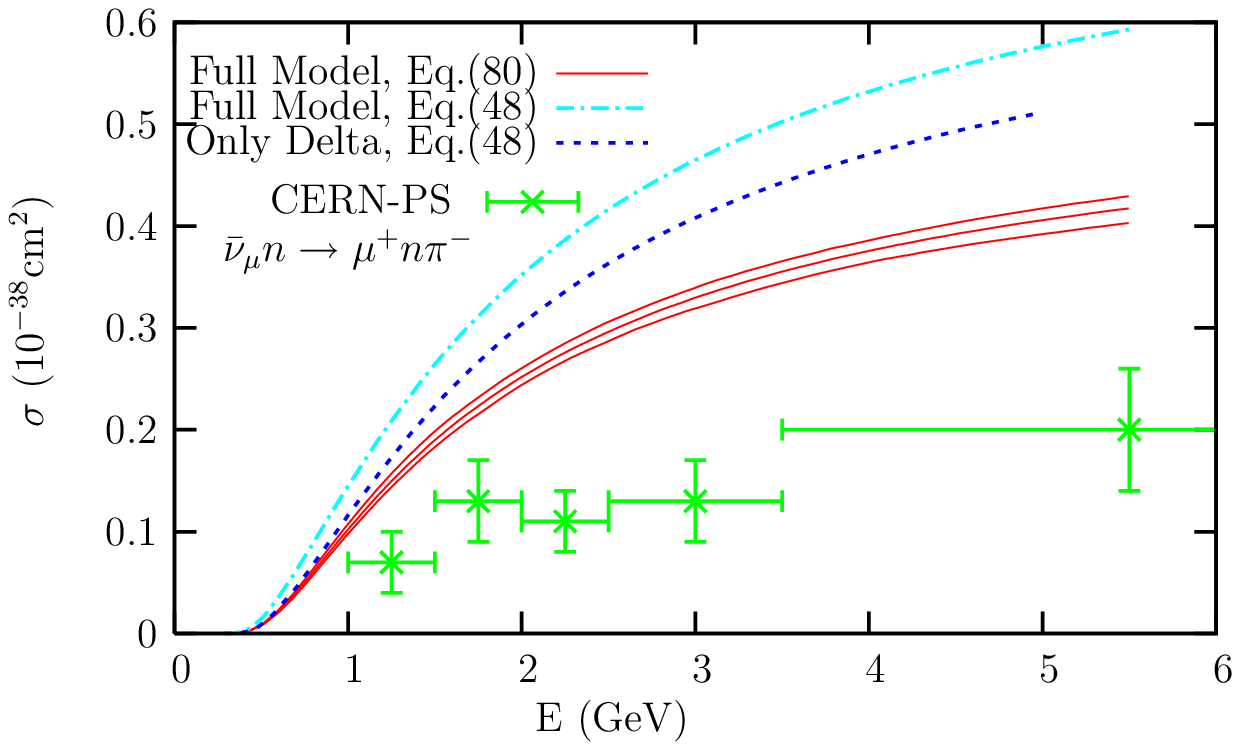}\hspace{-6.2cm}\includegraphics[scale=0.74]{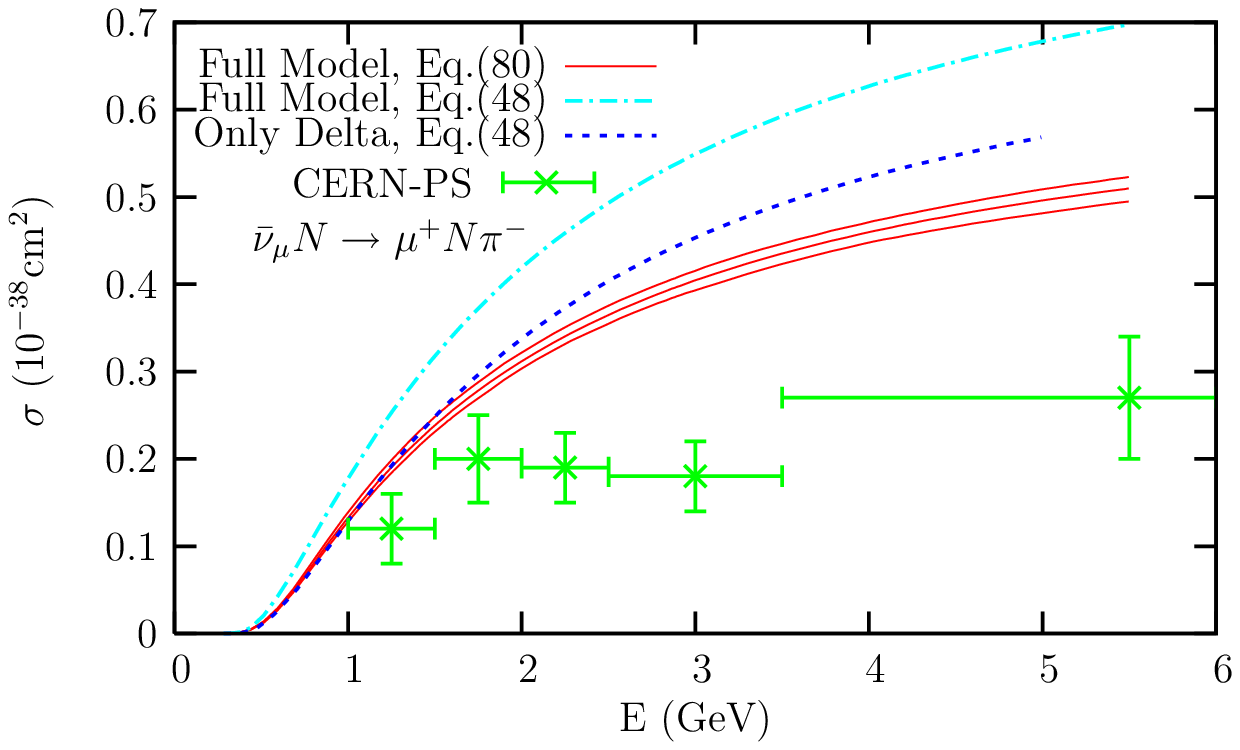}}
\end{center}
\vspace{-15cm}
\caption{\footnotesize Energy dependence of the muon antineutrino
  cross section, with the invariant mass cut $W\le 1.4$ GeV, for the
  $\bar \nu_\mu n \to \mu^+ \pi^- n$ (left panel)  and the $\pi^-$ exclusive
  production (right panel) reactions. 
  Data--points are taken from the CERN PS experiment of
  Ref.~\cite{cern1}. Dashed lines stand for the contribution of the
  excitation of the $\Delta$ resonance and its subsequent decay
  ($\Delta P$ mechanism) with $C_5^A(0)=1.2$ and $M_{A\Delta}= 1.05$
  GeV. Dashed--dotted and central solid lines are obtained when the
  full model of Fig.~\ref{fig:diagramas} is considered with
  $C_5^A(0)=1.2,\, M_{A\Delta}= 1.05$ GeV (dashed-dotted) and with our
  best fit parameters $C_5^A(0)=0.867,\, M_{A\Delta}= 0.985$ GeV
  (solid). In addition, we also show the 68\% CL bands (solid lines)
  deduced from the Gaussian correlated errors quoted in
  Eq.~(\protect\ref{eq:besfit}).}\label{fig:cern-ps}
\end{figure}

Next we show in Fig.~\ref{fig:anl-integrated}, the total ANL $\nu_\mu
p \to \mu^-p\pi^+$, $\nu_\mu n \to \mu^-p\pi^0$ and $\nu_\mu n \to
\mu^-n\pi^+$ cross sections, with the kinematical cut $W<1.4$ GeV, as
a function of the neutrino energy and the predictions of the three
schemes defined above: only $\Delta P$ contribution with
$C_5^A(0)=1.2,\, M_{A\Delta}= 1.05$ GeV and the full model derived in
this work, including background terms, with the latter set of
parameters for $C_5^A(q^2)$ and with that given in
Eq.~(\ref{eq:besfit}). As can be appreciated in the different plots of
the figure we achieve a reasonable description of data, finding the
largest discrepancies in the $\pi^+ n$ channel. The inclusion of the
chiral symmetry background terms derived in this work, brings in an
overall improved description of the three channels as compared to the
case where only the $\Delta P$ mechanism is considered. In the case of
$\pi^+ n$ and $\pi^0 p$ cross sections the reduction of the
contribution of this latter mechanism is compensated by the inclusion
of the background terms. Our results are similar in quality to those
obtained from the model of Ref.~\cite{SUL03}. We also display in the
various plots of this figure, BNL cross section data from
Ref.~\cite{bnlviejo} which do not include the $W<1.4$ GeV cut in the
$\pi N$ invariant mass. For neutrino energies below 1 GeV, the
effects of the  $\pi N$ invariant mass cut is almost negligible (see
Table III  of Ref.~\cite{anl}). Such effects become much more sizeable
for larger neutrino energies (see Table III  of Ref.~\cite{anl}) which
have prevented us to present BNL cross section data in the plots for neutrino
energies above 1 GeV. We observe some degree of inconsistency among
the ANL and BNL measurements.  The present model, including
non-resonant background terms, with a $C_5^A(q^2)$ form factor
consistent with the off diagonal Goldberger-Treiman relation
(Eq.~(\ref{eq:gt})) would lead to a better description of the BNL data
(see dashed--dotted lines).

In Fig.~\ref{fig:phi} we compare the pion azimuthal and pion-nucleon
invariant mass distributions (neutrino flux averaged) predicted by the
different models examined here with that measured in the ANL
experiment~\cite{anl,anl-cam73}. For both plots, we normalize the area
under the theoretical curve obtained when the full model of
Fig.~\ref{fig:diagramas} is considered with parameters
$C_5^A(0)=0.867,\, M_{A\Delta}= 0.985$ GeV, to that under the
experimental data. The inclusion of chiral background terms leads to a
more pronounced $\phi_\pi^*$ dependence improving in this way the
agreement with the observed event distribution in the ANL
experiment. In the right panel of Fig.~\ref{fig:phi} we show the
$W-$distribution of ANL events, which clearly shows   the
$\Delta (1232)$ peak.  The chiral background terms dominate the
distribution near the pion production threshold, and
they also produce a slight shift of the maximum of the distribution 
to lower invariant masses.

In Fig.~\ref{fig:cern-ps}, we compare the predictions of our model
with the CERN-PS muon antineutrino cross section data of
Ref.~\cite{cern1}. Our model provides larger cross sections than the
experiment, but nevertheless, we find here again a reasonable
description of the data, which is certainly better than when the
$\Delta P$ mechanism alone is considered.  We would like to remind
here that from isospin symmetry we have $\langle n \pi^- | j^\mu_{\rm
cc-}(0) | n \rangle = \langle p \pi^+ | j^\mu_{\rm cc+}(0) | p
\rangle$. Therefore the only dynamical difference between the $\nu_\mu
p \to \mu^- p \pi^+$, which we describe correctly (see
Fig.~\ref{fig:anl-integrated}), and the $\bar\nu_\mu n \to \mu^+ p
\pi^-$ reactions is the sign of the antisymmetric term of the lepton
tensor ($L_{\mu\sigma}^{(\bar\nu)} = L_{\sigma\mu}^{(\nu)}$ ). Thus,
if one neglects the parity--violating part of the antisymmetric
hadronic tensor $\left(W^{\mu\nu}_{{\rm CC} \pi}\right)^{{\rm
PV}}_{a}$ (see Eq.~(\ref{eq:w1}) in Appendix~\ref{sec:phi}), the cross sections
for both reactions will be the same but for differences in the
interference between the axial and vector contributions of the
hadronic current.  However, this vector--axial interference does not
affect the sum of the cross sections $\sigma(\nu_\mu p \to \mu^- p
\pi^+) + \sigma(\bar\nu_\mu n \to \mu^+ p \pi^-)$, except for its
contribution to the parity--violating part of the symmetric hadronic
tensor. For instance at $E=3$ GeV, the experimental value for this sum
of cross sections is around $0.78\times 10^{-38}$ cm$^2$ while our
best theoretical prediction for it is around 11\% bigger. Considering
the large errors in experimental data we think this difference is not
significative\footnote{We note the discrepancy is significantly larger
($\approx 60\%$) if one uses $C_5^A(0)=1.2$.}. That might suggest that the
discrepancy in the antineutrino cross section might be solved by the
inclusion of relative phases between the vector and the axial current
theoretical contributions.

Relative signs among the different background contributions are well
established, since all of them have been deduced from the same
lagrangian (Eq.~(\ref{eq:lint})) and the vector and axial currents
derived from it. One might think in a possible inconsistency between
the relative signs of background and resonant pieces. However, 
\begin{itemize}
\item These relative signs are consistent with those deduced from a quark model
picture~\cite{QM}. 
\item As already mentioned, the vector part of the model presented
here, reduces to that derived in Ref.~\cite{GNO97} for the $ e N \to
e' N \pi$ reaction. Note the relation 
\begin{equation}
C_3^V(q^2) = \frac{M}{m_\pi} \sqrt{\frac23} f_\gamma(q^2)/e
\end{equation}
with $e=\sqrt{4\pi\alpha}$ the proton charge, between $C_3^V$ and
the usual $\gamma N \Delta$ coupling, $f_\gamma$, used in pion
electroproduction reactions (see Eqs.(5) and (A-12) in
Ref.~\cite{GNO97}). This relation is obtained from the
non-relativistic reduction of the $C_3^V$ Dirac structure, which
dominates in that limit. Using $f_\gamma(0)\approx 0.122$ as in
Ref.~\cite{GNO97}, one obtains $C_3^V(0)\approx 2.2$ in good
agreement with Eq.~(\ref{eq:c3v}). The model of
Ref~\cite{GNO97} described\footnote{ A small relative phase between
  the $\Delta P$
and the background terms was also included in Ref.~\cite{GNO97}. This
phase is deduced from Watson's theorem~\cite{Watson}, it depends on
$q^2$ and $W$, and for the kinematics of interest in this work is
comprised in the range $10-20^{\rm o}$.} reasonably well
the available data for the $ e N \to e' N \pi$ reaction at low and
intermediate energies, including pion angular dependences. This makes
us confident on our election of relative signs between resonant and
non-resonant terms.
\item In what the axial part concerns, we use a consistent sign convention 
for both the diagonal and off-diagonal Goldberger-Treiman's relations (see the
choice of relative signs in Eqs.~(\ref{eq:gtN}) and
~(\ref{eq:gt})). 
\end{itemize}
Nevertheless, in the Appendix~\ref{app:phases} we have
examined the effect of including relative minus signs between the
axial and vector resonant contributions and also between the $\Delta
P$ and the background terms. Changing the relative sign between the
vector and axial contributions of the $\Delta$ mechanism is totally discarded
by the data, while modifying the relative sign between resonant and
non-resonant terms has a little effect, once the $W$ integration is
performed. This can be understood by looking at the $\pi N$ invariant
mass distribution of the right panel in Fig.~\ref{fig:phi}. At the
$\Delta$ peak resonant and non-resonant contributions do not
interfere, since the first one is purely imaginary while the latter
one is real, and thus in this region the relative sign between both
type of contributions is irrelevant. At lower and higher values of
$W$, where the resonant contribution takes also a real part and thus
it has a non-vanishing interference with the background terms, there
exists a constructive and destructive, respectively, interference. A
change of the relative sign between resonant and non-resonant terms
would reverse constructive into destructive interferences, and
vice-versa, but the net effect after integrating in $W$ is
greatly diminished.

After this discussion, we stress that one would need relative phases
(and not merely minus signs) between vector and axial resonant and non
resonant contributions to improve the combined description of neutrino
and antineutrino cross sections. However, very recently S.K. Singh and
collaborators have pointed out~\cite{singh} that nuclear medium
effects and pion absorption (note that the antineutrino data of
Ref.~\cite{cern1} presented in Fig.~\ref{fig:cern-ps} were measured on
freon-propane) were not properly taken into account in the original
work of Bolognese and collaborators~\cite{cern1}.  In this manner, the
apparent discrepancies highlighted by the former discussion might disappear
(see Figs. 17 and 18 of Ref.~\cite{singh}). More accurate data,
possibly available in the next future from the MiniBoone and T2K
experiments, in conjunction with Watson's theorem~\cite{Watson} might
shed light into this interesting issue.
\begin{figure}[t]
\begin{center}
\makebox[0pt]{\includegraphics[scale=0.7]{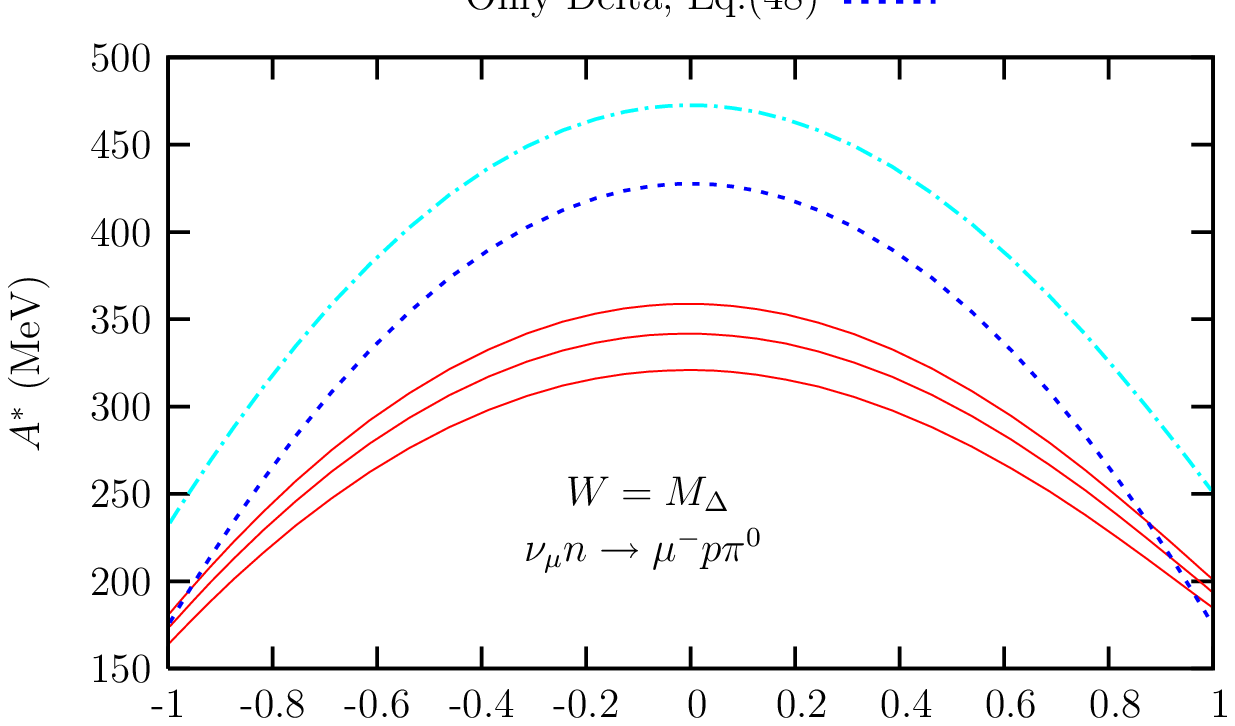}\hspace{-5.2cm}\includegraphics[scale=0.7]{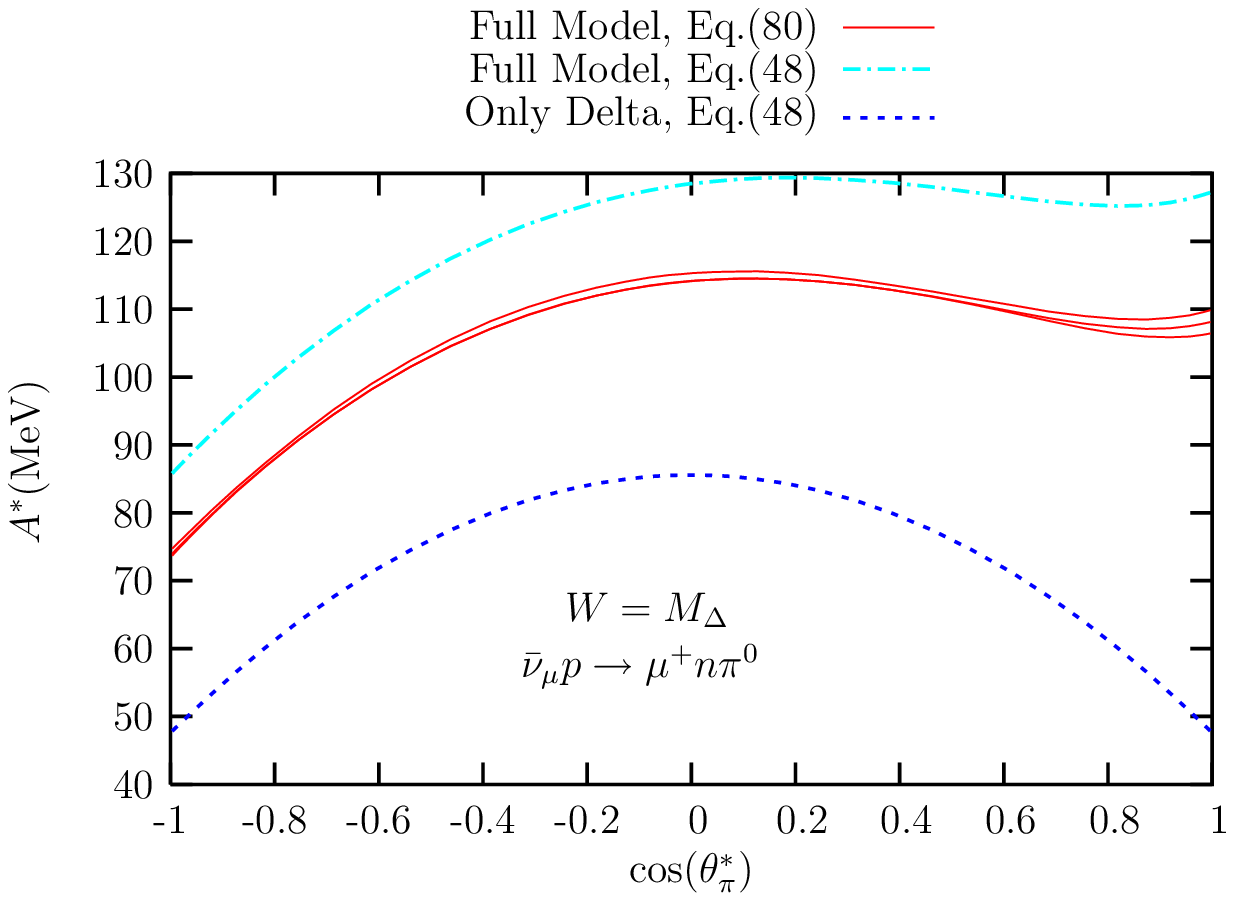}}\\\vspace{-15.6cm}
\makebox[0pt]{\includegraphics[scale=0.7]{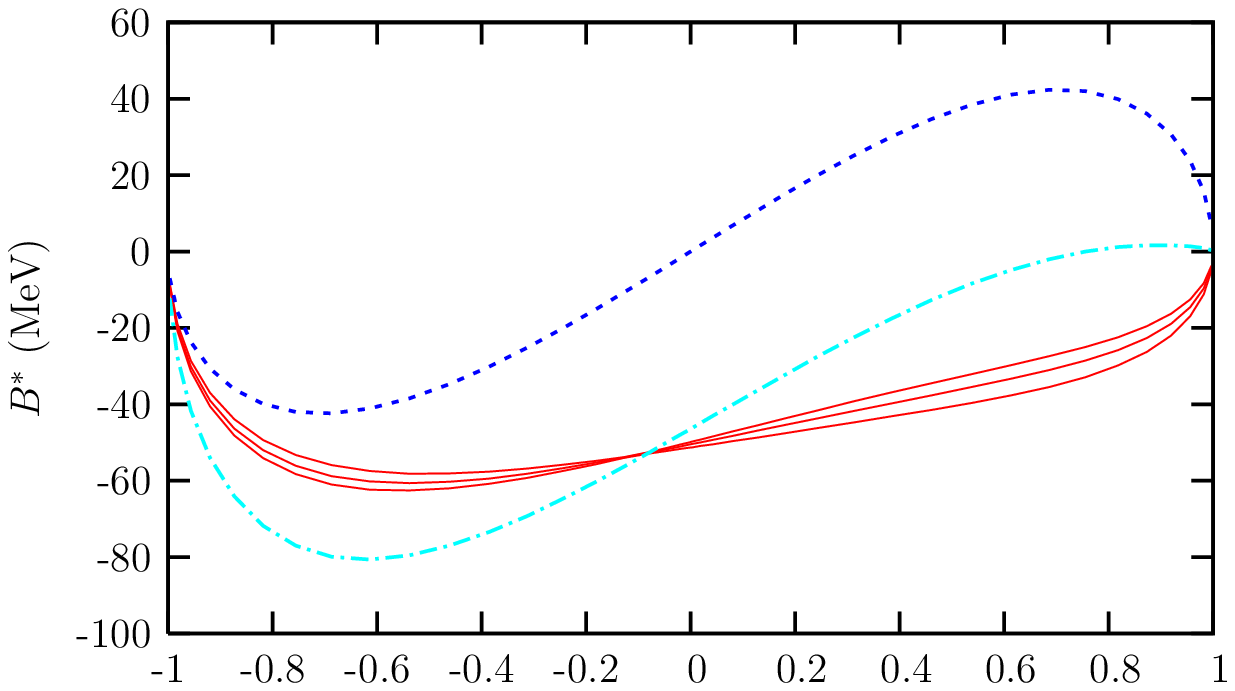}\hspace{-5.2cm}\includegraphics[scale=0.7]{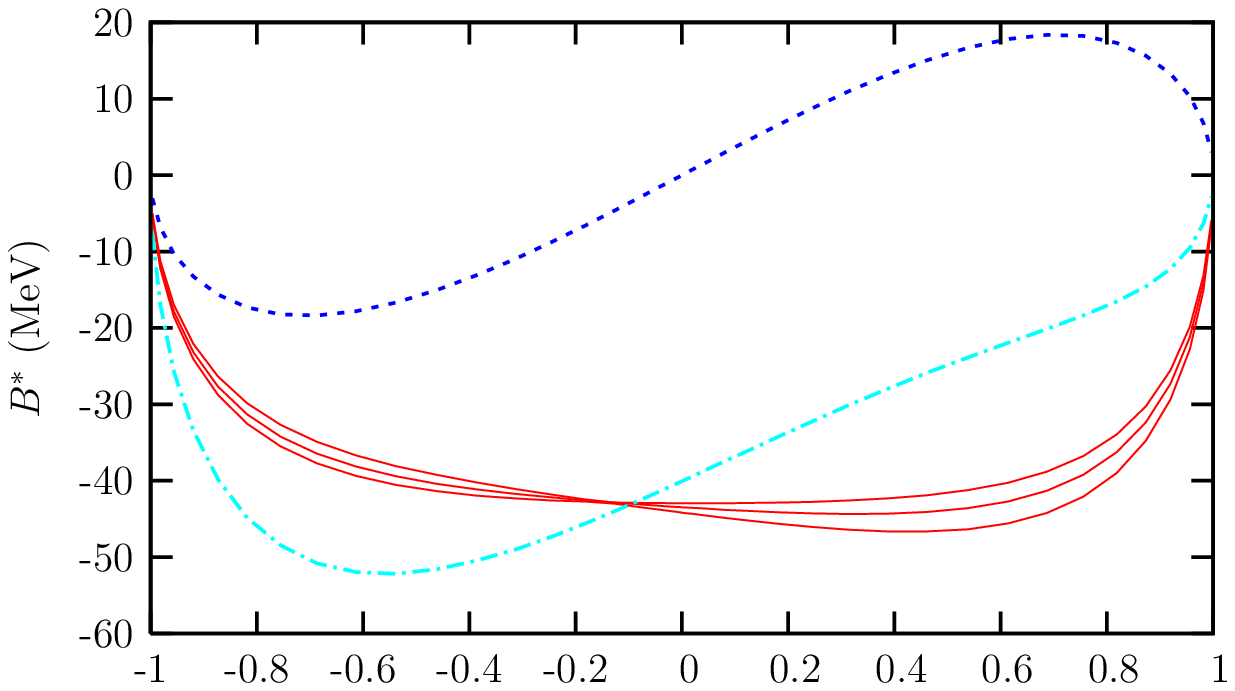}}\\\vspace{-15.6cm}
\makebox[0pt]{\includegraphics[scale=0.7]{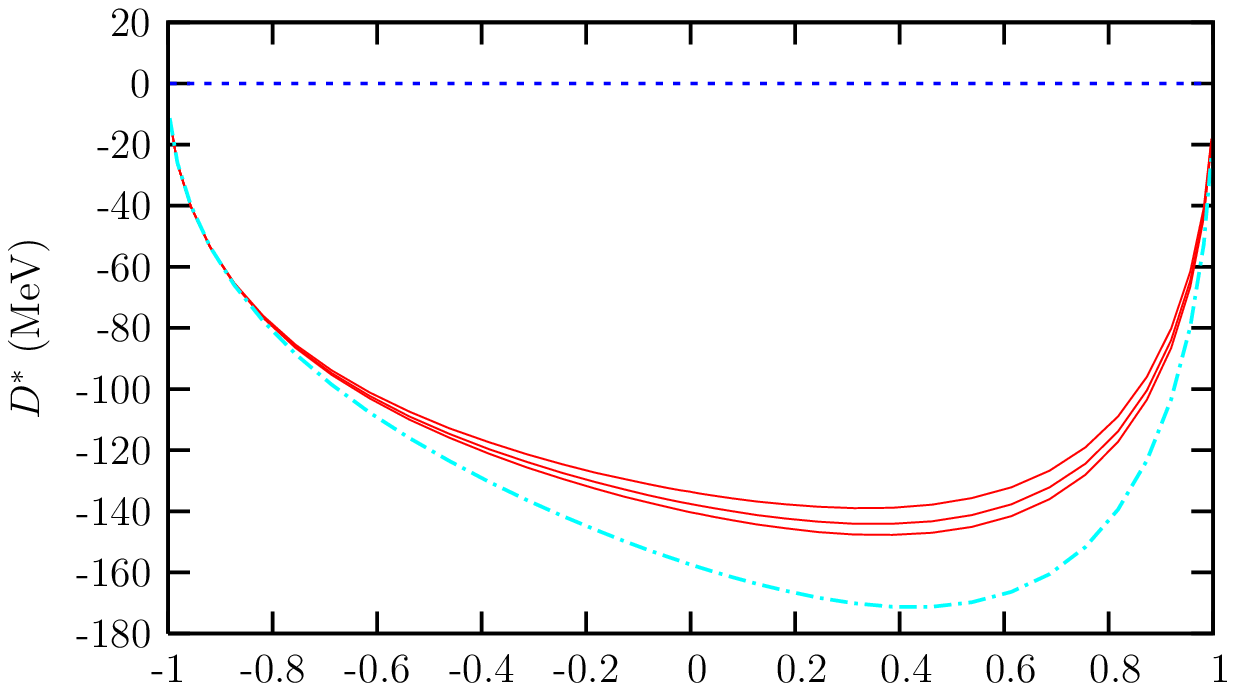}\hspace{-5.2cm}\includegraphics[scale=0.7]{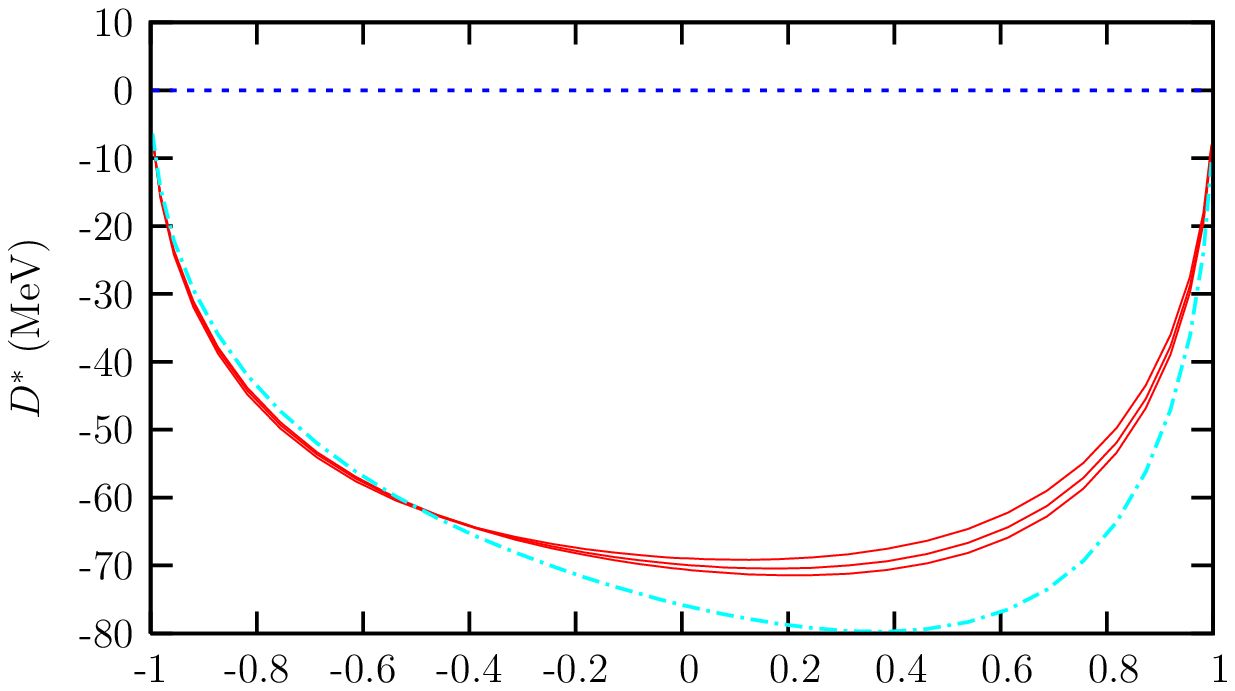}}\\\vspace{-15.6cm}
\makebox[0pt]{\includegraphics[scale=0.7]{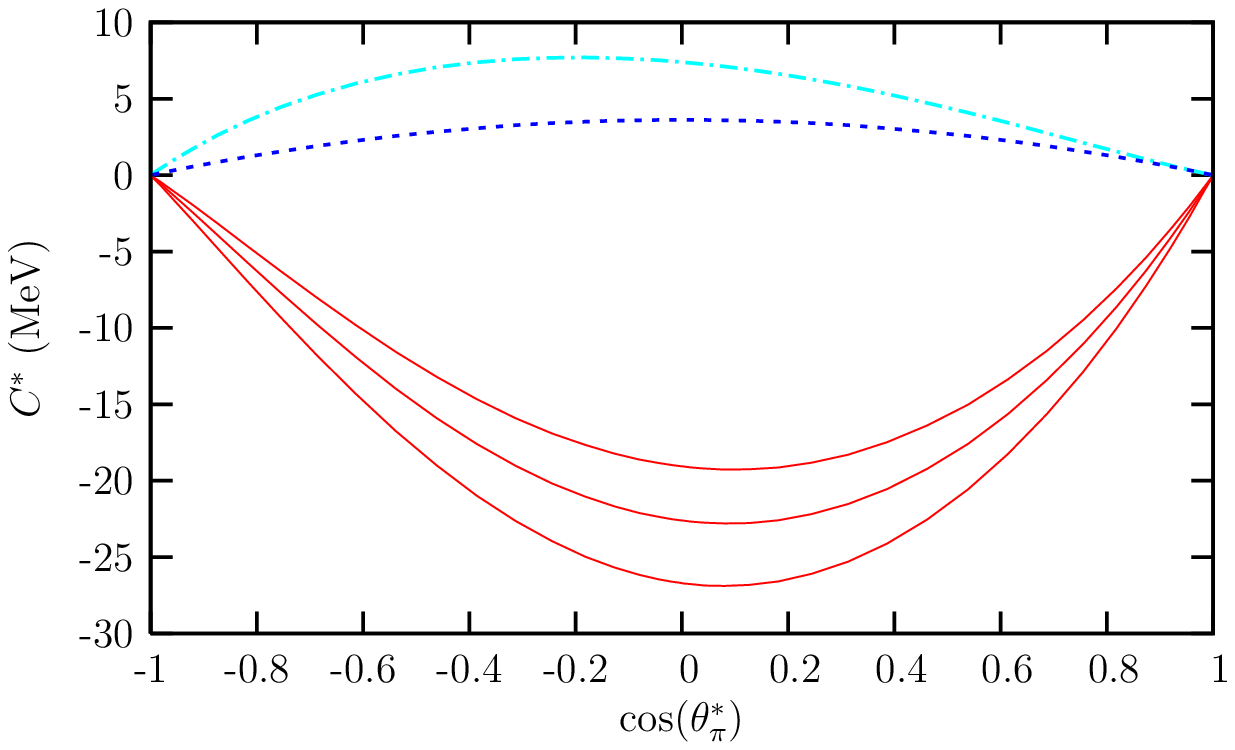}\hspace{-5.2cm}\includegraphics[scale=0.7]{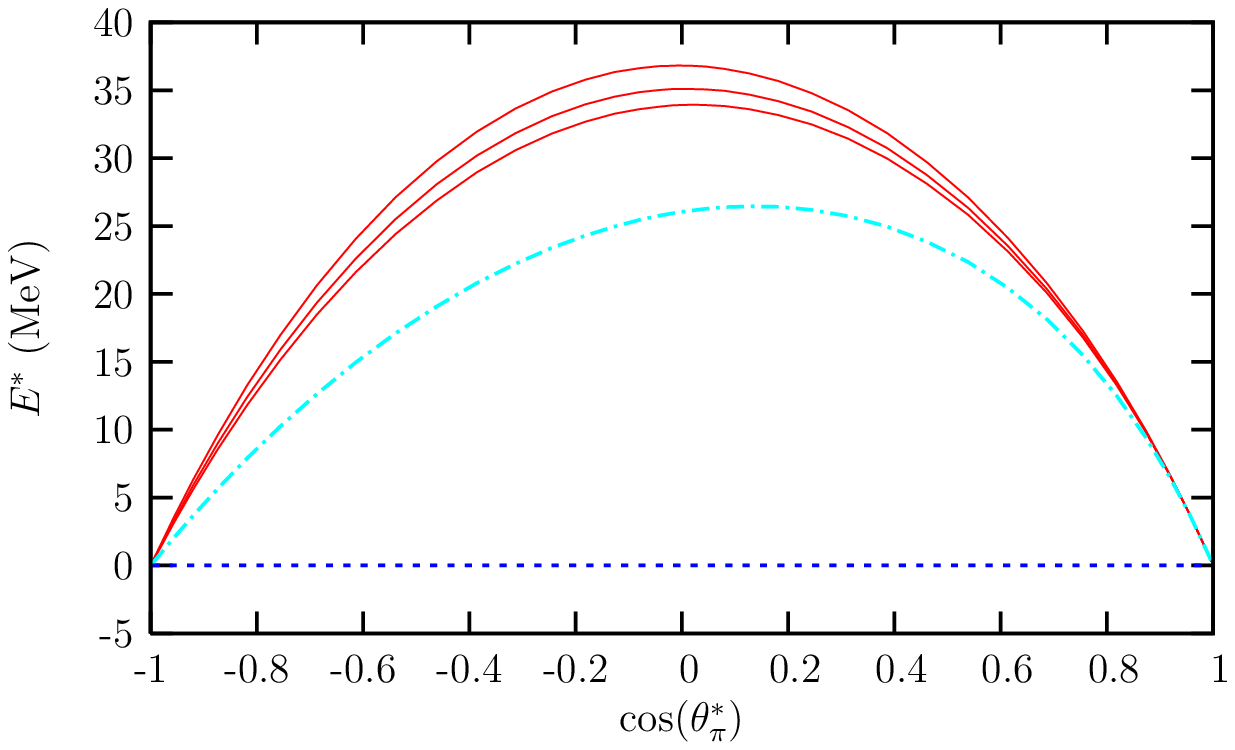}}\end{center}
\vspace{-14.5cm}
\caption{\footnotesize Pion polar angle dependence of the $\phi_\pi^*$
  structure functions defined in Eq.~(\ref{eq:phipi*}) for the
  $\nu_\mu n \to \mu^- p\pi^0$ and $\bar\nu_\mu p \to \mu^+ n\pi^0$
  reactions. The neutrino incoming energy is $E=1.5$ GeV,
  $q^2=-0.5$ GeV$^2$, $W=M_\Delta$ and the pion polar angle is
  referred to the $\pi N$ CM frame.  Neutrino (antineutrino) $A^*$,
  $B^*$ and $D^*$ structure functions are displayed in the three upper
  right (left) hand side plots. The $C^*$ and $E^*$ structure
  functions, which are equal for both reactions, are showed in the
  two panels of the last row. Dashed lines stand for the
  contribution of the excitation of the $\Delta$ resonance and
  its subsequent decay ($\Delta P$ mechanism) with $C_5^A(0)=1.2$ and
  $M_{A\Delta}= 1.05$ GeV.  Dashed--dotted and central solid lines are
  obtained when the full model of Fig.~\ref{fig:diagramas} is
  considered with $C_5^A(0)=1.2,\, M_{A\Delta}= 1.05$ GeV
  (dashed-dotted) and with our best fit parameters $C_5^A(0)=0.867,\,
  M_{A\Delta}= 0.985$ GeV (solid). In addition, we also show the 68\%
  CL bands (solid lines) deduced from the Gaussian correlated errors
  quoted in Eq.~(\protect\ref{eq:besfit}).}
\label{fig:ccw1-abcde}
\end{figure}
\begin{figure}[t]
\begin{center}
\makebox[0pt]{\includegraphics[scale=0.7]{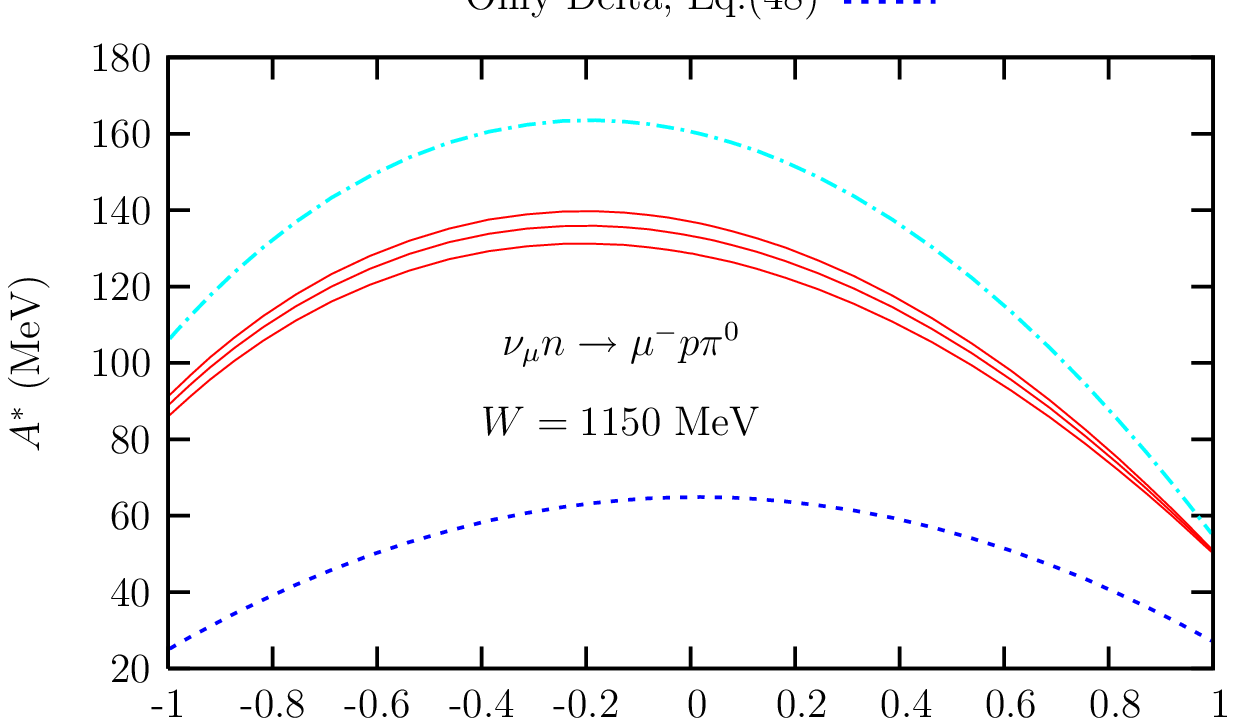}\hspace{-5.2cm}\includegraphics[scale=0.7]{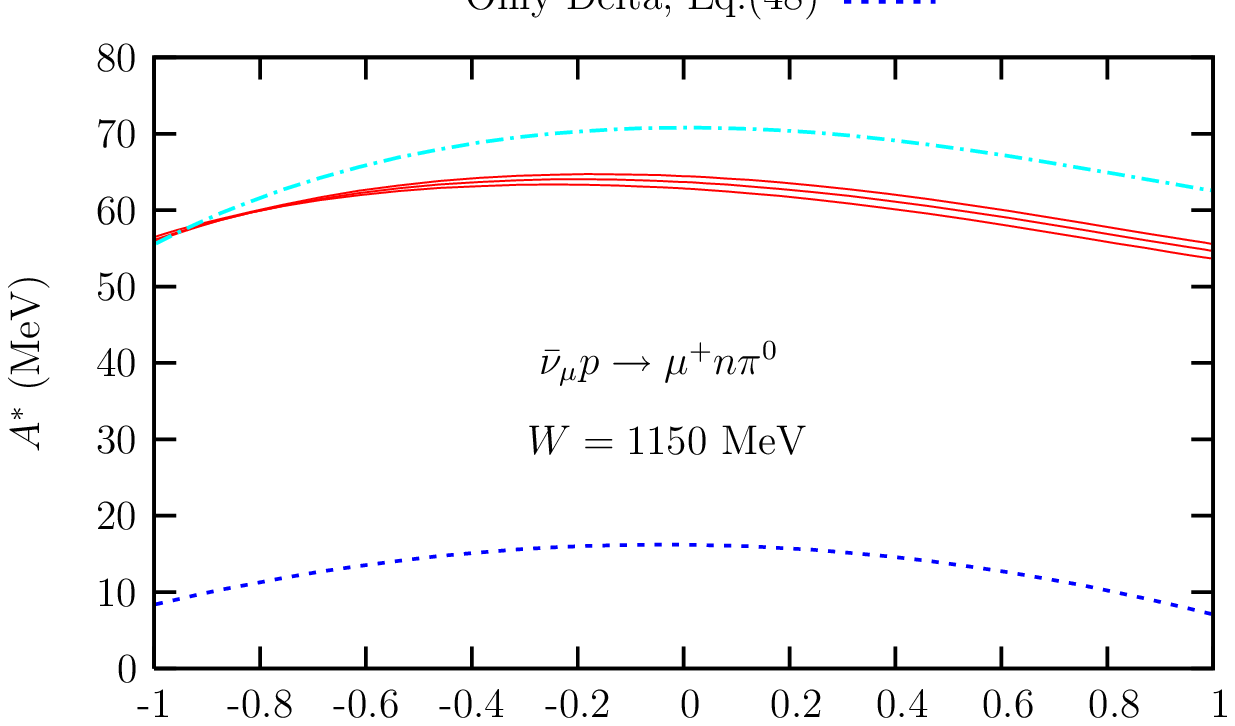}}\\\vspace{-15.6cm}
\makebox[0pt]{\includegraphics[scale=0.7]{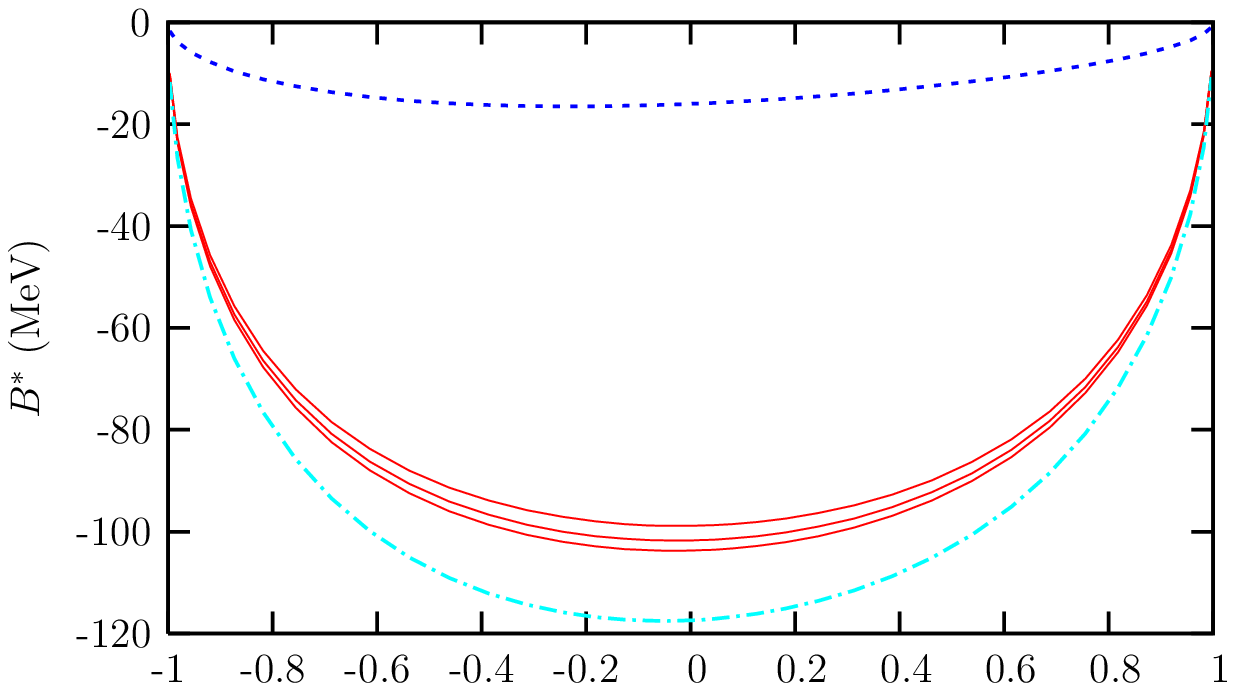}\hspace{-5.2cm}\includegraphics[scale=0.7]{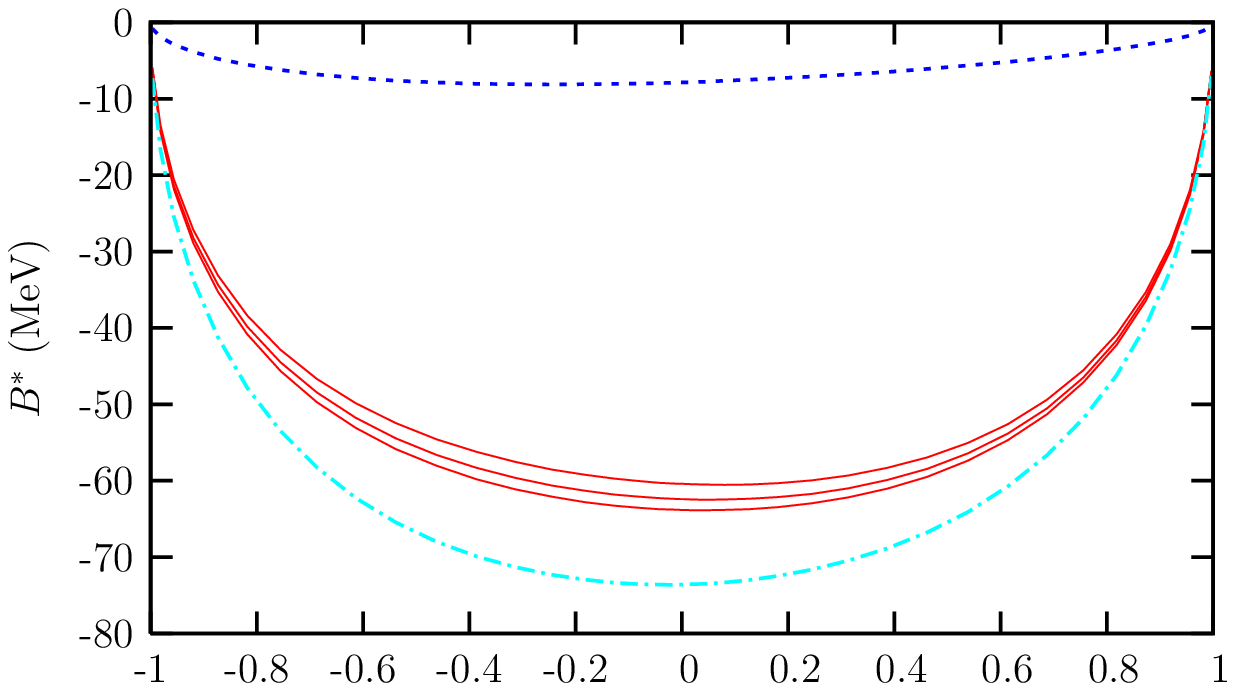}}\\\vspace{-15.6cm}
\makebox[0pt]{\includegraphics[scale=0.7]{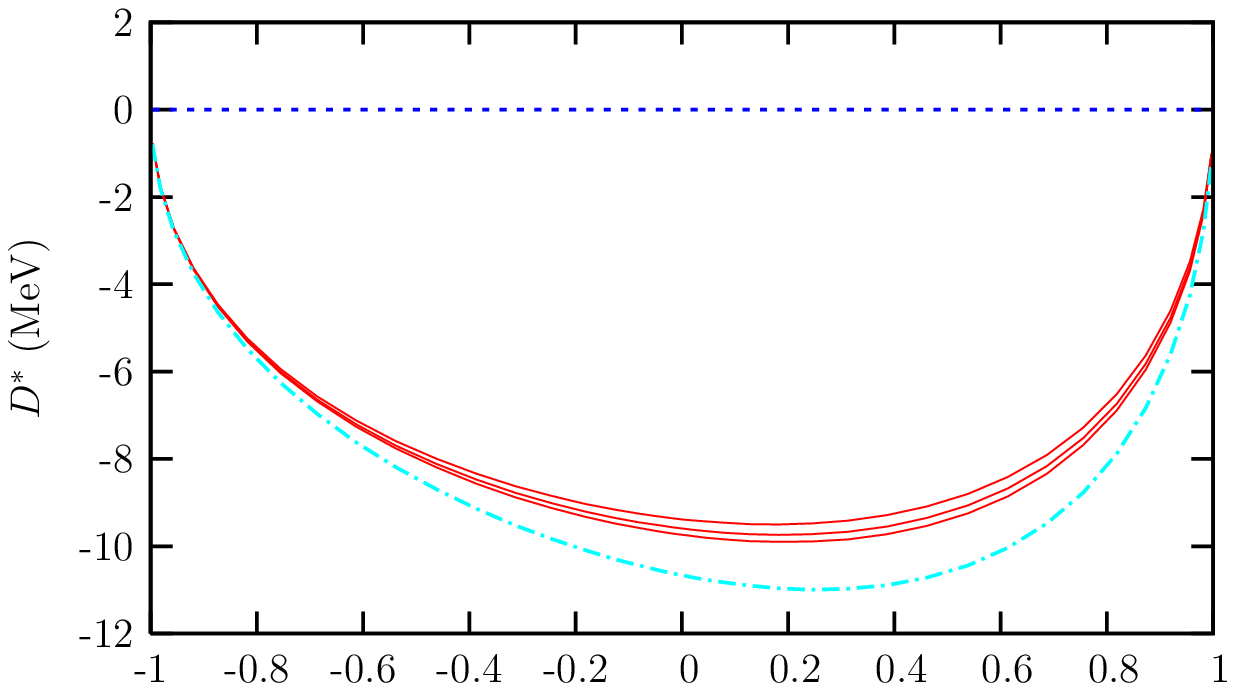}\hspace{-5.2cm}\includegraphics[scale=0.7]{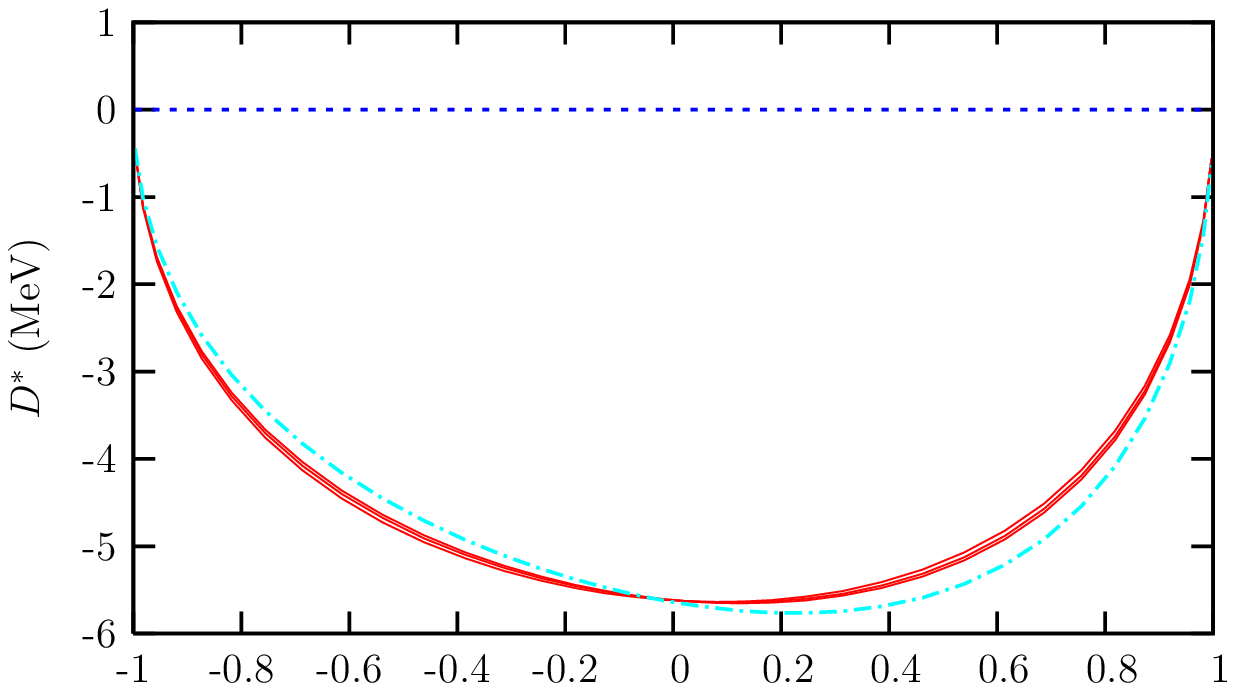}}\\\vspace{-15.6cm}
\makebox[0pt]{\includegraphics[scale=0.7]{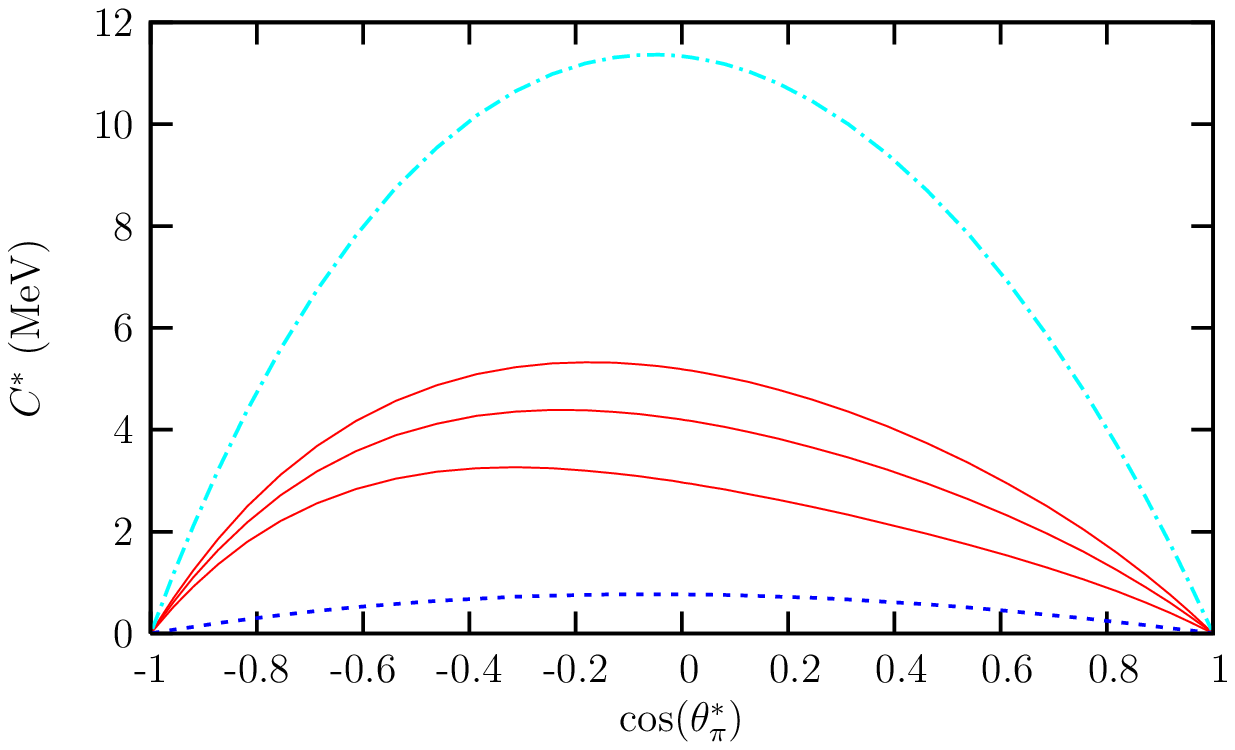}\hspace{-5.2cm}\includegraphics[scale=0.7]{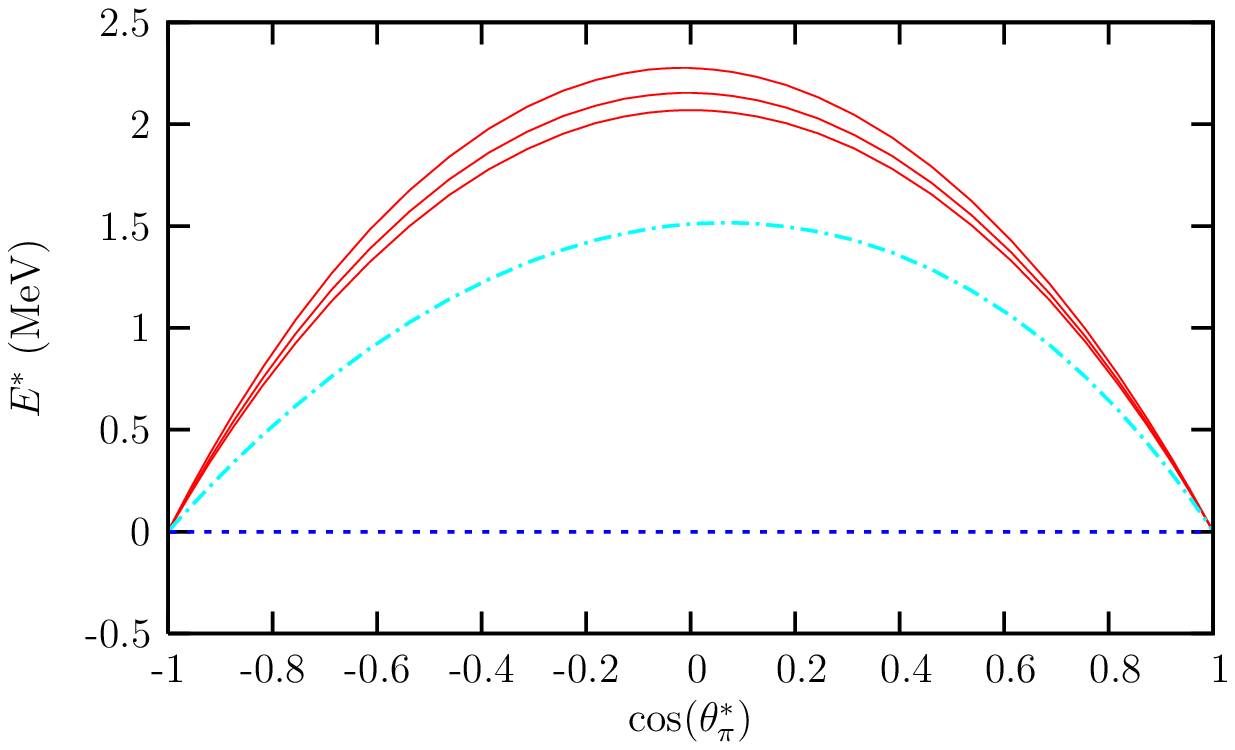}}\end{center}
\vspace{-14.5cm}
\caption{\footnotesize  Same as in Fig.~\ref{fig:ccw2-abcde} for
  $W=1150$ MeV.}
\label{fig:ccw2-abcde}
\end{figure}

Next we pay attention to the differential cross section decomposition
of Eq.~(\ref{eq:phipi*}) following the different allowed dependences
on the pion azimuthal angle. In Figs.~\ref{fig:ccw1-abcde}
and~\ref{fig:ccw2-abcde} we simultaneously compare results for the
$\nu_\mu n \to \mu^- p\pi^0$ and $\bar\nu_\mu p \to \mu^+ n\pi^0$
reactions.  Thanks to isospin symmetry (see Eq.~(\ref{eq:anti3})), the
hadronic tensor is the same for both processes and hence, they are
only distinguished by the different neutrino-- and
antineutrino--induced lepton vertices, which produces a change of sign
in the antisymmetric part of the leptonic tensor. Therefore, and
following Eqs.~(\ref{eq:phi-sym}) and~(\ref{eq:phi-antisym}), the
$C^*$ and $E^*$ structure functions are equal for both reactions,
while the antisymmetric contributions to the neutrino and antineutrino
$A^*$, $B^*$ and $D^*$ ones change sign.  We fix $E=1.5$ GeV and
$q^2=-0.5$ GeV$^2$, which naturally lies into the ANL kinematics, and
we consider two different $\pi N$ invariant masses, $W=1150$ MeV and
$W=M_\Delta$, to better understand the effect of the chiral symmetry
background terms on the structure functions defined in
Eq.~(\ref{eq:phipi*}). Results displayed in these plots show clearly
both the difference between neutrino and antineutrino structure
functions and the effect of the the chiral symmetry background terms
on them. For instance in the $W=M_\Delta$ case, neutrino and
antineutrino $A^*$ structure functions differ by about a factor of
three, which will provide a similar factor in the integrated cross
sections. In these channels, background terms have a greater influence
in the antineutrino induced process than in the neutrino one. On the
other hand, for $W=1150$ MeV the $\Delta P$ mechanism becomes
sub-dominant and the bulk of the structure functions is determined by
the background terms and their interferences with the $\Delta P$ one.

Besides, the interference between the $\Delta P$ mechanism and the
rest of the chiral background terms leads to non-vanishing $D^*$ and
$E^*$ structure functions. These functions provides dependences in
$\sin \phi_\pi^*$ and $\sin 2\phi_\pi^*$ and arise from the parity
violating terms in the hadronic tensor decomposition in
Eq.~(\ref{eq:w1}). These parity-violating contributions to the fifth
differential cross section $d^{\,5}\sigma_{\nu_l
l}/d\Omega(\hat{k^\prime})dE^\prime d\Omega(\hat{k}_\pi)$ disappear
when the pion solid angle integration is performed, as required by the
scalar, up to the factor $ |\vec{k}^\prime|/|\vec{k}~|$, nature of the
triple differential cross section $d^{\,3}\sigma_{\nu_l
l}/d\Omega(\hat{k^\prime})dE^\prime$.  Note that the coordinate system
used to define $d\Omega(\hat{k}_\pi)$ involves the pseudo-vector
$\vec{k} \times \vec{k}^\prime$ to set up the $Y-$axis, which induces
the non-parity invariant nature of
$d^{\,5}\sigma/d\Omega(\hat{k^\prime})dE^\prime d\Omega(\hat{k}_\pi)$.
In electropion production processes, the leptonic tensor is purely
symmetric, and the symmetric part of the hadronic one can not contain
terms involving the Levi-Civita tensor, since the electromagnetic
interaction preserves parity. Hence in that case
$d^{\,5}\sigma/d\Omega(\hat{k^\prime})dE^\prime d\Omega(\hat{k}_\pi)$
turns out to be a scalar under parity. Moreover, these terms also
induce $T-$odd correlations in the $(L^{(\nu,\bar\nu)})_{\mu\nu}
W^{\mu\nu}$ contraction (see Eqs.~(\ref{eq:w1})--(\ref{eq:time}) in the
Appendix~\ref{sec:phi}), which
do not imply a genuine violation of time reversal invariance because
of the existence of strong final state interaction
effects\footnote{Within our formalism, the inclusion of the $\Delta$
resonance width accounts partially for the strong final state
effects.}~\cite{KLS68,CLS70}.

\subsection{NC pion production cross sections}

There hardly exist~\cite{krenz78,anl1NC,anl2NC} NC experimental
measurements at intermediate energies.  In the Gargamelle
propane--freon experiment run at CERN~\cite{krenz78}, the NC neutrino
induced pion production cross sections, in all possible channels, were
measured at averaged neutrino energy of around 2.2 GeV, and given in
the form of NC/CC ratios. These data have been reanalyzed and absolute
cross sections, without imposing any cut in the pion-nucleon invariant
mass, have been recently provided~\cite{Hawker}.  Experiments using
the Argonne 12-foot deuterium bubble chamber were run over a neutrino
energy interval $(0.3 \le E \le 1.5)$ GeV~\cite{anl1NC,anl2NC}.  These
latter experiments gave results for the NC $\nu n \to \nu p \pi^-$
cross section~\cite{anl1NC}, as well as the NC to CC cross section
ratios $R_+=\sigma(\nu p\to \nu n \pi^+)/\sigma(\nu p\to \mu^-
p\pi^+)$, $R_0=\sigma(\nu p\to \nu p \pi^0)/\sigma(\nu p\to \mu^-
p\pi^+)$~\cite{anl2NC} and $R_-=\sigma(\nu n \to \nu p
\pi^-)/\sigma(\nu p \to \mu^- p \pi^+)$~\cite{anl1NC}.

The NC pion production reaction was proposed in Ref.~\cite{tau-nc} as
a potential tool to distinguish $\tau-$neutrinos from antineutrinos,
below the $\tau-$production threshold, but above the pion production
one. Model independent neutrino-antineutrino asymmetries both in the
totally integrated cross sections and in the pion azimuthal
differential distributions were predicted in Ref.~\cite{tau-nc}.
Results of the current model for these neutrino--antineutrino
asymmetries were presented there.  Here, we will focus on the
comparison of our results with the available experimental data as well
as emphasizing other aspects of the NC pion production processes.

In Table~\ref{tab:ratios} we compare our results for the $R_+,\,R_0$
and $R_-$ NC over CC ratios with the ANL experimental data.  Our
results are obtained for an incoming neutrino energy range of
$E=0.6-1.2$ GeV using our full model of Fig.~\ref{fig:diagramas} and
with our best fit parameters $C_5^A(0)=0.867,\, M_{A\Delta}= 0.985$
GeV. We find a fair agreement for all ratios, when the experimental
uncertainties are taken into account.
\begin{table}
\begin{tabular}{c c c} 
 & ANL & Our results \\\hline
$R_+=\sigma(\nu p \to \nu n \pi^+)/\sigma(\nu p \to\mu^-p\pi^+)$\hspace{1cm} &
  0.12 $\pm$ 0.04~\cite{anl2NC} & 0.12 -- 0.10 \\
$R_0=\sigma(\nu p \to \nu p \pi^0)/\sigma(\nu p \to\mu^-p\pi^+)$\hspace{1cm} & 
  0.09 $\pm$ 0.05~\cite{anl2NC} & 0.18 -- 0.14 \\
$R_-=\sigma(\nu n \to \nu p \pi^-)/\sigma(\nu p \to \mu^-p\pi^+)$\hspace{1cm} &
  0.11 $\pm$ 0.022~\cite{anl1NC} & 0.12 -- 0.09
\end{tabular}
\caption{NC to CC ($\nu p \to \mu^-p\pi^+$) cross section ratios. 
  Experimental data taken from the ANL analyses of Refs.~\cite{anl1NC}
  ($R_-$), and \cite{anl2NC} ($R_0$,$R_+$). Our results are 
   obtained for an incoming neutrino energy range of $E=0.6-1.2$ GeV
   using our full model of Fig.~\ref{fig:diagramas} 
and with our best fit parameters $C_5^A(0)=0.867,\,  M_{A\Delta}=
  0.985$ GeV. No kinematical cut in the $W$ invariant mass has been used.}
\label{tab:ratios}
\end{table}

In Table~\ref{tab:NCXsec} we present results for the total NC cross
sections in different channels at $E=$2.2 GeV and compare them with
the reanalysis done in Ref.~\cite{Hawker} of the original data by the
Gargamelle experiment at CERN~\cite{krenz78}.  We use our full model
of Fig.~\ref{fig:diagramas} with our best fit parameters
$C_5^A(0)=0.867,\, M_{A\Delta}= 0.985$ GeV with no upper limit in the
$W-$integration to allow a direct comparison with the experiment. The
agreement with data is good. However a word of caution is in order
here. As we discussed above, our model suffers from larger
uncertainties in the $W>1.4$ GeV region which will be accessible for
neutrinos of this energy.
\begin{table}
\begin{tabular}{c c c} 
 & \cite{Hawker} &  Our results \\\hline
$\sigma(\nu p \to \nu p \pi^0)$ \hspace{1cm} &
  0.130 $\pm$ 0.020 & 0.105$\pm$0.006 \\
$\sigma(\nu p \to \nu n \pi^+)$ \hspace{1cm} & 
  0.080 $\pm$ 0.020 & 0.091$\pm$0.003 \\
$\sigma(\nu n \to \nu n \pi^0)$ \hspace{1cm} &
  0.080 $\pm$ 0.020 & 0.104$\pm$0.006 \\
$\sigma(\nu n \to \nu p \pi^-)$ \hspace{1cm} &
  0.110 $\pm$ 0.030 & 0.082$\pm$0.003
\end{tabular}
\caption{NC cross sections in units of $10^{-38}$cm$^2$ for different
channels.  Data shown correspond to the results of a recent
reanalysis~\cite{Hawker} of the original data taken by the Gargamelle
experiment at CERN~\cite{krenz78}.  Our results are obtained for an
incoming neutrino energy of $E=2.2$ GeV using our full model of
Fig.~\ref{fig:diagramas} and with our best fit parameters
$C_5^A(0)=0.867\pm 0.075,\, M_{A\Delta}= 0.985 \pm 0.082$ GeV. No cut in the
pion--nucleon invariant mass $W$ has been applied and we Monte Carlo
propagate the latter errors to our results for the cross sections.
}
\label{tab:NCXsec}
\end{table}

In the left panel of Fig.~\ref{fig:neutral} we show our results for
the $\nu n \to \nu p \pi^-$ cross section as a function of the energy
and compare them with the ANL data of Ref.~\cite{anl1NC}. There, and
to better compare with the data, we also give results without the
$W\le 1.4$ GeV constraint. Up to incoming neutrino LAB energies of the
order of 1 GeV the implementation of this cut hardly changes the cross
section. We find a good description of the data. In the right panel of
Fig.~\ref{fig:neutral} we show the $W-$differential cross section for
the same channel and neutrinos of 1 GeV. There one can appreciate
clearly the $\Delta (1232)$ peak.  The chiral background terms
dominate the differential cross section near the pion production
threshold, and they also produce a slight shift of the maximum of the
cross section to lower invariant masses, as it also happened in the
distribution of CC events displayed in Fig.~\ref{fig:phi}. In both plots
we see that the reduction of the contribution of the $\Delta P$
mechanism, due to the diminution of the value of $C_5^A(0)$, is
partially compensated by the inclusion of the background terms.

\begin{figure}[tbh]
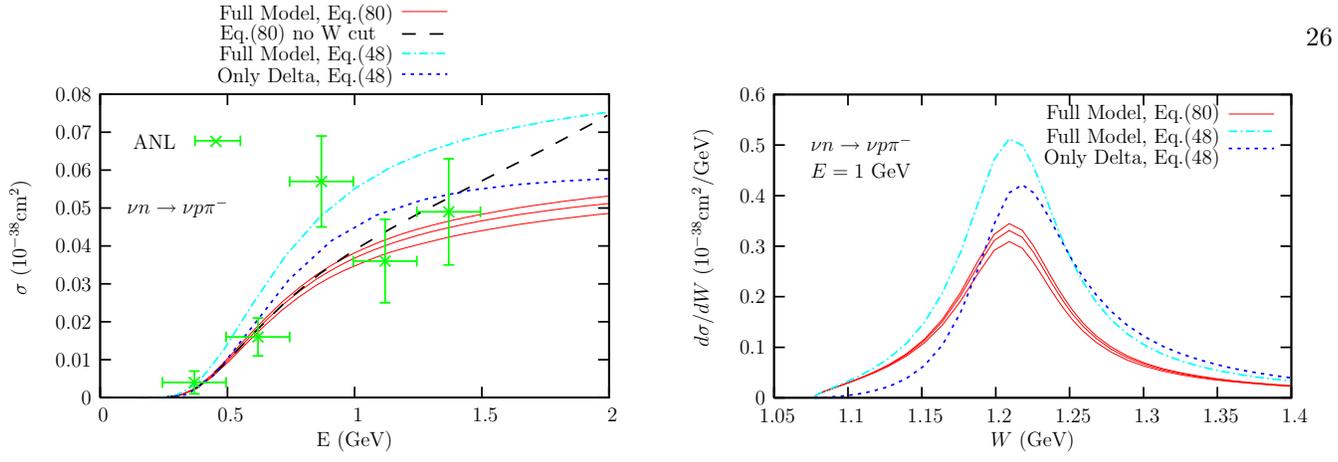

\begin{center}
\makebox[0pt]{\includegraphics[scale=0.65]{derrick.epsi}\hspace{1cm}\includegraphics[scale=0.65]{dwpim.epsi}}
\end{center}
\caption{\footnotesize Total (left) and $W-$differential (right) cross
  sections for the $ \nu n \to \nu p \pi^- $ reaction.  Left panel data
  are taken from Ref.~\cite{anl1NC} and have been measured without
  limiting the pion--nucleon invariant mass $W$. Dashed lines stand for the
  contribution of the excitation of the $\Delta$ resonance and its
  subsequent decay ($\Delta P$ mechanism) with $C_5^A(0)=1.2$ and
  $M_{A\Delta}= 1.05$ GeV.  Dashed--dotted and central solid lines are
  obtained when the full model of Fig.~\ref{fig:diagramas} is
  considered with $C_5^A(0)=1.2,\, M_{A\Delta}= 1.05$ GeV
  (dashed-dotted) and with our best fit parameters $C_5^A(0)=0.867,\,
  M_{A\Delta}= 0.985$ GeV (solid). In addition, we also show the 68\%
  CL bands (solid lines) deduced from the Gaussian correlated errors
  quoted in Eq.~(\protect\ref{eq:besfit}). In all these cases, we have
  limited the invariant mass phase-space ($W\le 1.4$ GeV). In the left
  hand side  panel the  long  dashed-line  stands for our full model results
  without including the $W\le 1.4$ GeV cut. The LAB incoming neutrino
  energy in the right hand side plot is 1 GeV. }\label{fig:neutral}
\end{figure}

\begin{figure}[tbh]
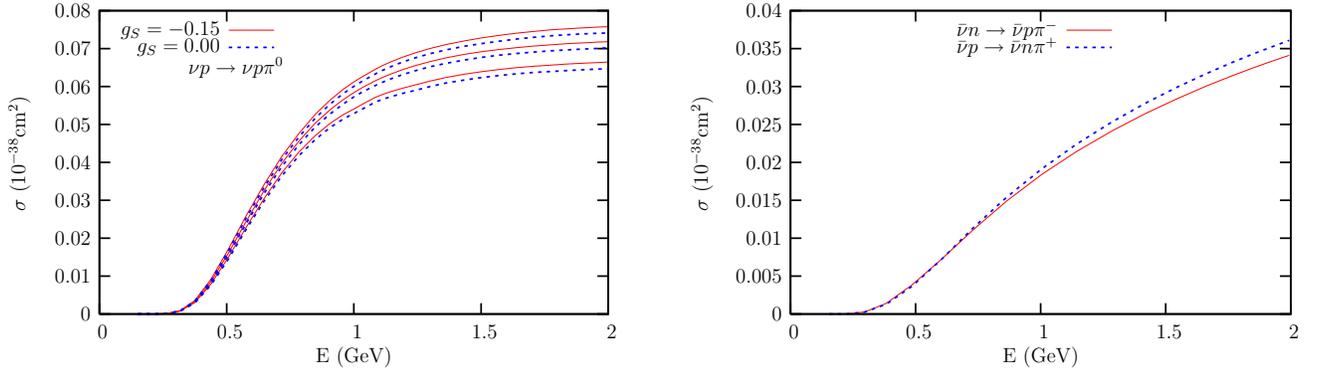

\begin{center}
\makebox[0pt]{\includegraphics[scale=0.65]{s_ppi0.epsi}\hspace{1cm}\includegraphics[scale=0.65]{anti.epsi}}
\end{center}
\caption{\footnotesize Total $\nu p \to \nu p \pi^0$ (left panel),
$\bar\nu n \to \bar\nu p \pi^-$ and $\bar\nu p \to \bar\nu n \pi^+$
(right panel) cross sections, with the $W\le 1.4$ GeV cut, as a
function of the neutrino or antineutrino energy. In the left panel we
show results obtained for two different values of the nucleon strange
content ($g_S$ in Eq.~(\ref{eq:gs})) 
with the full model of Fig.~\ref{fig:diagramas} and
our fitted $C_5^A(q^2)$ form factor. We also include the 68\% CL bands
inferred from Eq.~(\protect\ref{eq:besfit}). In the right panel we do
not give the CL bands and $g_S$ is set to
$-$0.15. }\label{fig:neutral-iso}
\end{figure}

Next, we study the effect of the strange content of the nucleon within
our model (left panel of Fig.~\ref{fig:neutral-iso}). We find that
effects are even smaller than the statistical uncertainties deduced
from the fit of the $C_5^A(q^2)$ form factor to the ANL data in
Eq.(\ref{eq:besfit}).  Similarly, results of the right panel of this
figure show that the isovector part of the NC completely dominates the
pion production reaction at the intermediate energies studied
here. This is because, as can be deduced from Eqs.~(\ref{eq:64}),
(\ref{eq:65}) and (\ref{eq:70}), both cross sections just differ in
the interference between the isovector and the isoscalar parts of the NC.
Finally, we would like to mention that antineutrino induced
cross sections are around a factor 2 or 3 smaller than neutrino
induced ones, as can be appreciated by comparing predictions for
$p\pi^-$ final state given in the left panel of Fig.~\ref{fig:neutral}
and the right panel of Fig.~\ref{fig:neutral-iso}.

\section{Conclusions}
\label{sec:concl}
We have developed a model for the weak pion production off the nucleon
driven both by CC and NC at intermediate energies, which improves most
of the existing ones.  Besides the $\Delta P$ mechanism, we have also
included some background terms, required by chiral symmetry.  Starting
from a SU(2) non-linear $\sigma$ model involving pions and nucleons,
which implements the pattern of spontaneous chiral symmetry breaking
of QCD, we derive the corresponding vector and axial currents (up to
order ${\cal O}(1/f_\pi^3)$)) which determine the structure of the
chiral non-resonant terms. Vector current conservation and PCAC are
also employed to establish some relations between the weak form
factors. In this way constructed, this model represents the natural
extension of that developed in Ref.~\cite{GNO97} for the $e N \to e'
N' \pi$ reaction.

As a result of the inclusion of the background contributions, we had
to re-fit the $C_5^A(q^2)$ form factor to the flux averaged $\nu_\mu p
\to \mu^-p\pi^+$ ANL $q^2-$differential cross section data with $W<
1.4$, finding a smaller contribution of the $\Delta P$ mechanism than
traditionally assumed in the literature (see
Eq.~(\protect\ref{eq:besfit})). We find a correction of the order of
30\% to the off diagonal Goldberger-Treiman relation
(Eq.~(\ref{eq:gt})), which we interpret is not in conflict with the
lattice QCD results shown in Figure 4 of Ref.~\cite{negele07}, if they
were extrapolated to realistic pion masses.  Within this scheme, we
have calculated several differential and integrated\footnote{There are
some inconsistencies among the ANL and BNL totally integrated cross
section data (Fig.~\ref{fig:anl-integrated}) and we would obtain a
better description of the BNL data with a $C_5^A(q^2)$ form factor
consistent with the off diagonal Goldberger-Treiman relation.}  cross
sections, including pion angular distributions, induced by neutrinos
and antineutrinos and driven both by CC and NC.  In all cases we find
that the background terms produce quite significant effects, and for
those quantities where there are experimental measurements, we find
that the inclusion of these terms brings in an overall improved
description of the data, as compared to the case where only the
$\Delta P$ mechanism is considered. We give 68\% CL bands for most of
the computed observables as deduced from the Gaussian correlated
errors quoted in Eq.~(\protect\ref{eq:besfit}). For NC reactions the
isoscalar contribution is quite small and in particular we find the
nucleon strange content effects are smaller than the statistical
uncertainties deduced from the fit of $C_5^A(q^2)$ to the ANL data.

At higher $\pi N$ invariant masses than those considered in this work,
heavier resonances than the $\Delta (1232)$ (as for example  $N(1440),
N(1535), N(1520) \cdots$) will certainly play an important
role. However, we might safely expect the contribution of
 these heavier resonances to be negligible at pion threshold,
where the chiral background terms computed in this work are dominant, and 
that it would remain quite small up to  the $\pi N$ invariant masses
around 1.3--1.4 GeV considered here~\cite{LPP06}.

We also show that the interference between the $\Delta P$ and the
background terms produces parity-violating contributions to the fifth
differential cross section $d^{\,5}\sigma_{\nu_l
l}/d\Omega(\hat{k^\prime})dE^\prime d\Omega(\hat{k}_\pi)$, which are
intimately linked to $T-$odd correlations in the
$(L^{(\nu,\bar\nu)})_{\mu\nu} W^{\mu\nu}$ contraction. However, these
latter correlations do not imply a genuine violation of time reversal
invariance because of the existence of strong final state interaction
effects.

The extension of this work to the study of the weak two-pion
production off the nucleon near threshold is natural and will be
presented elsewhere~\cite{twopion}.

\begin{acknowledgments}
We thank M.J. Vicente-Vacas and L. Alvarez-Ruso for useful
discussions.  This work was supported by DGI and FEDER funds, under
contracts FIS2005-00810, BFM2003-00856 and FPA2004-05616, by Junta de
Andaluc\'\i a and Junta de Castilla y Le\'on under contracts FQM0225
and SA104/04, and it is a part of the EU Integrated Infrastructure
Initiative Hadron Physics Project contract RII3-CT-2004-506078.
\end{acknowledgments}

\appendix 

\section{ Dependence of the neutrino differential cross section 
on the outgoing pion azimuthal angle }

\label{sec:phi}

The hadronic tensor is completely determined by up to a total of 19
Lorentz scalar and real, structure functions~\cite{tau-nc} $W_i(q^2,\,
p\cdot q,\, p\cdot k_\pi,\, k_\pi \cdot q)$,
\begin{eqnarray}
  \left(W^{\mu\nu}_{{\rm CC} \pi}\right)^{(\nu)}_{s,a} &=&
  \left(W^{\mu\nu}_{{\rm CC} \pi}\right)^{(\nu), {\rm PC}}_{s,a} +
  \left(W^{\mu\nu}_{{\rm CC} \pi}\right)^{(\nu),{\rm PV}}_{s,a} \label{eq:w1}
\end{eqnarray}
where (for simplicity from now on we drop the CC$\pi$ and ($\nu$)
labels in the notation of the hadronic tensor) 
\begin{eqnarray}
  \left(W^{\mu\nu}\right)^{\rm PC}_s &=& W_1 g^{\mu\nu} + W_2 p^\mu
  p^\nu + W_3 q^\mu q^\nu + W_4 k^\mu_\pi k^\nu_\pi + W_5 (q^\mu p^\nu
  + q^\nu p^\mu )+ W_6 (q^\mu k_\pi^\nu + q^\nu k_\pi^\mu )\nonumber
  \\ &+&W_7(p^\mu k_\pi^\nu + p^\nu k_\pi^\mu )  \\
  \left(W^{\mu\nu}\right)^{\rm PV}_s &=& W_8
  \left(q^\mu\epsilon^\nu_{.\alpha\beta\gamma}k_\pi^\alpha p^\beta
  q^\gamma + q^\nu\epsilon^\mu_{.\alpha\beta\gamma}k_\pi^\alpha p^\beta
  q^\gamma \right) + W_9
  \left(p^\mu\epsilon^\nu_{.\alpha\beta\gamma}k_\pi^\alpha p^\beta
  q^\gamma + p^\nu\epsilon^\mu_{.\alpha\beta\gamma}k_\pi^\alpha p^\beta
  q^\gamma \right)\nonumber\\ &+&
  W_{10}\left(k_\pi^\mu\epsilon^\nu_{.\alpha\beta\gamma}k_\pi^\alpha
  p^\beta q^\gamma +
  k_\pi^\nu\epsilon^\mu_{.\alpha\beta\gamma}k_\pi^\alpha p^\beta
  q^\gamma \right)\\ &&\nonumber \\ 
\left(W^{\mu\nu}\right)^{\rm PV}_a
  &=& W_{11} (q^\mu p^\nu - q^\nu p^\mu )+ W_{12} (q^\mu k_\pi^\nu -
  q^\nu k_\pi^\mu ) +W_{13}(p^\mu k_\pi^\nu - p^\nu k_\pi^\mu ) \\
  \left(W^{\mu\nu}\right)^{\rm PC}_a &=& W_{14}
  \epsilon^{\mu\nu\alpha\beta}p_\alpha q_\beta + W_{15}
  \epsilon^{\mu\nu\alpha\beta}p_\alpha k_{\pi\beta} +W_{16}
  \epsilon^{\mu\nu\alpha\beta}q_\alpha k_{\pi\beta}+W_{17}
  \left(q^\mu\epsilon^\nu_{.\alpha\beta\gamma}k_\pi^\alpha p^\beta
  q^\gamma - q^\nu\epsilon^\mu_{.\alpha\beta\gamma}k_\pi^\alpha p^\beta
  q^\gamma \right)\nonumber\\ &+&
  W_{18}\left(p^\mu\epsilon^\nu_{.\alpha\beta\gamma}k_\pi^\alpha
  p^\beta q^\gamma - p^\nu\epsilon^\mu_{.\alpha\beta\gamma}k_\pi^\alpha
  p^\beta q^\gamma \right)+
  W_{19}\left(k_\pi^\mu\epsilon^\nu_{.\alpha\beta\gamma}k_\pi^\alpha
  p^\beta q^\gamma -
  k_\pi^\nu\epsilon^\mu_{.\alpha\beta\gamma}k_\pi^\alpha p^\beta
  q^\gamma \right) \label{eq:w2}
\end{eqnarray}

The tensor $\left(W^{\mu\nu}\right)^{\rm
PV}=\left(W^{\mu\nu}\right)^{\rm PV}_s +~ {\rm i}
\left(W^{\mu\nu}\right)^{\rm PV}_a$  when contracted with the leptonic
one, $L_{\mu\nu}^{(\nu)}$, provides a pseudo-scalar quantity,
i.e., such contraction is not invariant under a parity
transformation. Indeed, under a parity transformation we have, 
\begin{equation}
L_{\mu\nu}^{(\nu)} \to \left(L^{\nu\mu}\right)^{(\nu)}, \qquad \left(W_{\mu\nu}\right)^{\rm
PV} \to -\left(W^{\nu\mu}\right)^{\rm PV} \label{eq:parity}
\end{equation}
whereas the tensor $\left(W^{\mu\nu}\right)^{\rm
  PC}=\left(W^{\mu\nu}\right)^{\rm PC}_s +~ {\rm i}
\left(W^{\mu\nu}\right)^{\rm PC}_a$ transforms as $\left
  (L^{\mu\nu}\right)^{(\nu)}$. This explains the origin of the adopted
labels PC and PV, which stand for parity violating and conserving
contributions to the fifth differential cross section
$d^{\,5}\sigma_{\nu_l l}/d\Omega(\hat{k^\prime})dE^\prime
d\Omega(\hat{k}_\pi)$, respectively. The triple differential cross
section $d^{\,3}\sigma_{\nu_l l}/d\Omega(\hat{k^\prime})dE^\prime$ is
a scalar, up to the factor $ |\vec{k}^\prime|/|\vec{k}~|$. Thus all
parity-violating contributions must disappear when performing the pion
solid angle integration.  A further remark concerns about the
time-reversal ($T$) violation effects apparently encoded in the
decomposition of the hadronic tensor in
Eqs.~(\ref{eq:w1}--\ref{eq:w2}). Under a time reversal transformation,
and taking into account the antiunitary character of the $T-$operator,
we have
\begin{equation}
L_{\mu\nu}^{(\nu)} \to \left(L^{\mu\nu}\right)^{(\nu)}, \qquad
\left(W_{\mu\nu}\right)_{\rm PC} \to \left(W^{\mu\nu}\right)^{\rm PC},
\qquad \left(W_{\mu\nu}\right)^{\rm PV} \to
-\left(W^{\mu\nu}\right)^{\rm PV} \label{eq:time}
\end{equation}
and therefore $(L^{(\nu)})_{\mu\nu} W^{\mu\nu}$ is not $T-$invariant
either, because of the presence of the PV terms in the hadronic
tensor\footnote{Note that transformations given in
  Eqs.~(\ref{eq:anti}), (\ref{eq:parity}) and (\ref{eq:time}) imply
  that the leptonic tensor, by itself, is invariant under CPT.}. This
does not necessarily means that there exists a violation of
$T-$invariance in the process because of the existence of
strong final state interaction effects~\cite{KLS68,CLS70}.  

With our election of
kinematics ($\vec{k},\vec{k}^\prime$ in the XZ plane), we find
$(L^{(\nu)}_s)_{0y} = (L^{(\nu)}_s)_{xy} =
(L^{(\nu)}_s)_{zy}=(L^{(\nu)}_a)_{0x}=(L^{(\nu)}_a)_{0z}=(L^{(\nu)}_a)_{xz}=0$,
and then
\begin{eqnarray}
\int_0^{+\infty}\frac{dk_\pi k_\pi^2}{E_\pi} (L^{(\nu)}_s)_{\mu\nu} 
W^{\mu\nu}_s &=& \int_0^{+\infty}\frac{dk_\pi k_\pi^2}{E_\pi} \Big\{ (L^{(\nu)}_s)_{00}W^{00}_s
+ 2(L^{(\nu)}_s)_{0x}W^{0x}_s + 
2(L^{(\nu)}_s)_{0z}W^{0z}_s + (L^{(\nu)}_s)_{xx}W^{xx}_s\nonumber \\ 
&+& (L^{(\nu)}_s)_{yy}W^{yy}_s + (L^{(\nu)}_s)_{zz}W^{zz}_s 
+  2 (L^{(\nu)}_s)_{xz}W^{xz}_s \Big \}\nonumber \\ 
 &=& A_s + B_s \cos\phi_\pi + C_s
\cos 2\phi_\pi+ D_s \sin\phi_\pi + E_s
\sin 2\phi_\pi \label{eq:phi-sym}\\
\int_0^{+\infty}\frac{dk_\pi k_\pi^2}{E_\pi}\ (L^{(\nu)}_a)_{\mu\nu} W^{\mu\nu}_a &=& 2 \int_0^{+\infty}\frac{dk_\pi k_\pi^2}{E_\pi}\Big\{ (L^{(\nu)}_a)_{0y}W^{0y}_a
+ (L^{(\nu)}_a)_{xy}W^{xy}_a + 
(L^{(\nu)}_a)_{yz}W^{yz}_a \Big\} \nonumber \\
&=& -A_a - B_a \cos\phi_\pi- D_a \sin\phi_\pi \label{eq:phi-antisym}
\end{eqnarray}
where we explicitly show the $\phi_\pi$ dependence\footnote{All
structure functions, $W_{i=1,\cdots 19}$, depend on the Lorentz scalar
$p\cdot k_\pi$ and $k_\pi \cdot q$ factors, which are functions of the
angle formed between the $\vec{q}$ and $\vec{k}_\pi$ vectors, and thus
they are independent of $\phi_\pi$, when $\vec{q}$ is taken along the
$Z-$axis. }. The PV term of the hadronic tensor has led to the parity
violating $\sin\phi_\pi$ and $\sin 2\phi_\pi$ contributions (all of
them proportional to $k_{\pi y})$. They disappear when the pion solid
angle integration is performed, as anticipated. The above equations 
automatically imply Eq.~(\ref{eq:phipi}).

\section{ Vector--axial resonant and resonant--nonresonant relative signs}

\label{app:phases}
Because of the poor description of the muon antineutrino cross section
data of Ref.~\cite{cern1} (see Fig.~\ref{fig:cern-ps}) achieved with
the different models studied in this work, we examined here the effect
of including relative minus signs between the axial and vector
resonant contributions and also between the $\Delta P$ and the
background terms. We have focused on the flux averaged $\nu_\mu p \to
\mu^-p\pi^+$ ANL $q^2-$differential cross section displayed in the
left panel of Fig.~\ref{fig:anl-bnlq2}, and on the total $\nu_\mu p
\to \mu^- p \pi^+$ and $\bar\nu_\mu n \to \mu^+ p \pi^-$ cross
sections given as a function of the neutrino/antineutrino energy in
the top and left panels of Figs.~\ref{fig:anl-integrated} of
~\ref{fig:cern-ps}, respectively. Results of our analysis are
presented in Fig.~\ref{fig:phases}. There, and in addition to the
results obtained from the three models presented up to now (excitation
of the $\Delta$ resonance and its subsequent decay, $\Delta P$
mechanism, with $C_5^A(0)=1.2$ and $M_{A\Delta}= 1.05$ GeV (blue
dashed--lines), and the full model of Fig.~\ref{fig:diagramas} with
$C_5^A(0)=1.2,\, M_{A\Delta}= 1.05$ GeV (cyan dashed-dotted lines)
and with our best fit parameters $C_5^A(0)=0.867,\, M_{A\Delta}=
0.985$ GeV (red solid lines)) with the set of signs deduced in our
scheme, we also show results from different choices of the relative signs.
Curves denoted by V (A) have been obtained by changing the sign of the
$WN\Delta$ vector (axial) form factors in Eq.~(\ref{eq:del_ffs}), and
similarly results denoted by VA have been obtained by including an extra
minus sign between the $\Delta P$ and the background contributions.
Changing the sign of either the vector or the axial contributions of the
$\Delta$ mechanism is strongly disfavored by the data, while
modifying the relative sign between resonant and non-resonant terms
has a little effect, as we already mentioned in the discussion of
Fig.~\ref{fig:cern-ps}.

These results provide a further confirmation of the sign convention
used in this work, and the muon antineutrino discrepancies of
Fig.~\ref{fig:cern-ps} point out to the existence of non--trivial
relative phases (and not merely minus signs) between vector and axial
resonant and non resonant contributions and/or more likely that some
nuclear medium effects were not properly discounted~\cite{singh} in
Ref.~\cite{cern1}, when providing cross sections off the nucleon from
measurements obtained in the liquid bubble chamber Gargamelle that was
filled with propane and a small admixture of heavy freon CF$_3$Br.

\begin{figure}[t]
\begin{center}
\makebox[0pt]{\includegraphics[scale=0.74]{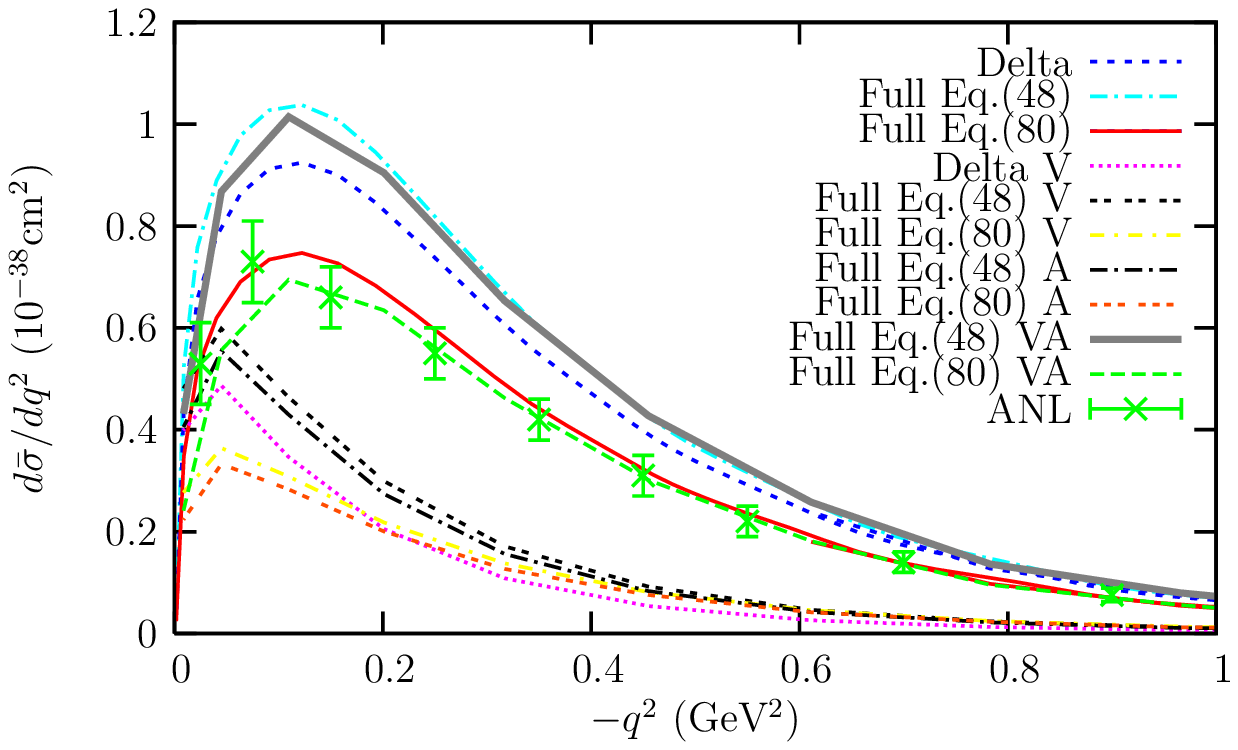}}\\
\makebox[0pt]{\includegraphics[scale=0.74]{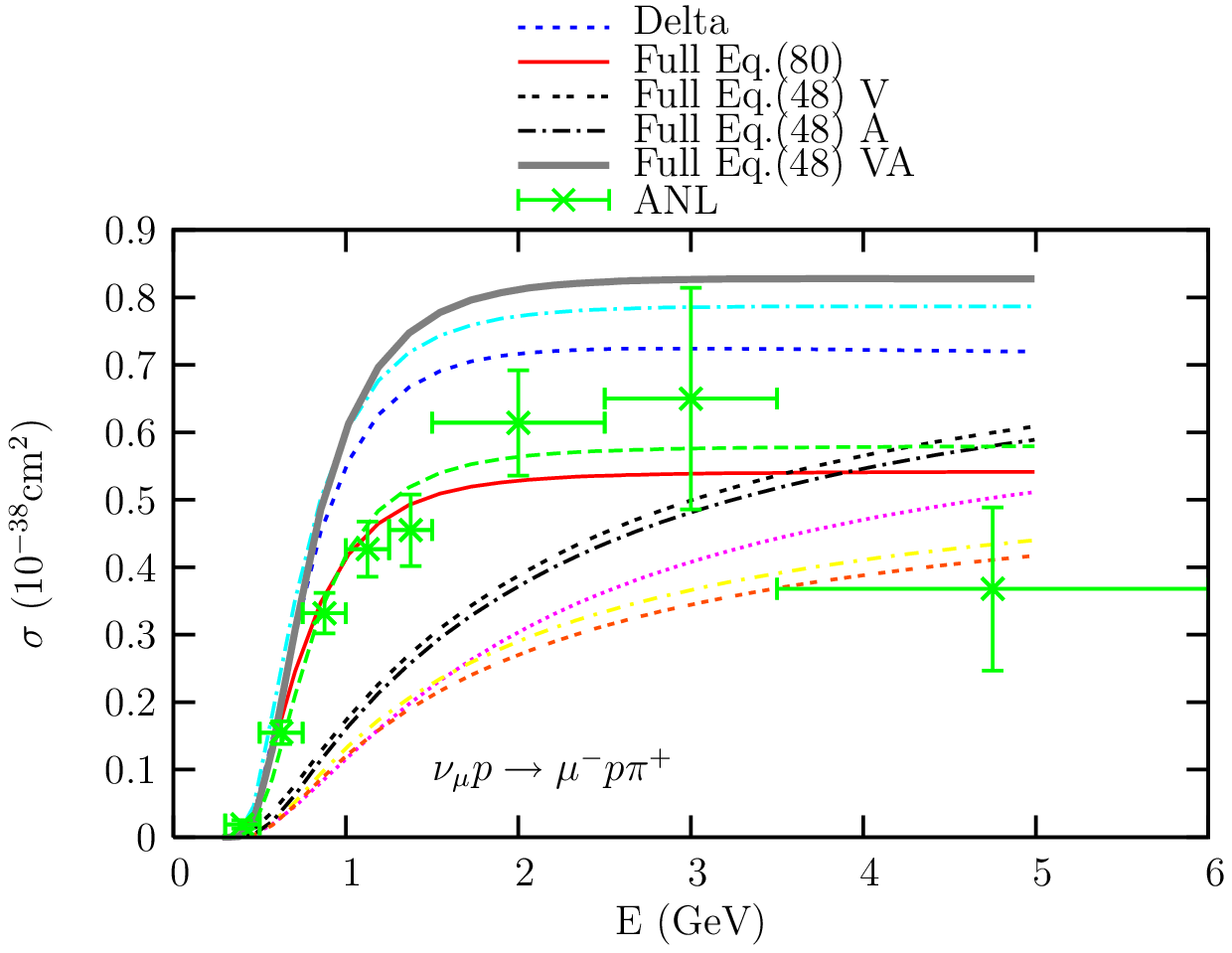}\includegraphics[scale=0.74]{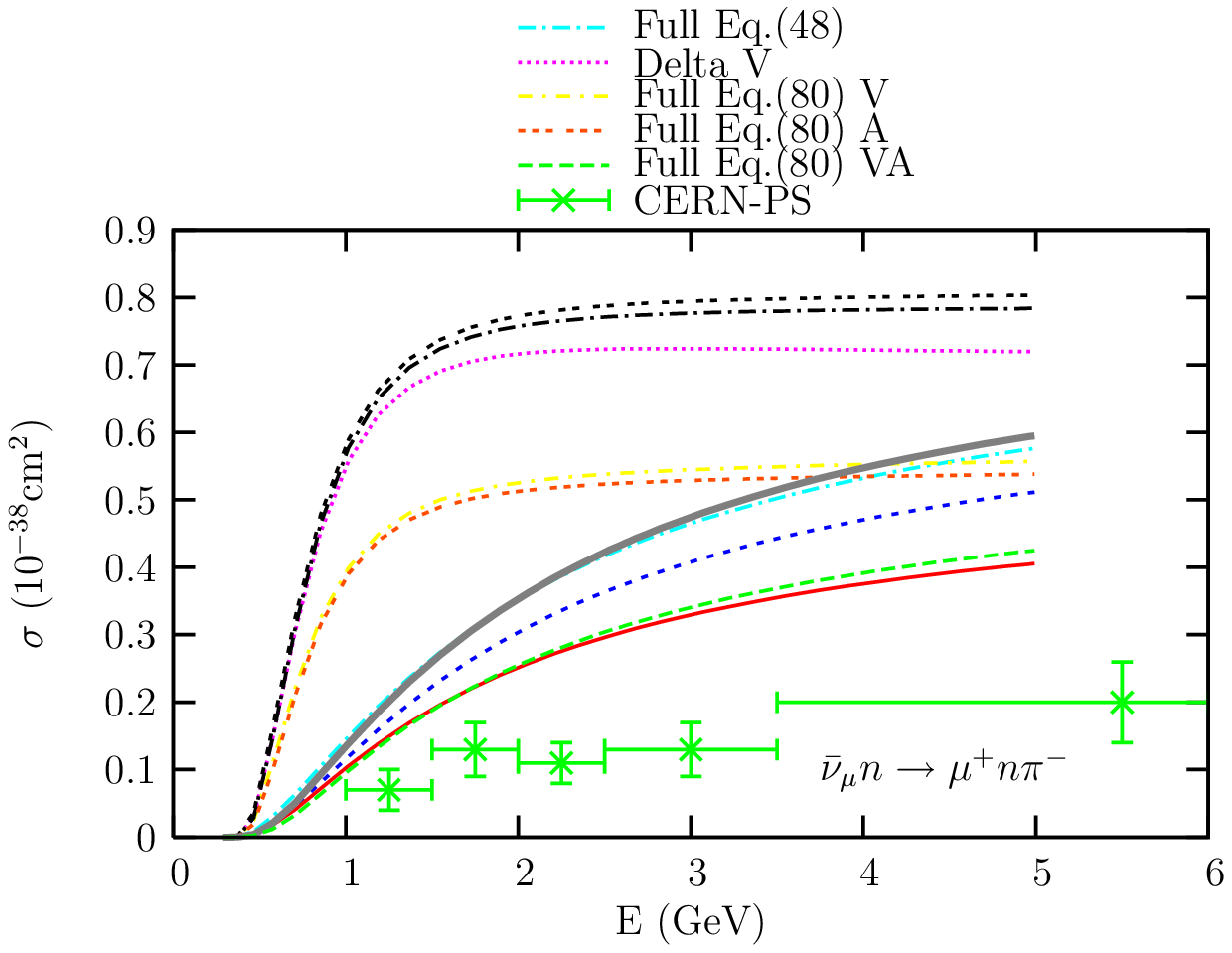}}
\end{center}
\caption{\footnotesize Flux averaged $\nu_\mu p \to \mu^-p\pi^+$ ANL
$q^2-$differential cross section (top) and total $\nu_\mu p \to \mu^-
p \pi^+$ and $\bar\nu_\mu n \to \mu^+ p \pi^-$ cross sections as a
function of the neutrino/antineutrino energy (bottom) from different
choices of the vector--axial resonant and resonant--nonresonant
relative signs (see text for details). }
\label{fig:phases}
\end{figure}

\end{document}